\documentstyle[twoside,epsf,10pt]{book}

\newif\ifcopyright \newif\iffigurespage \newif\iftablespage
\copyrighttrue \figurespagetrue \tablespagefalse

\begin{document}

\pagenumbering{roman}
\pagestyle{plain}
\thispagestyle{empty}%
\null\vskip1in%
\begin{center}
      {\Large\uppercase\expandafter{Topics in \\ Quantum Measurement and Quantum Noise}}
\end{center}
\vfill
\begin{center}
\rm By\\
\vskip.5in
Kurt Jacobs\\
\end{center}
\vskip1in
\begin{center}
      \sc A thesis submitted to the University of London\\
      for the degree of Doctor of Philosophy\\
      The Blackett Laboratory\\
      Imperial College\\
      June 1998\\
\end{center}
\vfill
\vskip.5in\newpage





\begin{minipage}{8cm}
\vspace{4cm}
\end{minipage}

\begin{center}
\begin{minipage}{8cm}
An Indian and a white man fell to talking. The white man drew a circle
in the sand and a larger circle
to surround it.

``See", he said,  pointing to the small circle ``that is what the Indian
knows, and that", pointing to the larger circle, ``is what the white man
knows."

The Indian was silent for many minutes, and then slowly he pointed with his
arm to the east and turned and waved to the west. 

``And that, white man", he said, ``is what neither of us knows".
\end{minipage}
\end{center}

\newpage


\addcontentsline{toc}{chapter}{Acknowledgments}

{\Huge\bf Acknowledgments}
\vspace{1cm}

I would like to thank Prof. P. L. Knight for his supervision, patience and encouragement during the entire length of my PhD research, and for providing helpful comments and suggestions on my thesis drafts. Working in the quantum optics group at Imperial College has been both challenging and enjoyable. I would like to thank also the many colleagues with whom I have worked throughout my time here, in particular Vlatko Vedral, Mike Rippin, Sougato Bose and Martin Plenio. Working with them has been a pleasure and a privilege. I would also like to thank Howard Wiseman, Gerd Breitenbach, Ilkka Tittonen, Sze Tan and Andrew Doherty. Their suggestions and advice, albeit mainly at a distance, was nevertheless valuable and greatly appreciated. 

There are many friends who have contributed to my time here in ways incalculable. In particular I would like to thank Peter, Nat, Nick, Brody and Dave for being there in spirit, and Mike, Brooke, Elaine, Rys and Jaimie for being there in person. Very special thanks go to my parents, Sandra, Ian and Lorraine, for always being there.

I would also like to thank the Association of Commonwealth Universities for the Commonwealth Scholarship which allowed me to study in England, and the New Zealand Vice-Chancellors' Committee for the Edward \& Isabel Kidson Scholarship and the L. B. Wood Travelling Scholarship, both of which provided very useful additional support for travel.


\newpage


\newcommand{\bq}{\begin{equation}}
\newcommand{\eq}{\end{equation}}
\newcommand{\bqa}{\begin{eqnarray}}
\newcommand{\eqa}{\end{eqnarray}}
\newcommand{\nn}{\nonumber}
\newcommand{\ms}[1]{\mbox{\scriptsize #1}}
\newcommand{\dg}{^\dagger}
\newcommand{\smallfrac}[2]{\mbox{$\frac{#1}{#2}$}}
\newcommand{\la}{\langle}
\newcommand{\ra}{\rangle}
\newcommand{\ket}[1]{| {#1} \ra}
\newcommand{\bra}[1]{\la {#1} |}
\newcommand{\ioh}{-\frac{i}{\hbar}}
\newcommand{\oh}{-\frac{1}{\hbar^2}}
\newcommand{\sch}{Schr\"odinger }
\newcommand{\hei}{Heisenberg }
\newcommand{\htn}{Hamiltonian }
\newcommand{\htnn}{Hamiltonian}
\newcommand{\half}{\smallfrac{1}{2}}
\newcommand{\bl}{{\bigl(}}
\newcommand{\br}{{\bigr)}}

\addcontentsline{toc}{chapter}{List of Publications}

{\Huge\bf List of Publications}
\vspace{1cm}

For this thesis I have selected from among my research those publications primarily concerned with quantum measurement theory. In the following chronological list, those on which this thesis is based are asterisked.

\newcounter{listnum}
\begin{list}{\arabic{listnum}.}{\usecounter{listnum}}

\item K. Jacobs, P. Tombesi, M. J. Collett and D. F. Walls,`A QND Measurement of Photon Number using Radiation Pressure', Phys. Rev. A {\bf 49}, 1961, (1994).

\item G. J. Milburn, K. Jacobs and D. F. Walls,`Quantum Limited Measurements with the Atomic Force Microscope', Phys. Rev. A {\bf 50}, 5256 (1994).

\item K. Jacobs, M. J. Collet, H. M. Wiseman, S. M. Tan and D. F. Walls,`Force Measurement via Dark State Cooling', Phys. Rev. A {\bf 54}, 2260 (1996).

\item *K. Jacobs and P. L. Knight, `Conditional Probabilities for a Single Photon at a Beam Splitter', Phys Rev. A {\bf 54}, R3738 (1996).

\item K. Jacobs, `A Model for the Production of Regular Fluorescent Light from Coherently Driven Atoms', J. of Mod. Optics {\bf 44}, 1475 (1997).

\item *K. Jacobs, P. L. Knight and V. Vedral, `Determining the State of a Single Cavity Mode from Photon Statistics', J. of  Mod. Optics, {\bf 44}, 2427 (1997).

\item S. Bose, K. Jacobs and P. L. Knight, `Preparation of non-classical states in a cavity with a moving mirror', Phys Rev. A {\bf 56}, 4175 (1997).

\item V. Vedral, M. B. Plenio, K. Jacobs and P. L. Knight, `Statistical Inference, Destinguishability of Quantum States, and Quantum Entanglement', Phys Rev. A {\bf 56}, 4452 (1997).

\item *K. Jacobs and P. L. Knight, `Linear Quantum Trajectories: Applications to Continuous Projection Measurements', Phys. Rev. A {\bf 57}, 2301 (1998).

\item S. Bose, K. Jacobs and P. L. Knight, `A Scheme to Probe the Decoherence of a Macroscopic Object', (in submission).

\item *I. Tittonen, K. Jacobs and H. M. Wiseman, `Quantum noise in the position measurement of a cavity mirror undergoing Brownian motion', (in submission).

\end{list}

\newpage


\addcontentsline{toc}{chapter}{Abstract}

{\Huge\bf Abstract}
\vspace{1cm}

In this thesis we consider primarily the dynamics of quantum systems subjected to continuous observation. In the Schr\"{o}dinger picture the evolution of a continuously monitored quantum system, referred to as a `quantum trajectory', may be described by a stochastic equation for the state vector. We present a method of deriving explicit evolution operators for linear quantum trajectories, and apply this to a number of physical examples of varying mathematical complexity.

In the Heisenberg picture evolution resulting from continuous observation may be described by quantum Langevin equations. We use this method to calculate the noise spectrum that results from a continuous observation of the position of a moving mirror, and examine the possibility of detecting the noise resulting from the quantum back-action of the measurement.

In addition to the work on continuous measurement theory, we also consider the problem of reconstructing the state of a quantum system from a set of measurements. We present a scheme for determining the state of a single cavity mode from the photon statistics measured both before and after an interaction with one or two two-level atoms.

\vspace{9cm}


\newpage





\newpage
        \tableofcontents
        \newpage
        \iftablespage
                \listoftables
                \addcontentsline{toc}{chapter}{List of Tables}
                \newpage
        \fi
        \iffigurespage
                \listoffigures
                \addcontentsline{toc}{chapter}{List of Figures}
                \newpage
        \fi
\mbox{} \newpage
        \pagenumbering{arabic}
        \pagestyle{headings}



\chapter{Introduction}
\label{chapter1}

In this chapter we introduce quantum measurement theory and explain why it is 
considered to be philosophically unsound even though it is extremely 
successful. We note that realistic measurements often consist of observing 
a system indirectly by performing a measurement on a second system with which 
the first interacts. We introduce the resulting formalism of generalised 
measurements, consisting of {\em operations} and {\em effects}. If the 
second system remains unobserved, then the first system undergoes {\em 
decoherence}, and we examine the relationship of this decoherence to the 
measurement problem. We go on to explain how dissipation and continuous 
measurement result when a small system interacts with a very large system, 
or {\em reservoir}. If we choose to monitor the reservoir, and hence the 
system, then the system undergoes an evolution referred to as a quantum 
{\em trajectory}. We show that the nature of the trajectory is determined by 
the way we choose to monitor the reservoir.

\section{Overview of the Thesis}

This thesis is concerned with the measurement of quantum systems. We focus primarily on the dynamics of quantum systems undergoing continuous measurement processes, and the resulting noise on the measurement. In addition to this main theme, we consider the problem of `state reconstruction', which involves determining a quantum state from a set of measurements, where each is performed on the state in question. In the first two chapters we introduce the aspects of quantum measurement theory, and the theory of open quantum systems that provide the background. The original work is then presented in Chapters 3, 4 and 5. Some supplementary material is also included in the three appendices.

In Chapter~\ref{chapter1} we introduce quantum measurement theory and the theory of open quantum systems. The purpose of the first part of this introduction is to present the key peculiarities of quantum mechanics, that of the measurement problem and non-locality, and to see how they are related to entanglement. The second part of the introduction presents the quantum trajectory formulation of the theory of open quantum systems in a physically motivated manner. That is, by considering a system undergoing a real continuous measurement process. As the theory of quantum trajectories developed in a number of separate ways, we also include a brief overview of the various approaches. 

In Chapter~\ref{cqtqn} we are concerned with the mathematics required to describe quantum trajectories. First we introduce stochastic calculus, which is required for the treatment of stochastic differential equations, and explain why it obeys rules that are different from those for standard calculus. We then show that quantum trajectories may be formulated in a simple manner in terms of the generalised measurements introduced in Chapter~\ref{chapter1}, and show how quantum trajectories may be written in equivalent ways as either linear or non-linear stochastic equations. In the last part of this chapter we consider the \hei picture formulation of open quantum systems, which results in quantum Langevin equations. We use these to derive the master equation which was introduced in Chapter~\ref{chapter1}.

In Chapter~\ref{evopch} we present a method for deriving evolution operators for linear quantum trajectories, and apply this to a number of physical examples. We are primarily concerned with trajectories involving the Wiener process, and which describe continuous projection measurements of physical observables. However, we also show that we may extend the method to provide a unified approach to quantum trajectories driven by both Wiener and Poisson processes.

In performing a measurement of a physical observable, the uncertainty in that observable is reduced. Due to the uncertainty principle the uncertainty in the conjugate observable is increased. If the dynamics couples the two observables together, then the increase in uncertainty feeds back into the measured observable, and the resulting noise is seen in the measurement. In Chapter~\ref{qnc} we examine an experimental realisation of a scheme for measuring this quantum back-action noise in a continuous measurement of the position of a mechanical harmonic oscillator. We consider various sources of experimental noise and examine how the back-action noise appears among these contributions.

In Chapter~\ref{cmm} we turn to the problem of the reconstruction, or measurement, of the state of a quantum system. This is possible if many copies of the state are available, so that independent measurements may be performed on each one. We present various schemes for determining the state of a single cavity mode, in which the photon statistics are measured both before and after the interaction of the mode with one or two two-level atoms.

In Chapter~\ref{conc} we conclude with a brief summary of the work presented in the thesis, and comment on the current state of quantum measurement theory, and directions for future work. 

We include, in addition, three appendices which contain material that we felt, while useful, was best separated from the main text. As we use the input-output formulation of open quantum systems, and as there do not seem to be any really detailed introductions to this topic in the literature, we have included an introduction in Appendix~\ref{appendix1}. In Appendix~\ref{apB} we include a calculation of the action of exponentials linear and quadratic in the single particle position and momentum operators, a result that we require in Chapter~\ref{evopch}. Finally, in Appendix~\ref{app3}, we present some expressions which are required for the discussion in Chapter~\ref{cmm}.

\section{Quantum Mechanics and Measurement}

\subsection{Breaking the Rules of Probability Theory}

In the early part of this century a new theory was developed to model the 
behaviour of atomic systems~\cite{EarlyQ}. Referred to as `The Quantum 
Theory' it broke from the inadequate classical theories which had 
preceded it by describing the world as existing in states that are sharply at 
variance with the states in which we perceive the world to exist. The 
relationship between quantum 
theory and measurement, or equivalently between quantum theory and our 
perception of the world, is still regarded today as a philosophical, if 
not a practical, problem.

A central peculiarity of quantum mechanics is that it breaks the rules 
of classical probability theory~\cite{Feynman}. A simple example of this is 
given by the electron interference experiment involving two 
slits~\cite{twoslit}. In this experiment an electron is fired at a barrier 
containing two adjacent slits which we will denote by {\bf slit 1} 
and {\bf slit 2}. If the electron passes through the slits it falls upon a 
screen where its position is recorded. First of all we repeat the experiment many times with {\bf slit 2} blocked 
to determine the position distribution of the electron resulting from 
passing through {\bf slit 1}. We may denote this distribution by 
$P(x|{\bf 1})$, which should be read as `the probability distribution for 
the electron to be at position $x$ on the screen {\em given} that it passed 
through {\bf slit 1}'. Similarly we may block {\bf slit 1} and determine 
$P(x|{\bf 2})$. Using the rules of probability theory, we may now determine 
the distribution of the electron on the screen when both slits are uncovered. 
Denoting the probability that the electron passes through {\bf slit 1} and 
{\bf slit 2} by $P({\bf 1})$ and $P({\bf 2})$ respectively, probability 
theory then tells us that the final probability distribution for the 
electron at the screen is given by
\bq
  P(x) = P({\bf 1})P(x|{\bf 1}) + P({\bf 2})P(x|{\bf 2}). \label{rpt}
\eq
Astonishingly, performing the experiment we obtain a completely different 
result for the distribution, one which contains the famous interference 
fringes. 
There is nothing wrong with probability theory, however. If each of the 
electrons passed through {\em either} one slit {\em or} the other, then the 
final distribution would {\em have} to be given by Eq.(\ref{rpt}). This is 
because in this case each electron could be labelled as having gone through 
either one slit or the other, and the two sets of electrons would possess the 
distributions $P(x|{\bf 1})$ and $P(x|{\bf 2})$. Then Eq.(\ref{rpt})  
follows immediately. If we set up a measuring apparatus to detect unambiguously 
which slit each electron passes through during the experiment, we discover, 
as we must, that this time the resulting distribution is indeed given by 
Eq.(\ref{rpt}). That is, the act of measuring the electrons during their 
passage disturbs them, and the experiment is altered. We are left with the 
conclusion that in the absence of the `which-path' detection apparatus, the 
electrons do {\em not} pass through one slit or the other, but must pass in 
some sense through both.

Quantum mechanics successfully models the world by allowing a physical system 
to be in a superposition of any of its many possible states. The coefficients 
in the superposition are called probability amplitudes, and the probability 
that the system is found to be in any given state is the square modulus of 
the respective probability amplitude. While in probability theory it is the 
conditional probability densities that obey the Chapman-Komolgorov equation, 
in quantum theory it is instead the conditional probability amplitudes which 
obey this equation. To describe the two-slit experiment, we assume that when 
the electron passes the slits, it exists in a superposition of being at 
{\bf slit 1} and {\bf slit 2}, with the probability amplitudes 
$\psi({\bf 1})$ and $\psi({\bf 2})$ respectively. Denoting the conditional 
probability amplitude density for the electron at the screen, after 
passing through {\bf slit 1}, as $\psi(x|{\bf 1})$, and after passing through 
{\bf slit 2} as $\psi(x|{\bf 2})$, we obtain the final probability amplitude 
density as
\bq
    \psi(x) = \psi({\bf 1})\psi(x|{\bf 1}) + \psi({\bf 2})\psi(x|{\bf 2}).
\eq
The final probability density, being the square of the probability amplitude 
density is then
\bqa
  P(x) & = & |\psi({\bf 1})|^2|\psi(x|{\bf 1})|^2 + |\psi({\bf 2})|^2 |\psi(x|{\bf 2})|^2 + 2 \mbox{Re}[ \psi({\bf 1})\psi({\bf 2})\psi(x|{\bf 1})\psi(x|{\bf 2}) ] \nonumber \\
       & = & P({\bf 1})P(x|{\bf 1}) + P({\bf 2})P(x|{\bf 2}) + 2 \mbox{Re}[ \psi({\bf 1})\psi({\bf 2})\psi(x|{\bf 1})\psi(x|{\bf 2}) ] .
\eqa
Comparing this with Eq.(\ref{rpt}) we see the appearance of an extra term, and 
it is this which provides the interference fringes.

To describe the world, quantum mechanics requires that physical systems exist 
in superpositions of their possible states. These superpositions are 
completely outside of our experience, as systems always appear to us to be in 
one state or another. It is this which lies at the heart of the quantum 
mechanical measurement problem. Somehow all the things which we sense in the 
world around us must `collapse' to one state or another, because this is how 
we experience them. But so far we have found no physical mechanism to bring 
this collapse about. In all quantum mechanical measurement theory we simply 
postulate that at some stage during the measurement process a physical 
system under observation collapses to one of its many possible states, with 
respective probabilities dictated by quantum mechanics. It is for this reason that quantum mechanical 
measurement theory is considered philosophical unsound even though is is 
extremely successful as a practical theory. So far it has predicted without fail every experiment undertaken, even though current experiments are now being performed at the level of a single atom~\cite{atomexp}. The answer to the measurement problem appears to hold some great secret as to the nature of the universe and our relation to it, but as yet no-one knows this answer.

The measurement problem is synonymous with Schr\"{o}dinger's famous thought 
experiment involving a cat~\cite{scat}. The crux of the matter is that while a 
microscopic particle appears to exist in superpositions, a macroscopic 
object such as a cat never does, and certainly we never do (have you ever 
experienced being in a superposition?). The problem is that there is no 
boundary separating a cat from a microscopic particle, there is only a 
difference of scale. So where do quantum superpositions end and macroscopic 
objects like cats begin? This is another statement of the quantum mechanical 
measurement problem. We will show below how quantum mechanical entanglement 
may be used to solve this problem in practice, although it does not provide 
a complete resolution.

\subsection{Entanglement and Decoherence} \label{EntDec}

When two or more quantum mechanical systems interact, the final state of one 
of the systems may well depend upon the final state of the others, and this 
situation is referred to as entanglement. To see how entanglement comes about 
we will need to first examine how to treat a system which is composed of 
subsystems.

Let us consider two two-state systems, denoted by $a$ and $b$, with basis 
states $|a_1\rangle$ and $|a_2\rangle$, and $|b_1\rangle$ and $|b_2\rangle$ 
respectively. The basis states of the composite system, which consists of 
both $a$ and $b$, are therefore given by the four combinations where system 
$a$ is in either of its basis states and $b$ is in either of its basis states. 
We may denote these four basis states by $|a_i\rangle|b_j\rangle\equiv 
|a_i,b_j\rangle$ where $i$ and $j$ may be 1 or 2. If we assume some 
arbitrary interaction between the two subsystems, then this is equivalent to 
an arbitrary Hamiltonian in the four dimensional space of the composite 
system which we will denote by $S=a\otimes b$. The resulting evolution will 
explore arbitrary states of $S$. In particular consider the state
\bq
  |B\rangle_{S} = \smallfrac{1}{\sqrt{2}}|a_1\rangle|b_1\rangle 
                   + \smallfrac{1}{\sqrt{2}}|a_2\rangle|b_2\rangle .
\label{bells1}
\eq 
In this case it is clear that if a measurement is performed on $a$, then the 
state of $b$ will depend upon the result. For the purposes of rigour let us 
pause to obtain the result formally. A measurement on $a$ alone corresponds 
to a projection onto a single state in the space of $a$, while projecting 
onto the entire space of $b$. The projection operator corresponding to a 
measurement resulting in the state $\ket{a_1}$ is therefore given by
\bq
   {\cal P}_a(a_1) = I_b\otimes \ket{a_1}\bra{a_1} ,
\eq
where $I_b$ is the identity operator in the space of system $b$. Because the action of $I_b$ does not alter system $b$, it is standard practice to drop it, a convention which simplifies the notation. We will follow this practice here, and write the projector simply as ${\cal P}_a(a_1)=\ket{a_1}\bra{a_1}$. Performing the measurement by operating on $|B\rangle_{S}$ with ${\cal P}_a(a_1)$, we 
obtain
\bq
   {\cal P}_a(a_1) |B\rangle_{S} = \smallfrac{1}{\sqrt{2}}|a_1\rangle|b_1\rangle .
\eq
The probability that this particular result is obtained is then the norm of 
this state (being 1/2). We see that if $a$ is found in state $|a_1\rangle$ 
then $b$ will be projected into state $|b_1\rangle$, and conversely. When the composite system is in a pure state, we can say that the two systems are entangled when a measurement on one system changes our state of knowledge about the other. When the composite system is in a mixed state, however, the definition of entanglement is more complex, and we refer the interested reader to~\cite{EntMeas}. 

We now wish to consider the relationship between the state of the combined 
system, $S$, and the states of the subsystems, $a$ and $b$. If we assume that 
$S$ is in the pure state
\bq
  |\psi\rangle_{S} = \sum_{ij}\alpha_{ij}\ket{a_{i}b_{j}}
                   = \ket{a_{1}}(\alpha_{11}\ket{b_{1}} + \alpha_{12}\ket{b_{2}})
                       + \ket{a_{2}}(\alpha_{21}\ket{b_{1}} + \alpha_{22}\ket{b_{2}}),
\label{psis}\eq
then $p_{ij}=|\alpha_{ij}|^2$ is the probability that $S$ will be found in 
the state $|a_i,b_j\rangle$. Therefore, a measurement of $a$ (in the basis 
$\{ |a_1\rangle,|a_2\rangle\}$) will result in the state 
$|a_1\rangle$ with probability $p_{11}+p_{12}$, and in the state 
$|a_2\rangle$ with probability $p_{21}+p_{22}$. If the result is to project 
$a$ into the state $|a_1\rangle$, then the state of $b$ is
\bq
  \frac{1}{\sqrt{p_{11}+p_{12}}}(\alpha_{11}\ket{b_{1}} + \alpha_{12}\ket{b_{2}}) ,
\eq
which is the state of $b$ `multiplying' $\ket{a_{1}}$ in Eq.(\ref{psis}), 
appropriately renormalised following the projection. Similarly the state of 
$b$ following a measurement of $a$ resulting in the state $|a_2\rangle$ is 
the term which `multiplies' $|a_2\rangle$ with appropriate renormalisation. 
Writing this in the density matrix formalism we have
\bq
  \rho_S = \ket{\psi}\bra{\psi}_S = \left( \begin{array}{cccc}
                 p_{11}      & c_{11}^{12} & c_{11}^{21} & c_{11}^{22} \\
                 c_{12}^{11} & p_{12}      & c_{12}^{21} & c_{12}^{22} \\
                 c_{21}^{11} & c_{21}^{12} & p_{21}      & c_{21}^{22} \\
                 c_{22}^{11} & c_{22}^{12} & c_{22}^{21} & p_{22}     \\
                \end{array}
         \right) ,
\eq
where  $c_{ij}^{kl}=\alpha_{ij}\alpha_{kl}^*$ is the `coherence' between the 
states $|a_i,b_j\rangle$ and $|a_k,b_l\rangle$. The states of $b$ resulting 
from a measurement of $a$ which produces the results $|a_1\rangle$ and 
$|a_2\rangle$, are, respectively,
\bq
     \rho_1 = \frac{1}{p_{11}+p_{12}}  \left(
              \begin{array}{cc}
                p_{11}      & c_{11}^{12} \\
                c_{12}^{11} & p_{12}     
              \end{array}  \right) \;\; , \;\;\;\;
     \rho_2 = \frac{1}{p_{21}+p_{22}}  \left(
              \begin{array}{cc}
                p_{21}      & c_{21}^{22} \\
                c_{22}^{21} & p_{22}     
              \end{array}  \right) .
\eq
That is, the top left hand corner of $\rho_S$, and the bottom right hand 
corner, respectively, both appropriately renormalised. If the measurement is 
made on $a$ by another observer who does not tell us the result, then we can 
only predict the final state with certain probabilities. We must in this 
case describe the state of $b$ as a mixture of the two possible states, each 
weighted by the probability that the measurement of the other observer produces 
them. The density matrix is therefore
\bq
   \rho =  \left(
              \begin{array}{cc}
                p_{11}      & c_{11}^{12} \\
                c_{12}^{11} & p_{12}     
              \end{array}  \right)
         + \left(
              \begin{array}{cc}
                p_{21}      & c_{21}^{22} \\
                c_{22}^{21} & p_{22}     
              \end{array}  \right) = \mbox{Tr}_a[\rho_S].
\eq
Note that the subsystems may have become separated by some arbitrary distance 
in between the interaction in which they became entangled, and the measurement 
on $a$. As a consequence, if the measurement by the (remote) observer on $a$ 
changes the state of $b$ from our point of view (that is, from the point of 
view of an observer at $b$), then this would constitute an instantaneous 
transfer of information from the observer at $a$ to the observer at $b$, and 
would be at odds with the theory of relativity~\cite{rel}. For quantum theory 
to be consistent with relativity it is therefore necessary for the observer 
at $b$ to be able to treat the state of $b$ as $\rho=\mbox{Tr}_a[\rho_S]$ 
regardless of whether or not a measurement has been performed on $a$. We will 
illustrate why this is the case using the example of the maximally entangled 
state given in Eq.(\ref{bells1}). First we note that if the state of system 
$S$ was $\ket{\psi}_S=1/\sqrt{2}(\ket{b_1}+\ket{b_2})\ket{a_1}$, then the 
density matrix would be
\bq
   \rho = \half \left(
                           \begin{array}{cc}
                               1 & 1 \\
                               1 & 1     
                           \end{array}  \right) .
\label{densup}
\eq
If, however, the combined state of $a$ and $b$ was $\ket{B}_{S}$ (given in 
Eq.(\ref{bells1})), then using $\rho=\mbox{Tr}_a[\rho_S]$ we would obtain 
the state of $b$ as
\bq
   \rho = \half \left(
                           \begin{array}{cc}
                               1 & 0 \\
                               0 & 1     
                           \end{array}  \right) .
\label{denmix}
\eq
While the density matrix in Eq.(\ref{densup}) describes a superposition of 
$\ket{b_1}$ and $\ket{b_2}$, the density matrix in Eq.(\ref{denmix}) 
describes a mixture in which $b$ is {\em either} in state $\ket{b_1}$ 
{\em or} in $\ket{b_2}$, each with a one-half probability. To see that this 
is the correct description as far as an observer at $b$ is concerned, even in 
the absence of a measurement upon $a$ so that $S$ is {\em still} in a 
superposition, consider a measurement performed on $b$ in the basis 
$\{\ket{+}_b,\ket{-}_b\}$, where $\ket{\pm}_b=(\ket{b_1}\pm\ket{b_2})/\sqrt{2}$. In the 
first case (Eq.(\ref{densup})), the result is always $\ket{+}_b$ because $b$ 
is simply in that state. Writing the projector as  ${\cal P}_b(+) = \ket{+}\bra{+}$ we obtain the probability for this result formally as
\bqa
   \bra{\psi} {\cal P}_b^{\dg}(+) {\cal P}_b(+) \ket{\psi}_S & = &
          \half \bra{a_1}(\bra{b_1}+\bra{b_2}) \;\; \ket{+}\bra{+} \;\;
          (\ket{b_1}+\ket{b_2})\ket{a_1} = \bra{a_1}\bra{+}\ket{+}\ket{a_1} 
    = 1 \nn .
\eqa
However, for the second case, we have   
\bqa
  \bra{B} {\cal P}_b^{\dg}(+) {\cal P}_b(+) \ket{B}_{S} & = &
        \half (\bra{a_1}\bra{b_1}+ \bra{a_2}\bra{b_2}) \;\; \ket{+}\bra{+} 
              \;\; (\ket{a_1}\ket{b_1}+ \ket{a_2}\ket{b_2}) \nn \\
  & = & \half \bra{a_1}\bra{b_1} \;\; \ket{+}\bra{+} \;\; \ket{a_1}\ket{b_1} 
      + \half \bra{a_2}\bra{b_2} \;\; \ket{+}\bra{+} \;\; \ket{a_2}\ket{b_2} \nn \\
  & = & \smallfrac{1}{4}+\smallfrac{1}{4} = \half  .
\eqa
We see that in the second line the orthogonality of $\ket{a_1}$ and $\ket{a_2}$ has cut the expression into two parts, each one corresponding to a different subspace of $a$. Consequently the projector onto the state $\ket{+}_b$ in $b$ acts 
separately in each subspace, there being no interference between 
subspaces. The probability of obtaining the state $\ket{+}_b$ is one-half 
in each subspace of $a$, and as the probability that the particle is in each 
of these subspaces is one-half, we obtain the total probability of one-half. 
This is, of course, precisely the result obtained with the density matrix 
given in Eq.(\ref{denmix}), which assumes that there is no interference 
between the states $\ket{b_1}$ and $\ket{b_2}$. 

We have seen now that entangling a quantum system, $b$, with another quantum 
system to which the observer has no access, destroys the interference, or 
{\em coherence} between the states of $b$ as far as the observer is 
concerned. This may be used for a partial solution of the quantum measurement 
problem, introduced in the previous section, in the sense that it explains 
why macroscopic objects do not appear to observers to exist in 
superpositions; they are entangled with many other systems constituting the 
environment in which they exist. This entanglement mechanism for a loss of 
coherence is referred to as environmentally induced decoherence (EID). 
However, EID does not actually solve the measurement problem. It does not 
explain how it is that the observer experiences one outcome or the other 
when performing a measurement on $b$. For a single outcome to result, the 
state of the composite system, $S$, must collapse, and as a result the 
measurement problem remains unsolved.

\subsection{Entanglement and Generalised Measurements}

In the previous section we examined the result of entangling two quantum 
systems from the point of view of an observer which has access to only one 
of the systems. An alternate question that we may ask is, what is the change 
in our state of knowledge about one of the systems given that we know the 
result of a measurement performed on the other. This is the theory of 
generalised measurements~\cite{Schumacher,Braginski,Kraus}. This formalism is general enough to describe every possible operation to which a system may be subjected, and is essential because real physical measurements are very rarely described accurately by direct projection measurements.

We will refer to the system which undergoes the direct projection measurement 
as the {\em probe}, and the other simply as the system. We assume that the 
system is initially in some unknown state, $\rho_i$, the probe is prepared in 
some known pure state, $\rho_{\mbox{\scriptsize p}}=\ket{\psi}\bra{\psi}_{\mbox{\scriptsize p}}$, 
and their interaction up until the measurement is described by the evolution 
operator $U$. The state of the combined system just prior to the measurement 
on the probe is then
\bq
   \rho_{S} = U\;\rho_i\otimes\rho_{\ms{p}}\;U^{\dg} .
\eq
The measurement consists of a projection onto a state in a chosen basis of 
the probe. Denoting this basis by $\{\ket{n}_{\ms{p}}\}$, where the index $n$ 
may be discrete or continuous, the unnormalised state of the system 
following the measurement is
\bq
   \tilde{\rho_n} = \bra{n}U\;\rho_i\otimes\rho_{\ms{p}}\;U^{\dg}\ket{n} = \bra{n}U\ket{\psi}_{\ms{p}} \;\rho_i\; \bra{\psi}_{\ms{p}}U^{\dg}\ket{n} = \Omega_n \;\rho_i\; \Omega_n^{\dg} ,
\label{rtor}
\eq
where we have defined the {\em measurement} operators 
$\Omega_n=\bra{n}U\ket{\psi}_{\ms{p}}$, and the tilde signifies that the 
density matrix is not normalised. The probability for obtaining this result 
is
\bq
   \mbox{Pr}[n] = \mbox{Tr}[\tilde{\rho}_n] = \mbox{Tr}[\Omega_n \;\rho_i\; \Omega_n^{\dg}] = \mbox{Tr}[\Omega_n^{\dg} \Omega_n \;\rho_i\; ] .
\eq
It follows immediately from this last equation that
\bq
  \sum_n \Omega_n^{\dg} \Omega_n = I ,
\label{DecUn}
\eq
as the sum of the probabilities for each of the possible results must be 
unity. In fact, it can be shown that this is the only restriction which the 
operators $\Omega_n$ must satisfy. The set of operators is referred to as a Positive Operator-Valued Measure, or POVM, because it associates a positive operator, $\Omega_n^{\dg} \Omega_n$, with every measurement outcome, rather than just a probability~\cite{Helstrom}. It is possible to show that given any set of operators 
$\Omega_n$ which satisfy Eq.(\ref{DecUn}), it is possible to find a 
corresponding indirect measurement which is described by that 
set~\cite{Schumacher,Kraus}. The action of the operators in the last line of 
Eq.(\ref{rtor}) transforms density matrices (bounded positive operators) to 
density matrices, and may therefore be called an {\em operation}. The positive
operator $W_n=\Omega_n^{\dg} \Omega_n$, which appears in the expression for 
the probability of detecting the result $n$, is referred to as the 
{\em effect} for that result. This general formulation of quantum 
measurements was known originally as the `theory of operations and effects'~\cite{Kraus}. However, it is common to refer to a generalised measurement simply as a `POVM'. Note that if all the $\Omega_n$ are chosen to be orthogonal projection operators, then this formalism reduces to that describing simple direct projection measurements.

The theory we have introduced above is not completely general however, 
because it assumes that we have complete information regarding the initial 
state of the probe, and complete information regarding the result of the 
measurement. A measurement for which this is the case is referred to by 
Wiseman~\cite{HMWPhD} as an {\em efficient} measurement. Because in this 
case $\rho_n$ (Eq.(\ref{rtor})) can be written as the outer product of state 
vectors, an efficient measurement on a system initially in a pure state 
results also in a pure state. To describe the more general case, in which we 
have only partial information about the initial probe state, and/or partial 
information about the result of the measurement, we need specify measurement 
operators subtended by two indices, $\Omega_{n,m}$. The first index 
specifies our possible states of knowledge after the measurement, and the 
second the various possible actual measurement results consistent with each 
particular state of knowledge. The state of the system after the measurement 
is then
\bq
   \rho_n = \sum_m \Omega_{n,m} \;\rho_i\; \Omega_{n,m}^{\dg} ,
\eq
and the set of effects for this kind of measurement is then
\bq
   W_n = \sum_m \Omega_{n,m}^{\dg} \Omega_{n,m} .
\eq
To see that this is also sufficient to describe the situation in which the 
initial state of the probe is mixed, we need only substitute a mixed state 
into Eq.(\ref{rtor}). If we were to make the measurement but ignore the 
result completely, then the final state of the system, $\rho_f$, would be a 
mixture of all the possible resulting states, weighted by their respective 
probabilities:
\bq
  \rho_f = \sum_n \mbox{Pr}[n] \frac{\tilde{\rho_n}}{\mbox{Tr}[\tilde{\rho_n}]} = \sum_n \tilde{\rho_n} = \sum_n \Omega_n \;\rho_i\; \Omega_n^{\dg} .
\eq
This is referred to as the {\em non-selective} evolution generated by the 
measurement. We note finally that the non-selective evolution, that is, 
$\rho_i$ and $\rho_f$, is not enough to determine the measurement operators. 
In particular, any two sets of measurement operators which are related by a 
unitary transformation will generate the same non-selective evolution. That 
is, given
\bq
  \Omega_m' = \sum_n c_{m,n} \Omega_n \;\; , \;\;\; \sum_l c_{l,m}c^*_{l,n} = \delta_{m,n} \; ,
\eq
we have
\bq
  \sum_m \Omega_m' \rho_i \Omega_m'^{\dg} = \sum_{m,n,l} c_{m,n} \Omega_n \rho_i c_{m,l}^* \Omega_l^{\dg} = \sum_{n,l} \delta_{n,l} \Omega_n \rho_i \Omega_l^{\dg} = \sum_{n} \Omega_n \rho_i \Omega_n^{\dg} .
\eq

\subsection{Entanglement and Non-Locality}

The existence of superpositions leads to another incredible feature of 
quantum mechanics: that it is non-local. By this we mean that the results it 
predicts cannot be explained by a theory that is real, causal, and 
{\em and also} local. The term {\em real} in this context is used to mean that 
probability distributions are non-negative, and by {\em causal} it is meant 
that later events cannot influence earlier ones. By {\em local} we mean that 
events at spatially separated points cannot affect each other instantaneously, 
in that any effect from one to the other will be delayed by at least the time 
that light takes to travel the intervening distance. However, the 
non-locality of quantum mechanics coexists peacefully with special relativity 
because it does not imply superluminal signalling. That is, it is not 
possible to use the non-locality inherent within quantum mechanics to 
communicate faster than the speed of light. 

The reason we know that the predictions of quantum mechanics cannot be 
explained by a real, causal and local theory is due to the work of Bell, who 
showed that the correlations between quantities measured at remote points 
must obey certain inequalities if they are to be explained by such a 
theory~\cite{Bell}. Because it is the correlations between remote 
measurements which are non-local in quantum mechanics, the separated 
observers must communicate their results to each other in order to see any 
non-local effect, and it is for this reason that this kind of non-locality 
cannot be used for superluminal communication.

Bell's argument runs as follows. Assume that two particles (or equivalently 
two physical systems) interact. The state of the combined systems is 
represented by some quantity ${\bf \lambda}$, and at the end of the 
interaction we allow this state to be uncertain so that it is given by some 
probability distribution $P({\bf \lambda})$. The particles are then 
separated, and a measurement is made on each. We allow the measurement at 
each location to depend on local variables $\theta$ and $\phi$. The 
assumption of locality implies that the result of the measurement on the 
first particle may depend upon the local experimental arrangement at the 
location of the first particle, given by $\theta$, but not upon the 
arrangement at the location of the second particle, which is given by $\phi$. 
Bell was able to show that under these conditions certain inequalities govern 
the correlation functions of the measurement results.

We examine now briefly a quantum optical example which violates Bell's 
inequalities and requires just a single photon~\cite{SP,Hardy}. The 
configuration consists of a single photon incident upon a beam splitter, in 
which the other input port is the vacuum. The two output modes, denoted by 
$a$ and $b$ respectively, are subjected to homodyne detection. The combined 
state of the output modes following the arrival of the photon is given by
\bq
   \ket{\psi} = \smallfrac{1}{\sqrt{2}}(\ket{0,1} + i\ket{1,0}) ,
\eq
where the ket $\ket{n,m}$, denotes $n$ photons in mode $a$ and $m$ photons in 
mode $b$. The photon can have either been reflected or transmitted, and the 
actual state is a superposition of these two possibilities. From 
subsection~\ref{EntDec} we see that this is an entangled state of the 
two modes. 
Let us now set the homodyne detector at $A$ to measure the quadrature 
$\hat{x}_\theta = \half (ae^{i\theta} + a^\dagger e^{-i\theta})$ and the 
detector at $B$ to measure $\hat{y}_\phi = \half (be^{i\phi} + b^\dagger 
e^{-i\phi})$. In this case $\theta$ and $\phi$ are the local variables upon 
which the measurement result depends. Calculating the joint probability 
distribution for the measurement results $x$ and $y$, given a choice of 
$\theta$ and $\phi$, we have 
\bq
   P(x,\theta ;y,\phi ) = |\langle \psi | x,\theta , y, \phi \rangle |^2 = \frac{4}{\pi}(x^2+y^2+2xy\sin(\theta-\phi))e^{-2(x^2+y^2)} ,
\eq
where $\ket{x,\theta , y, \phi}$ is a joint eigenstate of $\hat{x}_\theta$ 
and $\hat{y}_\phi$. The third term in the expression for the probability 
distribution shows that $x$ and $y$ are correlated, and this is a direct 
result of the interference between the two states in the superposition which 
makes up $\ket{\psi}$. We have, therefore, a correlation between measurements 
which are performed at spatially separated points. However, the crucial point 
is whether these correlations can be described by a local theory, and for 
this we turn to Bell's inequalities.

Each homodyne detector consists of a beam splitter with which the field to be 
measured is mixed with a local coherent field. The intensity of each output 
port is measured and the difference between them is proportional to the 
measured quadrature. Which quadrature is measured is determined by the phase 
of the local coherent field. By considering the four intensities which are 
measured in the two homodyne detectors as the quantities on which locality 
is imposed in the manner described above, it is possible to show that the 
correlation function~\cite{ReidW}
\bq
  E(\theta,\phi) = \frac{\langle \hat{x}_\theta \hat{y}_\phi \rangle}
                        {|\alpha |^2 + 1} ,
\eq
where $\alpha$ is the amplitude of the local coherent field, must satisfy the 
inequality
\bq
  -2 \leq E(\theta,\phi) - E(\theta,\phi') + E(\theta',\phi) + E(\theta',\phi') \leq 2 .
\eq
Evaluating the correlation function $\langle \hat{x}_\theta \hat{y}_\phi 
\rangle$ using $P(x,\theta ;y,\phi )$, we obtain
\bq
   E(\theta,\phi) = \frac{\sin(\theta-\phi)}
                        {|\alpha |^2 + 1} ,
\eq
and it is readily verified that for small $\alpha$ (in particular for 
$|\alpha |^2 < \sqrt{2} - 1$) the inequality is violated.

The investigation of non-locality does not end with Bell's inequalities 
however. By considering both direct photo-detection along with homodyne 
detection, Hardy has shown that non-locality may be seen with this system in 
a much more direct way without the need to use inequalities~\cite{Hardy}. In 
addition, Greenberger, Horne and Zeilinger have shown that when entangled 
states of {\em three} systems are used, it is possible to demonstrate 
non-locality in a single measurement, rather than with the many runs which 
are required to determine the correlation functions used in Bell's 
inequalities~\cite{GHZ}.

Recently it has been realised that remote entangled systems may be used to 
transfer quantum states from one place to another, referred to as `quantum 
teleportation'~\cite{QuanTel}, and for secure communication, or `quantum 
cryptography'~\cite{QuanCryp}. It has also been discovered that many weakly 
entangled remote systems may be used to obtain much fewer highly entangled 
systems by local operations and classical communication, a process which is 
referred to as `state purification', or `entanglement 
distillation'~\cite{EntDis}. These recent discoveries have shown us that our 
current understanding of entanglement is far from complete~\cite{EntMeas}, 
and has lead to the exciting and rapidly expanding field of quantum 
information theory.

\section{Open Quantum Systems: Master Equations}\label{d1me}

A quantum system is referred to as {\em open} when it is coupled to a large 
system possessing many degrees of freedom. In the limit in which the large 
system contains a continuum of natural frequencies, we find that the 
evolution of the reduced density matrix of the system is irreversible. We 
can understand this by considering a classical pendulum, $S$, coupled to a 
set of classical pendulums with various frequencies of oscillation. If 
pendulum $S$ is set in motion, and is coupled to just one other pendulum, 
then the energy initially possessed by $S$ oscillates back and forward between 
$S$ and the other pendulum. If $S$ is coupled to a number of pendulums with 
different frequencies, then it takes longer for all the energy initially in 
$S$ to return. At first the energy flows to the other pendulums, and will 
return back to $S$ from each at different times due to the different 
oscillation frequencies. Thus for all the initial energy to return to $S$ we 
must wait until the various oscillations have come back in synchronisation. 
Clearly as the number of oscillators increases and their respective 
frequencies become closer together the time that it takes for the energy to 
return to $S$ increases. In the limit of a continuum of frequencies the 
initial energy never returns to $S$ and we have an irreversible dissipative 
evolution. In addition, we saw in subsection~\ref{EntDec} that the entanglement 
of the small system with the large system, which is the result of the 
interaction, will cause the reduced density matrix of $S$ to decohere. Hence 
open systems undergo both irreversible dissipation and decoherence. In the 
theory of open systems the large system is referred to as a (heat) {\em bath} 
or {\em reservoir}, and models the environment of the small system. Naturally 
all systems interact with their environment, and the study of open systems is 
important in all cases in which this interaction cannot be ignored.
 
When the correlation time (memory) of the reservoir is much shorter than the time scale of the dynamics of $S$, a Markovian equation may be derived for the reduced density matrix of $S$ by tracing over the reservoir, and this is referred to as a {\em master equation}. The exact form of the master equation 
will depend upon the nature of the system-reservoir interaction. In 1976 Lindblad showed that in order to preserve the trace and positivity of the density matrix, a master equation must be able to be written in the form
\bq
  \dot{\rho} = \ioh [H,\rho] + \sum_{n}{\cal D}[c_n]\rho ,
\label{lme}
\eq
where ${\cal D}[c_n]$ is the superoperator defined as
\bq
  {\cal D}[c_n] \equiv c_n \rho c_n^{\dg} - \half c_n^{\dg} c_n \rho - \rho \half c_n^{\dg} c_n ,
\eq
and the $c_n$ are any bounded operators~\cite{Lindblad}. This is now referred to as the Lindblad form. Each superopertor ${\cal D}[c_n]$ represents a source of decoherence, and may also represent a source of loss. The master equation describing a single mode of an optical cavity in which one of the end mirrors is partially transmitting, and therefore possesses one source of loss, is
\bq
  \dot{\rho} = \ioh [H_c,\rho] + \gamma {\cal D}[a]  ,
\label{cme2}
\eq
where $H_c=\hbar\omega_c a^\dagger a$ is the free Hamiltonian of the cavity mode, $a$ is the annihilation operator describing the cavity mode, $\omega_c$ is the cavity mode frequency, and $\gamma$ is the rate at which photons leave the cavity via the partially transmitting mirror. To see that this master equation does describe the irreversible loss of photons from the cavity, we may use it to obtain the equation of motion for the average number of photons in the cavity, which is
\bq
  \frac{d}{dt} \langle a^{\dg} a \rangle = \mbox{Tr}[\dot{\rho}a^{\dg} a] =  -\gamma \langle a^{\dg} a \rangle ,
\eq 
and this clearly identifies $\gamma$ as the photon decay rate. We have now said all we wish to say about master equations at this point. We will derive the master equation for a lossy optical cavity in section~\ref{sqn} in the next chapter, via the use of quantum Langevin equations. Standard derivations using the \sch picture may be found in references~\cite{Carm,BHE,WandM}. 

\section{Open Quantum Systems: Quantum Trajectories}
\subsection{Direct Detection}

Master equations, which we introduced in the previous section, are adequate to describe the evolution of an open system if the observer has no access to the environment. In this section we describe how we may determine the evolution of the state of the system when we have information about the environment. For the case of a damped cavity this means performing measurements on the light that leaks out. In this description the state of the system will therefore be conditioned on the results of our observation of this output light. We describe in what follows the treatment of Carmichael~\cite{Carm,Cetc}, and continue to use the damped cavity as an example.

Consider an open system consisting of a cavity which is damped through just one end-mirror. The light travelling away from the system may be written as the sum of a contribution emitted by the cavity through the end-mirror, and the reflection of the incident field. In particular, if we examine the field just outside the lossy mirror, we may write
\bq
  b_{\ms{out}}(t) = b_{\ms{sys}}(t) - b_{\ms{in}}(t) ,
\eq
where $b_{\ms{out}}(t)$ is the negative frequency part of the electric field 
operator for the field travelling away from the system, $b_{\ms{in}}(t)$ is 
the equivalent operator for the field travelling toward the system (the input 
field), and $b_{\ms{sys}}$ is the contribution radiated by the system. In particular we have $b_{\ms{sys}}(t)=\sqrt{\gamma} a(t)$. For a detailed derivation of this relation, the reader is referred to Appendix A. Glauber's theory of photo-detection gives the average photo-counting rate as proportional to the expectation value of the product of the positive and negative frequency components of the electric field operator, and the input and output field operators in the above equation have been scaled such that the detection rate for a perfectly efficient detector is given by $\langle b_{\ms{out}}(t)b^{\dg}_{\ms{out}}(t) \rangle$.

We want to consider the detection of the light radiated by the system, so we 
will choose the external field to be in the vacuum state, and as a
consequence the photon detection rate is given by the system contribution
alone. Let us consider now the exclusive probability densities for the 
detection of photons. By this we mean the probability density that in a time interval $[t,t+T]$ we detect a photon at the times $t_1,t_2,\ldots t_m$ and no times in between, and let us write these as $p_m(t_1,t_2,\ldots t_m;[t,t+T])$. Carmichael was able to show, using photo-detection theory~\cite{Saleh,Gl}, that the exclusive photo-detection probability densities may be written as 
\bq
  p_m(t_1,t_2,\ldots t_m;[t,t+T]) = Tr[e^{({\cal L-J})(t+T-t_m)}{\cal J}\ldots {\cal J}e^{({\cal L-J})(t_2-t_1)}{\cal J}e^{({\cal L-J})(t_1-t)}\rho(t)] ,
\label{pftraj}
\eq
where ${\cal L}$ is the superoperator which gives the time evolution of the 
reduced density matrix of the small system, (for the cavity this is the 
master equation given by Eq.(\ref{cme2})), and ${\cal J}$ is the 
superoperator defined by ${\cal J}\rho\equiv b_{\ms{sys}}(0)\rho 
b_{\ms{sys}}(0)$, which, for the damped cavity is given by $\gamma a \rho 
a^\dagger$. Here, and in what follows, operators without a time argument denote \sch picture operators. This expression leads directly to a treatment of the evolution 
of the system in the following way. Let us assume that we have the density 
operator for the state of the system at time $t$, given that we have detected 
photons at times $t_1,t_2,\ldots t_m$ in the interval $[0,t)$, and let us 
call this density operator $\rho_{\ms{c}}$. The probability density for the 
detection of a photon at time $t$ should, by the formula presented above, be
given by $p_m(t|t_1,\ldots,t_m)=\mbox{Tr}[{\cal J}\rho_{\ms{c}}]$. We can now
find $\rho_{\ms{c}}$ by calculating this conditional probability density by 
using the ratio of two probability densities:
\bq
  p_m(t|t_1,\ldots,t_m) = \frac{p_{m+1}(t_1,t_2,\ldots t_m,t;[0,t))}{p_m(t_1,t_2,\ldots t_m;[0,t))}  = \mbox{Tr}[{\cal J}\rho_{\ms{c}}] .
\eq
From this we obtain the conditioned density matrix at time $t$, which is 
$\rho_{\ms{c}} = (\tilde{\rho}_{\ms{c}})/(\mbox{Tr}[\tilde{\rho}_{\ms{c}}])$, where the unnormalised density matrix is given by
\bq
   \tilde{\rho}_{\ms{c}} = e^{({\cal L-J})(t-t_m)}{\cal J}\ldots {\cal J}e^{({\cal L-J})(t_2-t_1)}{\cal J}e^{({\cal L-J})t_1}\rho(0) .
\label{rhoun}\eq
We see that we may interpret this equation for the conditioned density matrix 
in the following way. In between photo-detections the density matrix evolves 
according to the evolution super-operator $\exp[({\cal L-J})\tau]$, where 
$\tau$ is the time between detections, and upon a detection the density 
operator `jumps' by the action of the super-operator ${\cal J}$. In this 
case we have perfect detection, and the two types of evolution preserve the
purity of an initially pure state. As a consequence, choosing an initially pure 
state allows us to write a trajectory evolution for the system state vector. 
It is readily shown that the evolution by $\exp[({\cal L-J})\tau]$ is 
equivalent to evolution by the non-Hermitian \htn, $H_{\ms{eff}}=H - i\hbar 
b^{\dg}_{\ms{sys}} b_{\ms{sys}}/2$ and the quantum `jump' is given by the 
action of $b_{\ms{sys}}$. At time $t$ the probability per unit time for a jump to occur is given by $\mbox{Tr}[\rho_{\ms{c}}(t)]=
\langle b_{\ms{sys}}(t)b^{\dg}_{\ms{sys}}(t)\rangle$. This gives us a method
with which to simulate the evolution of the conditioned density operator. We
may chop time into small intervals, and during each calculate the probability
that a detection occurs. If it does we apply the jump operator and renormalise
the state vector. If it does not we evolve using the non-Hermitian \htn and
also renormalise appropriately. In fact, we may use this simulation process 
to cast the 
description of the evolution of the state vector into quite a different form 
by deriving a stochastic \sch equation as follows. We define a stochastic
increment, $dN$ which is usually zero, but takes the value unity at a 
random set of discrete points. This is described as a {\em point process}. 
The probability per unit time for $dN$ to be unity is given by 
$\langle b_{\ms{sys}} b^{\dg}_{\ms{sys}} \rangle$. If $dN$ is 
zero then in an infinitesimal time interval $dt$ the change in the state
vector is
\bq
  |\psi(t+dt)\rangle = |\psi(t)\rangle + dt \left( \frac{\langle b^{\dg}_{\ms{sys}}(t) b_{\ms{sys}}(t) \rangle}{2} \ioh H - \frac{b^{\dg}_{\ms{sys}} b_{\ms{sys}}}{2} \right) |\psi(t)\rangle ,
\eq
where $H$ is the \htn of the system, and to derive this expression we have
simply written down the re-normalised state after evolution through $dt$ 
using the non-Hermitian \htn and expanded to first order in $dt$. In the 
case that $dN=1$ the new state vector is
\bq
  |\psi(t+dt)\rangle = \frac{b_{\ms{sys}}|\psi(t)\rangle}{\sqrt{\langle b^{\dg}_{\ms{sys}}(t) b_{\ms{sys}}(t)\rangle}} .
\eq
The stochastic \sch equation may now be written as the combination of these    
two possibilities:
\bq
 d|\psi(t)\rangle = \left[ dt \left( \frac{\langle b^{\dg}_{\ms{sys}}(t) b_{\ms{sys}}(t) \rangle}{2} \ioh H - \frac{b^{\dg}_{\ms{sys}}b_{\ms{sys}}}{2} \right) + dN(t)\left( \frac{b_{\ms{sys}}|\psi(t)\rangle}{\sqrt{\langle b^{\dg}_{\ms{sys}}(t) b_{\ms{sys}}(t)\rangle}} - 1 \right)\right] |\psi(t)\rangle .
\label{se1} 
\eq
Note that this equation is nonlinear because $\langle b_{\ms{sys}}(t) 
b^{\dg}_{\ms{sys}}(t) \rangle$ is a non-linear function of 
$|\psi(t)\rangle$. This stochastic equation gives the evolution of the 
state vector conditioned on the measurement record, and the equivalent master  
equation gives the unconditional evolution of the density matrix, being the 
average over all possible measurement records. Consequently, the master
equation is readily recovered from the stochastic \sch equation by averaging
over the possible values of $dN$ for each infinitesimal time interval $dt$. 

Before we complete this section, we wish to note that the equation for the 
unnormalised conditional density matrix, 
Eq.(\ref{rhoun}), provides an alternative formulation of quantum trajectories
in terms of a {\em linear} stochastic \sch equation. We see from this 
equation that the unnormalised conditional density matrix is given by the
action of linear operators on the initial density matrix, and we can therefore
write a linear stochastic \sch equation for the equivalent unnormalised 
state vector using the same procedure as for the derivation of the 
non-linear version above, but this time not bothering to normalise at
each step. The resulting stochastic \sch equation is
\bq
 d|\tilde{\psi}(t)\rangle = \left[ dt(-iH/\hbar -b^{\dg}_{\ms{sys}}b_{\ms{sys}}/2)
                                 + dN(t)\left( b_{\ms{sys}} - 1 \right) 
                            \right] |\psi(t)\rangle ,
\eq
where we use a tilde to denote the unnormalised state vector. While this
equation now looks linear, a non-linearity remains in that the 
probability per unit time for the stochastic increment, $dN(t)$, still depends 
in a non-linear manner upon the state vector. However, it turns out that we 
may also eliminate this non-linearity by simply choosing a fixed probability
per unit time for the stochastic increment. This means that the stochastic 
equation will no longer choose the trajectories with the correct 
probabilities. Nevertheless, this equation provides a complete description of 
the quantum trajectories because the correct probabilities can be obtained
from the norm of the final unnormalised state vector, as indicated by Eq.(\ref{pftraj}). The linear formulation is useful as it is suggests methods of solution which are obscured in the non-linear formulation.

In this section we have shown how a consideration of direct photo-detection 
leads to a formulation of the evolution of a system in terms of a quantum
trajectory, which is said to {\em unravel} the master equation. In this case
we could write a stochastic \sch equation for the state of the system in 
terms of a point process which describes the photo-detections. In the 
following section we show how a different kind of measurement process leads to  
a different unravelling of the master equation in terms of a Gaussian
stochastic process.
  
\subsection{Homodyne Detection}
\label{HD}
Homodyne detection consists of mixing the field to be measured with a local
coherent field at a beam splitter, and then performing photo-detection on the
resulting field. The photo-current then contains information about the 
quadrature operators rather than just the field intensity, as is the case
with direct photo-detection. The beam splitter is arranged so that the field
to be measured is transmitted into the output which is detected, and the
local oscillator is reflected into this output. The detected field is then
given by 
\bq
  \varepsilon(t) = \sqrt{\gamma}(\sqrt{\eta}a(t) - i\sqrt{1-\eta}\alpha) ,
\eq
where $\eta$ is the transmission coefficient for the beam splitter, and we
have been able to replace the cavity mode operator for the local oscillator by
the coherent amplitude $\alpha$ because a coherent state is a eigenstate of
this operator. We now assume the limit in which both the transmittivity of the  
beam splitter and the coherent amplitude of the local oscillator are large so
that we may approximate $\eta\approx 1$ and set $\beta=\sqrt{1-\eta}\alpha$. 
The jump operator for the quantum state is now $C= \sqrt{\gamma}(a(t) 
+ i\beta)$, and the average rate of photo-detection is 
\bq
  \langle C^\dagger(t)C(t) \rangle = \gamma \beta^2 + \gamma \beta \langle Y(t) \rangle + \gamma \langle a^\dagger(t)a(t) \rangle ,
\eq
where we have chosen the phase of the local oscillator so that $\beta$ is real. With this choice it is the phase quadrature $Y=-i(a-a^{\dg})$ that is measured. Selecting an arbitrary phase for the local oscillator allows the measurement of an arbitrary quadrature. The effective \htn which generates the smooth evolution between jumps is
\bq
  H_{\ms{eff}}=H - i\hbar\gamma(a^\dagger a - 2i\beta a + \beta^2)/2 .
\eq

The photo-detection rate due to the local oscillator 
($\beta$) is taken to be much larger than the rate for the measured field, so
that the term $\langle a^\dagger(t)a(t)\rangle$ may be ignored in the 
photo-current signal. In this case, apart from the constant rate of detection 
due to the local oscillator (which may be subtracted), the photo-current 
signal is proportional to the phase quadrature of the measured field.
Naturally the measurement process could be simulated as in the previous
section, with the jump operator applied at each photo-count. However, as we
have chosen the local oscillator to have a much larger intensity than the measured
field, most of the photo-counts are due to the local oscillator, and cause
only a small change in the state of the system. On the time scale upon which 
the system changes, the photo-counts look therefore more like a continuously
varying signal. It would also be more advantageous to represent the 
photo-current as a continuous, albeit stochastic, signal, rather than as a 
series of photo-detections. The method we now use to achieve this was due 
originally to Carmichael, and was re-worked more rigorously by Wiseman and 
Milburn~\cite{SHM}.

Consider the evolution of the state vector over a time increment $\Delta t$,
in which the number of photo-detections, $m$, is much greater than unity. In 
particular we have $\langle m\rangle \approx \gamma\beta^2\Delta t \gg 1$. After 
this time step the unnormalised state of the system is given by the action
of $m$ jump operators, and $m+2$ smooth evolution operators for a time 
totalling $\Delta t$. As we wish to take the limit in which 
$\Delta t\rightarrow 0$, (but in which we scale $\beta$ with $\Delta t$ 
so that $\langle m\rangle$ tends to infinity), we may swap the order of the
smooth evolution and jump operators because they all commute to first order
in $\delta t$. This allows us to write
\bq
  |\tilde{\psi}(\Delta t)\rangle = \left[ \left(\frac{i\sqrt{\gamma}}{\beta}\right)^m e^{-\gamma\beta^2\Delta t/2} \right]  
  e^{-[iH/\hbar+\gamma a^\dagger a/2 - i\gamma\beta a]\Delta t }\left(1 - i\frac{a}{\beta} \right)^m |\psi(0)\rangle .
\eq
Careful inspection of the first part of this expression (the part in the 
square brackets) shows that it does not go to unity as $\langle m\rangle$
goes to infinity. However, as it is just a number which affects the 
normalisation and overall phase we may discard it. The rest of the expression
operating on the initial state tends to unity as $\Delta t\rightarrow 0$
so long as we scale $\gamma \Delta t$ and $\beta$ to ensure that while 
$\langle m\rangle \approx \gamma\beta^2\Delta t \rightarrow \infty $, we have
$\gamma\beta\Delta t \rightarrow 0$. Now we need to consider the statistics
of the number of counts, $m$. This is naturally governed by a Poisson 
distribution which may be well approximated by a Gaussian when the mean is 
large. In particular we may write 
\bq
  m=\gamma \beta^2 \Delta t + \gamma \beta \langle Y(t) \rangle \Delta t + \sqrt{\gamma}\beta \Delta W ,
\eq
where $\Delta W$ is Gaussian distributed with zero mean and variance 
$\langle \Delta W^2 \rangle = \Delta t$. Expanding the expression for 
$|\tilde{\psi}(\Delta t)\rangle$ to first order in $\gamma\beta \Delta t$, and 
replacing the increments with infinitesimals, we obtain a stochastic \sch 
equation for the unnormalised state vector. This is
\bq
  |\tilde{\psi}(t + dt)\rangle = \left[ 1 - \left( \frac{i}{\hbar}H + \frac{\gamma}{2}a^\dagger a + ia\gamma\langle Y\rangle  \right) dt - ia\sqrt{\gamma} dW  \right] |\tilde{\psi}(t)\rangle ,
\eq
where the integrated photo-current is 
\bq
 N(t) = \gamma \beta \int_0^t (\beta + \langle Y(t')\rangle) \; dt' + \sqrt{\gamma}\beta\int_0^t dW .
\eq
In these equations $dW$ is the Wiener increment, and requires the use of  
stochastic (Ito) calculus which will be introduced in the next chapter. Note
that this equation is for the unnormalised state vector. However, an
equation for the normalised state vector, the equivalent of Eq.(\ref{se1}) 
for homodyne detection, is easily derived by including in the equation a 
renormalisation at each time increment. Calculating the norm of 
$|\tilde{\psi}(\Delta t)\rangle$ we have
\bq
 \langle \tilde{\psi}(\Delta t)|\tilde{\psi}(\Delta t)\rangle = 1 + \gamma \langle Y \rangle^2 \Delta t + \sqrt{\gamma}\langle Y \rangle \Delta W .
\eq
Dividing $|\tilde{\psi}(\Delta t)\rangle$ by the square root of the norm, and 
expanding to first order in $\Delta t$ we obtain the stochastic \sch equation
for the normalised state vector as
\bqa
  |\psi(t + dt)\rangle & = & \left[ 1 - \left( \frac{i}{\hbar}H + \frac{\gamma}{2}(a^\dagger a + \langle Y(t) \rangle^2/4) + ia\gamma\langle Y(t)\rangle  \right) dt - (ia + \langle Y(t) \rangle/2)\sqrt{\gamma} dW  \right] |\psi(t)\rangle . \nn
\eqa
As with the stochastic \sch equation based on the Poisson process 
(Eq.(\ref{se1})) the density matrix at time $t$ is given by the average over
the states generated by all the possible trajectories. We may readily recover
the master equation by averaging $|\psi(t + dt)\rangle\langle\psi(t + dt)|$
over the possible values of the stochastic increment $dW$.

\subsection{Other Approaches}
In presenting quantum trajectories we have used the treatment of 
Carmichael to make it clear that they may be obtained from physical 
considerations such as the measurement of the field radiated by the 
system. However, the treatment of systems in terms of quantum trajectories 
did not develop in this way. We complete our discussion of this topic 
by giving a brief overview of the various approaches which have been used 
to arrive at quantum trajectories and stochastic \sch equations.

The beginning of the quantum jump formulation can be traced back to the Photo-detection theory of Srinivas and Davies~\cite{SD} in 1981, to the calculation of waiting-time distributions for the emission of photons in resonance fluorescence by Cohen-Tannoudji and Dalibard~\cite{CD} in 1986, and the more general calculation of exclusive probability densities for this case by Zoller {\em et al.}~\cite{ZMW} in 1987. The approach of Srinivas and Davies was to develop a description of photo-detection from the theory of generalised measurements. The quantum jump description was essentially completely captured by their theory, although it was not clear how it was connected to a physical measurement process, and they did not employ it as a procedure to simulate the master equation. The approach of Zoller {\em et al.} was quite different. In 1986, Cohen-Tannoudji and Dalibard~\cite{CD} had suggested using waiting-times rather than second order correlation functions to examine resonance fluorescence. Zoller {\em et al.} built on this by using a method for treating resonance fluorescence developed by Mollow~\cite{Mollow} to derive expressions for the exclusive emission probability densities. They found that the expressions they derived had the mathematical form of the Srinivas and Davis theory. In 1992 Hegerfeldt and Wilser~\cite{HW} and Dalibard {\em et al.}~\cite{DCM} showed that it was possible to simulate a master equation using a stochastic wavefunction evolution in which jumps occurred at the emission times (ie. a wavefunction equivalent of a Monte-Carlo method), and the quantum jump formulation proper was born. Very quickly this became widely used because it meant that a pure state could be employed in the simulation rather than a density matrix, reducing the required memory.

The formulation in terms of wavefunction stochastic equations driven by the Wiener process was first formulated by Gisin~\cite{Gisin84} in 1984, using formal measurement theory arguments, and a little later by Diosi~\cite{DiosiM}, Belavkin~\cite{Belvkn} and Barchielli~\cite{Barch}. The latter applied the theory to homodyne and heterodyne detection, and so in this sense his treatment is closest to Carmichael's formulation. Rather than just interpreting quantum trajectories as the evolution of a system during a measurement process, Percival~\cite{GP} has also suggested that stochastic equations might be used to solve the measurement problem. If they were viewed as a fundamental theory of quantum dynamics, they would provide an intrinsic source of collapse.

In 1992 Carmichael showed that the detection theory of Srinivas and Davies, and hence the quantum jump formulation, could be derived from the standard theory of photo-detection~\cite{Saleh,Gl}, and in a sense this completed the theory. Carmichael's treatment has been filled out and made more rigorous by Wiseman and Milburn~\cite{SHM}, and in the last few years the theory has been applied by many authors~\cite{PK}. 

\subsection{A Continuous Measurement of Position}
In Chapter~\ref{evopch} we will be concerned with deriving evolution operators
for stochastic \sch equations, and in particular with those corresponding to 
continuous measurements of physical observables. In this section we would like 
to give an example of a real physical scheme for the measurement of position
which is described by such an equation. This will also serve to show that the 
experimental arrangement considered in Chapter~\ref{qnc} corresponds to a
continuous measurement of position.

The measurement scheme consists of bouncing a light ray off a mirror and  
measuring the phase of the reflected light to determine the position of
the mirror. To treat the problem using the language of open quantum systems
discussed above, we will choose the mirror in question to be one side of an
optical cavity. The other mirror of the cavity is fixed, and is not perfectly
reflecting so that the cavity may be driven by the light beam through this 
fixed mirror. In the bad cavity limit we expect that the phase of the output 
light will provide a continuous measurement of the position of the movable
mirror. Allowing the position co-ordinate of the movable mirror to be 
confined by some arbitrary potential given by $H_{\ms{m}}$, the \htn of the
cavity-mirror system is~\cite{MJW}
\bq
  H_{0} = H_{\ms{m}} + \hbar\omega_{0}a^\dagger a - \hbar ga^\dagger aQ - \hbar iE(a e^{i\omega_{0}t} - a^\dagger e^{-i\omega_{0}t}) ,
\eq
where $\omega_0$ is the frequency of the cavity mode, $g$ is the coupling
constant between the cavity mode and the mirror co-ordinate (the expression 
for which need not concern us here), $Q$ is the position operator for the 
mirror and the term proportional to $E$ results from the coherent driving 
($E^2/\gamma$ gives the rate of photons impinging on the cavity). Dropping the free Hamiltonian of the cavity mode by moving into the interaction picture, the master equation for the cavity-mirror system becomes
\bq
  \dot{W} = \ioh[H_{\ms{m}},W] - [E(a-a^\dagger),W] + i[ga^\dagger aQ,W] + \gamma {\cal D}(a) ,
\eq
where $\gamma$ is the decay rate of the cavity. We are interested in the 
dynamics of the mirror when the light is subjected to phase sensitive 
detection. We are also interested in the limit of large cavity damping, 
and so it is convenient that in this regime we may adiabatically eliminate 
the cavity mode and derive a master equation for the mirror alone. To proceed
we note that the steady state of the driven cavity in the absence of the
interaction with the mirror, is the coherent state $|\alpha_0\rangle$ where
$\alpha_0=2E/\gamma$. Allowing the strength of the interaction to be much 
less than the damping rate, we transform the state of the system using
\bq
  W' = D(-\alpha_0)WD^\dagger(-\alpha_0) ,
\eq
where $D$ is the displacement operator~\cite{WandM}, so that the 
steady state of the system is very close to the vacuum. Transforming the
master equation using this `displacement' picture, we obtain
\bq
  \dot{W}' = \ioh [H_{\ms{m}},W'] + ig[(a^\dagger a + |\alpha_0|^2)Q,W'] + 
ig\alpha_0[(a-a^\dagger)Q,W'] + \gamma {\cal D}(a)W' .
\eq
Writing the density operator explicitly in the number basis we may use an
approximate solution of the form~\cite{SHM}
\bq
  W' = \rho_0\ket{0}\bra{0} + (\rho_1\ket{1}\bra{0} + \mbox{H.c.}) + \rho_2\ket{1}\bra{1} + (\rho_2'\ket{2}\bra{0} + \mbox{H.c.}) .
\label{apw}
\eq
Writing out the equations for $\rho_0,\rho_1,\rho_2$ and $\rho_2'$ we find that
the off-diagonal elements $\rho_1$ and $\rho_2'$ may be adiabatically 
eliminated (slaved to the on-diagonal elements). The resulting master equation
for the density matrix of the mirror, given by $\rho=\rho_0+\rho_2$, 
becomes
\bq
  \dot{\rho} = \ioh [H_{\ms{m}}-\hbar g |\alpha_0|^2Q, \rho] - k[Q,[Q,\rho]] ,
\eq
where $k=2g^2|\alpha_0|^2/\gamma$. The second is just a linear potential that the mirror experiences due to the pressure from the light. The last term, 
however, is the classic form for a continuous measurement, the effect of 
which is to diagonalise the density matrix in the position basis. To complete 
our analysis we need to see how the measurement proceeds during a quantum
trajectory. For this we require the jump operator corresponding to  
a photo-detection. As it is the phase of the output light that provides a readout of the position of the mirror we will use homodyne detection. The jump
 super-operator when written in terms of the full density matrix is therefore
${\cal J}W=\gamma(a + i\beta)W(a^\dagger - i\beta)$. In keeping with the
adiabatic approximation we wish to refer only to the density matrix of the
mirror, and we therefore want to find the equivalent jump super-operator which
acts on the mirror density matrix alone. To do this we slave $\rho_2$ to $\rho_0$, and this allows us to write W' in terms of $\rho$ by substituting into Eq.(\ref{apw}). The requisite jump super-operator is then given by 
\bq
  {\cal J}_{\ms{m}}\rho = \mbox{Tr}_{\ms{c}}[{\cal J}W] = \mbox{Tr}_{\ms{c}}[\gamma(a + \alpha_0 + i\beta)W'(a^\dagger + \alpha_0 - i\beta)] = C\rho C^\dagger , 
\eq
where
\bq
 C = (\sqrt{\gamma}\beta - i\sqrt{\gamma}\alpha_0 + \sqrt{2k} Q ) .
\eq
We can see that if we choose direct photo-detection ($\beta=0$) the average count rate is $\langle C^\dagger C\rangle = \gamma |\alpha_0|^2(1+(4g^2/\gamma^2)\langle Q^2\rangle)$, which contains no information about the average value of the mirror position (only the second moment). Using homodyne detection, however, and defining $\tilde{\beta}$ appropriately, we have
\bq
  C = (\tilde{\beta} + \sqrt{2k} Q ) ,
\eq
so that the average rate of photo-detection is
\bq
   \langle C^\dagger C\rangle = \tilde{\beta}^2 + 2\sqrt{2k}\tilde{\beta}\langle Q\rangle + 2k\langle Q^2\rangle ,
\eq
and provides a continuous measurement of the position of the mirror. Taking the homodyne detection limit as in section~\ref{HD}, we achieve our goal, which was to obtain a stochastic \sch equation for the mirror co-ordinate undergoing a continuous position measurement:
\bq
  |\tilde{\psi}(t + dt)\rangle = \left[ 1 - \left( \frac{i}{\hbar}H_{\mbox{\scriptsize m}} + kQ^2 \right) dt + [ 4k\langle Q\rangle dt  + \sqrt{2k} dW]Q  \right] |\tilde{\psi}(t)\rangle .
\eq
This is an equation for the unnormalised state vector, although it may be readily converted to an equation for the normalised state vector in the manner of section~\ref{HD}.

\chapter{Quantum Trajectories and Quantum noise}\label{cqtqn}

\section{Stochastic Calculus}\label{stca}

\subsection{The Wiener Process}
Stochastic calculus is the calculus required to handle differential equations which are driven by random processes. We have seen in the previous chapter that quantum trajectories are described by differential equations of this nature. We now examine these equations in some detail, what they mean, and how to handle them. Our treatment is not intended to be mathematically rigorous, but rather is aimed at providing the unfamiliar reader with an understanding of this often rather obscure topic.

A general stochastic equation for a variable $x$ includes both a deterministic increment proportional to $dt$, and a random increment~\cite{Gardiner1}. For the purposes of 
introducing these equations it is best to examine a small time step $\Delta t$
and take the infinitesimal limit later. We therefore write the general 
stochastic differential equation as
\bq
  \Delta x = f(x,t)\Delta t + g(x,t)\Delta W ,
\label{we1}
\eq
where $f(x,t)$ and $g(x,t)$ are arbitrary functions of $x$ and $t$, and $\Delta W$ is the random increment which will take a different value at 
each time step. We will choose $\Delta W$ to be Gaussian distributed with a
zero mean (any non-zero mean may be absorbed into the deterministic part in any case), so that
\bq
  P(\Delta W) = \frac{1}{\sqrt{2\pi\sigma^2(\Delta t)}}\exp\left\{-\frac{(\Delta W)^2} {2\sigma^2(\Delta t)}\right\} ,
\eq
where $\sigma^2(\Delta t)$ is the variance, being a function of $\Delta t$. Now, how should this variance depend on $\Delta t$? Consider the stochastic increment for a time step of $\Delta t/n$, so that the variance is $\sigma^2(\Delta t/n)$. We now want to make the stochastic increment over a time step $\Delta t$ consistent in the sense that it is given by the sum of $n$ of these stochastic increments. Probability theory tells us that when we add an arbitrary number of random variables, the variance of the sum is the sum of the individual variances. The variance for a time step $\Delta t$ should therefore satisfy the relation $\sigma^2(\Delta t) = n\sigma^2(\Delta t/n)$. This linearity condition is satisfied if $\sigma^2(\Delta t)$ is proportional to $\Delta t$. Because scaling the variance may be achieved by simply scaling the increment $\Delta W$, we may set $\sigma^2(\Delta t)=\Delta t$ without loss of generality. The result is the Wiener increment, with probability distribution given by
\bq
  P(\Delta W) = \frac{1}{\sqrt{2\pi\Delta t}}\exp\left[-\frac{(\Delta W)^2}{2\Delta t}\right].
\eq
In fact, it is easy to see that any other dependence of the variance on the time step makes no sense in the continuum limit. Let us assume that we take $\sigma^2(\Delta t)=\Delta t^\alpha$ for some $\alpha$, and calculate the variance of the finite increment over a time $t$ in the limit as $\Delta t$ tends to zero. The finite increment is given by
\bq
  W(t) = \int_0^t dW(t') \equiv  \lim_{N\rightarrow\infty} \sum_{n=1}^N \Delta W_n ,
\eq
where $\Delta W_n$ is the increment over the $n^{\ms{th}}$ interval of duration $t/N$. The variance of $W(t)$ is then
\bq
  \langle W(t)^2 \rangle = \lim_{N\rightarrow\infty} \sum_{n=1}^N \langle \Delta W_n \rangle = \lim_{N\rightarrow\infty} N \left(\frac{\Delta t}{N}\right)^\alpha = \lim_{N\rightarrow\infty} N^{1-\alpha}\Delta t^\alpha .
\eq
We see now that it is essential to choose $\alpha=1$ to avoid the result being zero or tending to infinity, both of which are absurd. 

To be able to manipulate the stochastic equation we must know how the stochastic increment acts when raised to some power. For example, if we want to calculate the equation governing $x^2$ in the case that the equation for $x$ is given by Eq.(\ref{we1}), then we may write
\bqa
   \Delta(x^2) & = & (x+\Delta x)(x+\Delta x) - x^2 \nonumber \\
               & = & 2x\;\Delta x + (\Delta x)^2 = 2xf\Delta t + 2xg\Delta W + f^2(\Delta t)^2 + 2fg\Delta t \Delta W + g^2(\Delta W)^2 .
\eqa
Usually the effect of second order terms vanishes in the infinitesimal limit, and the result is that given by the chain rule of ordinary calculus: $dx^2 = 2xdx$. However, note that the expectation value of $(\Delta W)^2$ is $\Delta t$, which suggests that terms second order in $\Delta W$ may contribute as terms first order in $\Delta t$. In fact, it turns out that in the infinitesimal limit we have precisely $(dW)^2=dt$. This appears especially strange because $dW$ is a stochastic increment while $dt$ is deterministic. We now examine why this is so. 

First note that the random variable $\Delta Z=(\Delta W)^2$ is non-negative, unlike $\Delta W$ which is positive and negative with equal probability. In particular, the probability density for $\Delta Z$ is
\bq
   P(\Delta Z) = \frac{e^{-\Delta Z/(2\Delta t)}}{\sqrt{2\pi\Delta t \Delta Z}} .
\eq
To find out how $\Delta Z$ behaves in the continuum limit, we must examine what happens when we sum many of them together. Let us define a variable, $Z_N(t)$, which is just such a sum over a time from $0$ to $t$,
\bq
  Z_N(t) = \sum_{n=1}^N \Delta Z_n \;\; , \;\;\; P(\Delta Z_n) = \frac{e^{-\Delta Z_n/(2t/N)}}{\sqrt{2\pi(t/N) \Delta Z_n}} ,
\label{we2}
\eq
and let us denote this variable in the continuum limit (as $N$ tends to infinity) as $Z(t)$. Clearly the average value of $Z_N(t)$ is constant for any value of $N$, and is given by $\langle Z(t)\rangle=t$. To obtain the result $dZ=dt$, however, we must show that in the infinitesimal limit, when $N\rightarrow\infty$, the fluctuations of $Z_N(t)$ about the mean value $t$ become smaller and smaller so that $Z(t)=t$. We could do this by calculating the variance of $Z_N(t)$, but it is possible to obtain the full distribution for $Z(t)$ which we now do using the characteristic function.

The characteristic function for a random variable is defined as a Fourier transform of the probability density. The characteristic function is useful for our purposes here because when adding a number of random variables, the characteristic function for the resulting random variable is given by multiplying together the characteristic functions for the random variables in the sum~\cite{Probtext}. To calculate the probability density for $Z_N(t)$ we therefore first obtain the characteristic function for $\Delta Z_n$ by taking the Fourier transform of the probability density given in Eq.(\ref{we2}):
\bq
  \chi_n(s) = \int_0^\infty \frac{e^{\Delta Z_n[is-N/(2t)]}}{\sqrt{2\pi(t/N) \Delta Z_n}} d(\Delta Z_n) = \frac{1}{\sqrt{\left( 1 - ist/(2N) \right)}} .
\eq
The characteristic function for $Z(t)$ is therefore
\bq
  \chi(s) = \lim_{N\rightarrow\infty}\left[\frac{1}{\sqrt{\left( 1 - ist/(2N) \right)}}\right]^N = \lim_{N\rightarrow\infty} \left( 1 + ist \left(\frac{1}{N/2}\right) \right)^{N/2} = e^{ist} ,
\eq
where we have used the binomial approximation in the second step because $N$ is large. Taking the inverse transform we obtain the probability distribution for $Z(t)$ as
\bq
  P(Z(t)) = \frac{1}{2\pi}\int_{-\infty}^{\infty}e^{-is(Z-t)} ds = \delta(z-t).
\eq
The distribution for $Z(t)$ is a delta function centred at $Z=t$, and therefore $Z(t)=t$. Over any time step the integral over the squares of the Wiener increments is equal to the time step, and this is true for any time step, in particular an infinitesimal one. As a consequence we may use the shorthand rule $dW^2=dt$, which is known as the Ito calculus relation. It may be shown that all higher powers of the stochastic increment contribute nothing in the continuum limit, and this is also true for $dtdW$~\cite{Gardiner1}. 

Now that we understand the rules required to handle the Wiener process, we will take a look at the solutions to these equations. Before we do however, let us justify our choice of a Gaussian random variable for the stochastic increment. Note that we are interested in the continuum limit, so that in each arbitrarily small time interval the system gets a kick which is uncorrelated with the kick received in the previous time interval. This is a good approximation if the noise driving the system is uncorrelated on the time scale of the system dynamics. If this is the case then the central limit theorem tells us that if the statistics of the kicks come from any distribution with finite moments, when we sum enough of them together the result will be kicks with a Gaussian distribution. We can then say that the Wiener process provides a good model for any reasonably behaved source of continuous physical noise in which the correlation time is much smaller than the dynamical time scale of the system. Physically this means that the noise which is driving the system of interest contains very high frequency fluctuations compared to the time scale of the system dynamics. 

In the remainder of this section we examine the solutions of two examples of Eq.(\ref{we1}) which will be relevant in later chapters. For our first example we take $f(x,t)=kx$, where $k$ is some complex number, and take $g$ to be constant, which gives us the {\em Ornstein-Uhlenbeck} process~\cite{Gardiner1}
\bq
  dx = kx \; dt + g \; dW .
\eq
This is simply an ordinary linear differential equation with a driving term which is the Wiener process. In this case the singular nature of the Wiener increment is not important. To see this we may find the solution by changing variables using $y=xe^{-kt}$ so that
\bq
  dy = d(xe^{-kt}) = dx \; e^{-kt} + x \; d(e^{-kt}) + dx\;d(e^{-kt}) = ge^{-kt} \; dW .
\eq
In this process of changing variables there are no contributions from terms involving $dW^2$. Consequently the solution we obtain is correct for any continuous driving. The solution for $y$ is now
\bq
   y(t) \; = \; y(0) + g \int_0^t e^{-kt'} \; dW(t') \; \equiv \; y(0) + g \lim_{N\rightarrow\infty} \sum_{n=1}^N e^{-k(n-1)\Delta t} \; \Delta W_n .
\eq
Using $x=ye^{kt}$ we obtain the solution for the Ornstein-Uhlenbeck process, which is
\bq
  x(t) = e^{kt}y(0) + g e^{kt}\int_0^t e^{-kt'} \; dW(t') .
\eq
While this solution is no different than that for any ordinary driving term, in this case the integral over the driving is a random variable. This random variable is simply a sum over Gaussian variables and is therefore itself Gaussian. The variance is easily evaluated to obtain
\bq
  \lim_{N\rightarrow\infty} \left\langle \left[ \sum_{n=1}^N e^{-k(n-1)\Delta t} \; \Delta W_n\right]^2 \right\rangle = \sum_{n=1}^N e^{-2k(n-1)\Delta t} \; \langle \Delta W_n^2 \rangle = \int_0^t e^{-2kt'} dt' ,
\eq
where we have used the independence of the Wiener increments which gives $\langle \Delta W_n \Delta W_m\rangle = \delta_{nm} \Delta t$.

The Wiener process, due to being of order $\sqrt{dt}$, is not differentiable (it is not of bounded variation) even though it is continuous. Nevertheless, it is still possible to write the Ornstein-Uhlenbeck equation in the form 
\bq
  \dot{x} = kx + g\varepsilon(t) ,
\eq
where we have used the derivative rather than the differential. It is possible to show that solutions to this equation are equivalent to solutions to the differential form if the noise term, $\varepsilon(t)$, is delta correlated. That is, if $\langle\varepsilon(t)\varepsilon(t')\rangle = \delta(t-t')$~\cite{Gardiner1}. In particular the moments of $\int_0^t\varepsilon(t')dt'$ are equal to those of $\int_0^t dW(t')$ as required. When written in this form the equation is usually referred to as a {\em Langevin} equation. The delta correlated (Langevin) noise source is naturally highly singular (its variance at any time is infinite), a fact which reflects the non-differentiability of the Weiner process. Because spectrum of the noise is the Fourier transform of the correlation function, the spectrum of $\varepsilon(t)$ is flat. For this reason the Wiener process is said to describe {\em white noise}.

While we have concentrated above on a differential equation involving a single variable, the Ornstein-Uhlenbeck process is easily generalised to a multi-variable linear equation, which may have multiple independent noise sources. We will use equations of this form for our analysis in Chapter~\ref{qnc}.

For our second example we consider the time independent linear stochastic equation given by
\bq
  dx = [f\; dt + g\; dW]x ,
\eq
in which $f$ and $g$ are constant. This differs from the Ornstein-Uhlenbeck process in that the Wiener increment now multiplies $x$, and for this reason it is referred to as linear multiplicative white noise. This equation is also readily generalised to multiple variables. In this case $f$ and $g$ become matrices, and $x$ the vector of the variables. It is this generalisation which describes linear quantum trajectories, and with which we will be concerned in Chapter~\ref{evopch}. For the case of a single variable, however, it is easily solved. The standard method of solution is to change variables to $y=\ln(x)$, so that 
\bq
  dy = \frac{1}{x}dx - \frac{1}{2x^2}(dx)^2 = [f - (g^2/2)] dt + g\; dW .
\eq
We are now able to integrate directly to obtain
\bq
  y(t) = y(0) + [f - (g^2/2)]t + g W(t) ,
\eq
and the full solution is then $x(t)=e^{y(t)}$. It is possible to obtain the solution using a different method however. This is useful because, as we show in Chapter~\ref{evopch}, it may be generalised to equations containing multiple variables, and hence to quantum trajectories. In this approach we write the equation in the form
\bq
  x(t+\Delta t) = [1 + f\; \Delta t + g\; \Delta W]x(t) = e^{(f - (g^2/2))\Delta t + g\Delta W}x(t) ,
\eq
where the last equality is easily shown by expanding the exponential to first order in $\Delta t$ and using $\Delta W^2=\Delta t$. The exponential now acts as an operator which propagates $x(t)$ forward in time by $\Delta t$. The solution may therefore be obtained using
\bqa
  x(t) & = & \lim_{N\rightarrow\infty}\left(\prod_{n=1}^N e^{(f - (g^2/2))\Delta t + g\Delta W_n}\right) x(0) \nn \\
       & = &  \exp\left\{ \lim_{N\rightarrow\infty} \left[(f - (g^2/2))\sum_{n=1}^N\Delta t + g\sum_{n=1}^N\Delta W_n \right] \right\} x(0) \nn \\
  & = & e^{[f - (g^2/2)]t + g W(t)} x(0) ,
\eqa
which is indeed the solution obtained previously. This concludes our discussion of stochastic equations driven by the Wiener process. For a comprehensive treatment of stochastic equations and stochastic calculus the reader is referred to reference~\cite{Gardiner1}.

\subsection{The Poisson Process}

The Poisson process consists of a series of instantaneous events occurring at random times, and processes of this nature are referred as {\em point processes}. For the Poisson process there is a constant probability per unit time for an event to occur. We saw in the previous chapter that in the right limit the Poisson process may be well approximated by the Wiener process. However, we are not always interested in this limit. A general stochastic equation driven by the Poisson process is
\bq
  dx = f(x,t)dt + g(x,t)dN ,
\eq
where this time $dN$ is the Poisson increment. The absence of an event in the time interval $dt$ is signified by $dN=0$, while an event is signified by $dN=1$. The probability for more than one event to occur in any time interval is higher than first order in the time interval, so we need only consider these two possibilities. Denoting the probability per unity time for an event by $\lambda$, the probabilities for time interval $dt$ are
\bq
  P(dN=0) = 1 - \lambda dt \;\;\; , \;\;\;\; P(dN=1) = \lambda dt .
\eq
The Poisson increment satisfies $dN^2=dN$ due to the fact that it takes only the values zero or one. The total number of events that have occurred in the interval $[0,t]$ is the integral of the Poisson increments over that time. More generally,  the integral of a function of time over the Poisson increment is
\bq
  \int_0^t f(t') dN(t') = \sum_{k=1}^{N(t)} f(t_k)
\eq
where $N(t)$ is the total number of events up until time $t$, and $t_k$ is the time of the $k^{\ms{\it th}}$ event. The evolution of $x$ is clearly smooth except when interrupted by discontinuous jumps at the times at which $dN=1$.

\section{Quantum Trajectories}
\subsection{Quantum Trajectories and Generalised Measurements}

In Chapter~\ref{chapter1} we introduced master equations by considering a real open quantum system (an optical cavity) and motivated quantum trajectories by considering a real continuous measurement (photo-detection) performed on this system. However, it is possible to start with the theory of generalised measurements, which we also introduced in Chapter~\ref{chapter1} and, using this to formulate a continuous measurement, obtain both master equations and quantum trajectories in a simple manner. To this end consider a generalised measurement in an infinitesimal time interval $dt$. The non-selective evolution caused by this measurement process is
\bq
  \rho(t+dt) = \sum_n \Omega_n(dt) \rho(t) \Omega_n^{\dg}(dt) .
\eq
We now choose a measurement with two possible results, where the measurement operators are
\bqa
  \Omega_0 & = & 1 - ( \smallfrac{i}{\hbar}H + \half c^{\dg}c) dt , \label{mop00} \\
  \Omega_1 & = & \sqrt{dt}c .
\label{mop01}
\eqa
The resulting non-selective evolution is
\bq
  \rho(t+dt) = \rho(t) -\smallfrac{i}{\hbar} [H,\rho(t)]dt + {\cal D}(c)\rho(t)dt.
\eq
Here ${\cal D}(c)$ is the Lindblad super-operator. Recall from Chapter~\ref{chapter1} that any master equation may be written as a sum over an arbitrary number of Lindblad super-operators, ${\cal D}(c_n)$, with different measurement operators $c_n$. From the above result we see that each Lindblad operator corresponds to a continuous measurement process, alternatively referred to as a decay channel. It is clear that this continuous measurement process is a point process, as the probability for the measurement to return the result 1 in an interval $dt$ is $\mbox{Tr}[c^\dagger c\rho(t)]dt$, being proportional to $dt$. The result of the measurement is therefore usually zero, and is unity only at a discrete series of random times.

If we choose to look at the selective rather than the non-selective evolution, then we obtain a quantum trajectory rather than a master equation which is the average over all the possible trajectories. Clearly the point process we have considered above corresponds to photo-detection of the light output from a lossy optical cavity, where a result of zero signifies that there was no detection, the result of unity corresponding to a detection. This measurement process may be modelled explicitly by an interaction with a two state probe system for time $dt$, where the probe is measured at the end of the interaction~\cite{Ueda}. Denoting the two states of the probe as $\ket{0}$ and $\ket{1}$, and defining a probe transition operator, $\sigma = \ket{0}\bra{1}$, we choose the interaction \htn to be
\bq
   H_{\mbox{\scriptsize int}} = \hbar g (c^{\dg}\sigma +\sigma^{\dg}c ) .
\eq
This choice of the interaction \htn means that the system will cause a transition in the probe such that the rate of transition depends upon $c$. A subsequent measurement of the probe tells us whether a transition has occurred, and this provides information about the system. 
Taking the initial state of the probe to be $\ket{0}$ (in fact the interaction is symmetric in the probe state so either initial state is acceptable), and evolving the system and meter for a time $\Delta t$, we obtain, to second order in $\Delta t$
\bqa
  \rho_{\ms{tot}}(t+\Delta t) & = & \rho_{\ms{tot}} - \smallfrac{i}{\hbar} [H_{\ms{tot}},\rho_{\ms{tot}}] \Delta t - \smallfrac{1}{2\hbar^2}[H_{\ms{tot}},[H_{\ms{tot}},\rho_{\ms{tot}}]] \Delta t^2 + \ldots \nn \\
 & = & - \smallfrac{i}{\hbar}[H,\rho(t)]\otimes \ket{0}\bra{0}\Delta t + (c\rho\otimes \ket{1}\bra{0} + \rho c^{\dg}\otimes \ket{0}\bra{1}) g\Delta t \nn \\
 & & \!\!\!\! + \half (2c\rho c^\dagger \otimes \ket{1}\bra{1} - c^\dagger c\rho \otimes \ket{0}\bra{0} - \rho c^\dagger c \otimes \ket{0}\bra{0} )(g\Delta t)^2 + \ldots
\eqa
where we have written the total Hamiltonian for the system plus probe as $H_{\ms{tot}}= H + H_{\mbox{\scriptsize int}}$.
At the end of the interval we perform a projection measurement onto the probe states $\ket{0}$ and $\ket{1}$. The state of the system upon obtaining the result $\ket{0}$ is
\bq
  \rho(t+\Delta t) = \rho(t) - \smallfrac{i}{\hbar}[H,\rho(t)]\Delta t - (1/2)(c^\dagger c\rho + \rho c^\dagger c)(g\Delta t)^2 ,
\eq
and upon obtaining the result $\ket{1}$ is
\bq
  \rho(t+\Delta t) = c\rho(t) c^\dagger (g\Delta t)^2.
\eq
Taking the limit as $\Delta t\rightarrow dt$, and allowing $g$ to scale as $1/\sqrt{\Delta t}$, this is just the result obtained by applying the measurement operators introduced in Eqs.(\ref{mop00}) and (\ref{mop01}). 

The set of measurement operators given by Eqs.(\ref{mop00}) and (\ref{mop01}) is not the only set which will generate the correct non-selective evolution. In Chapter~\ref{chapter1} we noted that applying a unitary transformation to the measurement operators leaves the non-selective evolution unaffected. A transformation of particular interest is that given by
\bq
    \left( \begin{array}{c} \Omega_0' \\ \Omega_1' \end{array} \right) = 
    \left( \begin{array}{cc} 1-\half |\tilde{\beta}|^2 dt & -\tilde{\beta}^*\sqrt{dt} \\ \tilde{\beta}\sqrt{dt} & 1-\half |\tilde{\beta}|^2 dt \end{array} \right) 
    \left( \begin{array}{c} \Omega_0 \\ \Omega_1 \end{array} \right) .
\eq
This is equivalent to the transformation
\bq
  c \rightarrow c + \tilde{\beta} \;\;\;\;\;\;\;\;\;\;\; H \rightarrow H - \half i\hbar(\tilde{\beta}^* c - \tilde{\beta} c^\dagger) .
\eq
In the limit in which $|\tilde{\beta}|^2\gg \langle c^\dagger c\rangle$ this corresponds to homodyne detection, in which the field to be measured is first mixed with a coherent field at a beam splitter before being subjected to photo-detection (see section~\ref{HD}). It is also possible to generate the same non-selective evolution using a completely different set of measurement operators, although we are not concerned with these here. For an example see reference~\cite{MilCav}. In the following section we will see how we may formulate linear and non-linear stochastic equations to describe the selective evolution, or {\em quantum trajectories}, generated by generalised measurement operators. 


\subsection{Linear and Non-linear Formulations}

We have seen in the previous section how the non-selective evolution produced by a Lindblad super-operator may be unravelled by the continuum limit of a generalised measurement process. In each time interval, $\Delta t$, a generalised measurement is performed, and the manner in which the state of the system is altered depends on the result of the measurement. This is repeated in the next time interval and so on. We want to examine now how we might simulate this process. In particular we are interested in (i) the probabilities of the possible measurement sequences, and (ii) the state of the system during and after the measurement process. In the following discussion we use the approach of Wiseman~\cite{qsco}.

Let us assume that the initial state is pure, and denote it by $\ket{\psi(t)}$. In this case the state of the system given the result $r$ is
\bq
  \ket{\psi(t+\Delta t)} = \frac{\Omega_r(\Delta t)\ket{\psi(t)}} {\sqrt{P_r(t)}} ,
\label{ns1}
\eq
where $P_r(t)$ is the probability for obtaining the result $r$, being
\bq
  P_r(t) = \bra{\psi(t)}\Omega_r^\dagger(\Delta t)\Omega_r(\Delta t)\ket{\psi(t)} .
\label{pr1}
\eq
To simulate this procedure, we could calculate the probabilities for the possible results using using Eq.(\ref{pr1}), and choose one at random using these probabilities. We then calculate the new state using Eq.(\ref{ns1}), and repeat the procedure for this new state. The probability of obtaining a particular final state at the end of a sequence of measurements is then given by the probability that we actually obtain that state at the end of our simulation. Call this method A. We may formulate this simulation procedure as a stochastic differential equation which turns out to be non-linear. Applying this to the measurement process described by the measurement operators $\Omega_0'$ and $\Omega_1'$, we have two possibilities at each infinitesimal time step, and we may use the Poisson process introduced in the previous section. By considering the change in the state vector for each of the two possibilities the stochastic equation is readily found to be
\bqa
  \ket{\psi(t+dt)} & = & \left[ 1 + \left( \frac{c+\tilde{\beta}}{\sqrt{\langle(c^{\dg}+\tilde{\beta}^*)(c+\tilde{\beta})\rangle}} - 1 \right) dN  \right. \nn \\
    & & \;\;\;\;\; + \left. \left( \frac{\langle c^\dagger c \rangle}{2} - \frac{c^\dagger c}{2} + \frac{\langle c^\dagger\tilde{\beta} + \tilde{\beta}^*c \rangle}{2} - \tilde{\beta}^*c - \smallfrac{i}{\hbar} H\right) dt \right] \ket{\psi(t)} .
\label{nsec2}
\eqa
It is possible to use another method to simulate the measurement procedure. Instead of choosing the outcomes at random using the correct probabilities, we choose them with fixed probabilities independent of the state. Following reference~\cite{qsco} we will refer to these as the {\em ostensible} probabilities. In this case the actual probability for obtaining a given final state is not the probability that we obtain that state at the end of the simulation. However, we can preserve this information by omitting to normalise the state vector at each step. In this case the state after a measurement is given by
\bq
   \ket{\tilde{\psi}(t + \Delta t)} = \Omega_r(\Delta t)\ket{\psi(t)} ,
\eq
where we use the tilde to denote an unnormalised state. Re-writing Eq.(\ref{pr1}) we see that the probability for obtaining the final state is simply the norm of that final state:
\bq
  P_r(t) = \langle \tilde{\psi}(t + \Delta t)\ket{\tilde{\psi}(t + \Delta t)} .
\eq
Following this process through a series of measurements we see that the probability of obtaining a particular final state is just the norm of that final state. This simulation process is now linear because the final (unnormalised) state is just the initial state transformed by the action of a series of linear operators. Call this method B. This method is not quite sufficient for our purposes, however. Note that in method A we divided the final state by the square root of the probability of obtaining that state, and this resulted in normalisation. In order to take the infinitesimal limit to obtain a continuous measurement (and in our case, a Poisson process), we will need to maintain this procedure of dividing the final state by the probability with which we choose it, although this time the probability will be independent of the initial state, as in method B. Thus we choose among the possible final states at random, using probabilities chosen to be independent of the initial state, this time taking the final state to be
\bq
  \ket{\tilde{\psi}(t + \Delta t)} = \frac{\Omega_r(\Delta t) \ket{\psi(t)}}{\sqrt{\Lambda_r(t)}},
\eq
where $\Lambda_r(t)$ is the ostensible probability with which we (artificially) choose the outcome $r$. Call this method C. Clearly the true probability of obtaining the result $r$ is still contained in the norm of the final state. In this case it is
\bq
  P_r(t) = \langle \tilde{\psi}(t + \Delta t)\ket{\tilde{\psi}(t + \Delta t)} \Lambda_r(t),
\eq
and the true probability for a given final state at the end of a series of measurements is
\bq
  P(t+\tau) = \langle \tilde{\psi}(t + \tau)\ket{\tilde{\psi}(t + \tau)} \tilde{P}(t) ,
\eq
where $\tilde{P}(t)$ is the ostensible probability that we obtain the given final state at the end of our simulation procedure. This method will allow us to write a linear stochastic equation for the quantum state.

To derive the linear stochastic equation describing photo-detection performed on the light emitted from a damped optical cavity, the measurement operators are $\Omega_0$ and $\Omega_1$ with $c=\sqrt{\gamma}a$. Choosing the ostensible probabilities $\Lambda_1(t) = \gamma dt$, and $\Lambda_0(t) = 1-\Lambda_1(t)$, to give a Poisson process with a constant average rate of events $\gamma$, the infinitesimal increment of the state at time $t$ becomes
\bqa
   \ket{\tilde{\psi}_1(t+dt)} & = & \frac{\Omega_1'(dt)
             \ket{\tilde{\psi}(t)}}{\sqrt{\Lambda_1(t)}}  =  a \ket{\tilde{\psi}(t)} , \\
   \ket{\tilde{\psi}_0(t+dt)} & = & \frac{\Omega_0'(dt)
             \ket{\tilde{\psi}(t)}}{\sqrt{\Lambda_0(t)}}  =  \left[ 1 - \left(\smallfrac{i}{\hbar}H + \smallfrac{\gamma}{2}a^\dagger a - \smallfrac{\gamma}{2}\right) dt \right] \ket{\tilde{\psi}(t)} .
\eqa
The equivalent stochastic equation is therefore
\bq
  \ket{\tilde{\psi}_1(t+dt)} = \left[ 1 - \left(\smallfrac{i}{\hbar}H + \smallfrac{\gamma}{2}a^\dagger a - \smallfrac{\gamma}{2}\right)dt + (a-1)dN \right] \ket{\tilde{\psi}_1(t)} .
\eq
We will examine how solutions to this equation may be found in the next chapter.

To derive a linear stochastic equation corresponding to homodyne detection we use the measurement operators $\Omega_0'$ $\Omega_1'$ and choose the ostensible probabilities $\Lambda_0(t) = 1-|\tilde{\beta}|^2 dt$, and $\Lambda_1(t) = |\tilde{\beta}|^2 dt$. The infinitesimal increment of the state at time $t$ becomes
\bqa
   \ket{\tilde{\psi}_1(t+dt)} & = & \frac{\Omega_1'(dt)
             \ket{\tilde{\psi}(t)}}{\sqrt{\Lambda_1(t)}}  =  \left[ 1 + \frac{c}{\tilde{\beta}}\right] \ket{\tilde{\psi}(t)} \\
   \ket{\tilde{\psi}_0(t+dt)} & = & \frac{\Omega_0'(dt)
             \ket{\tilde{\psi}(t)}}{\sqrt{\Lambda_0(t)}}  =  \left[ 1 - (\smallfrac{i}{\hbar} H + \half c^\dagger c + \tilde{\beta}^* c)dt \right] \ket{\tilde{\psi}(t)} ,
\eqa
and the stochastic equation is
\bq
  \ket{\tilde{\psi}_1(t+dt)} = \left[ 1 - \left( \smallfrac{i}{\hbar}H + \half c^{\dg}c + \tilde{\beta}^*c \right) dt + \frac{c}{\tilde{\beta}} dN  \right] \ket{\tilde{\psi}_1(t)} .
\eq
Note that this is much simpler than its non-linear equivalent Eq.(\ref{nsec2}). To obtain an equation for homodyne detection with a local phase set so as to measure the phase quadrature, we set $c=\sqrt{\gamma}a$ and $\tilde{\beta}=i\sqrt{\gamma}\beta$ where $\beta$ is real. We then take the Wiener process limit of the Poisson process as discussed in Chapter~\ref{chapter1}. The result is the linear equivalent of the non-linear stochastic equation derived for homodyne detection in that chapter,
\bq
  \ket{\tilde{\psi}_1(t+dt)} = \left[ 1 - \left( \smallfrac{i}{\hbar}H + \smallfrac{\gamma}{2} a^{\dg}a \right) dt -ia\sqrt{\gamma}dW  \right] \ket{\tilde{\psi}_1(t)} .
\eq
More generally, given a master equation of the form Eq.(\ref{lme}), a non-linear unravelling in terms of the Wiener process is given by
\bq
  |\psi(t+dt)\rangle = \left[ 1 -\smallfrac{i}{\hbar}Hdt - \sum_{n=1}^{N} (\half c^\dagger_n
 c_n - \langle c^\dagger_n \rangle c_n + \half \langle c^\dagger_n \rangle \langle c_n \rangle )dt + ( c_n - \langle c_n \rangle)dW_n(t) \right] |\psi(t)\rangle ,
\eq
where the $dW_n$ are independent Wiener increments. A linear unravelling for the same master equation is given by
\bq
  |\tilde{\psi}(t+dt)\rangle = \left[ 1 -\smallfrac{i}{\hbar}Hdt - \sum_{n=1}^{N} (\half c^\dagger_n
 c_n dt - c_n dW_n(t) ) \right] |\tilde{\psi}(t)\rangle .
\eq
In the linear case, the true probability measure for the system to have evolved to a particular state at time $t$ is given by~\cite{GandG}
\begin{equation}
 \langle \psi(t)|\psi(t)\rangle_{\mbox{\scriptsize w}} 
\mbox{d}P_{\mbox{\scriptsize w}} ,
\end{equation}
where $\mbox{d}P_{\mbox{\scriptsize w}}$ is the Wiener measure (the measure for the ostensible probabilities). That is, it represents integration over the joint probability density for all the random variables that appear in the expression for $|\psi(t)\rangle_{\mbox{\scriptsize w}}$. It follows therefore that moments of system operators calculated with the equivalent master equation at time $t$ are given by the expression
\begin{equation}
\langle {\cal O}\rangle = \int\langle \psi(t)|{\cal O}
|\psi(t)\rangle_{\mbox{\scriptsize w}} \mbox{d}P_{\mbox{\scriptsize w}}.
\label{avo}
\end{equation}
where ${\cal O}$ is the system operator in question, and 
$\mbox{d}P_{\mbox{\scriptsize w}}$ represents integration over all possible values of the random variables. In Chapter~\ref{evopch} we will show how expressions for final states and their corresponding probability densities may be obtained.

\section{Quantum Noise}\label{sqn}

So far we have considered the dynamics of an open quantum system using the \sch picture. However, it also is possible to use the \hei picture. In this case the system operators obey equations of motion driven by a quantum version of the Wiener process. We now examine this formulation which we use for our analysis in Chapter~\ref{qnc}.

We will consider here the specific example of a lossy optical cavity damped through one mirror, and will therefore require the part of the external field propagating along the axis of the cavity, and on one side of the cavity only. That is, a one-dimensional field in half-space. For a detailed discussion of the representation of this field in terms of mode operators the reader is referred to appendix A, where we examine the input-output relations for the cavity. Here we need merely know the Hamiltonian which gives the coupling between the cavity mode and the field mode operators, and this is
\begin{equation}
   H = H_{\ms{sys}} - 
ig\!\! \int_{-
\infty}^{\infty}\!\!\!\!\!\! (b^\dagger(k)a - a^\dagger 
b(k))\;\mbox{d}k + H_{\ms{field}} .
\end{equation}
Here the Hamiltonian for the cavity mode is denoted by $H_{\ms{sys}}$, and that for the external free field is denoted by $H_{\ms{field}}$. In the expression for the interaction Hamiltonian, $g$ is the coupling constant between the mode and the external field, $a$ is the mode annihilation operator and $b(k)$ is the field annihilation operator for the field mode with angular wave number $k$. In order to arrive at this interaction, which clearly models the exchange of photons between the cavity mode and the field, the rotating wave approximation has been employed, and this is discussed in appendix A. Using the above Hamiltonian we may calculate the equations of motion for an arbitrary system operator $c$, and for the field mode operators $b(k)$. These are
\bqa
   \dot{c} & = & \ioh [c,H_{\ms{sys}}] + g[c,a^\dagger]\left( \int_{-\infty}^{\infty} \!\!\!\!\!\! b(k,t)dk \right) - g\left( \int_{-\infty}^{\infty} \!\!\!\!\!\! b^\dagger(k,t)dk \right) [c,a] , \\
   \dot{b}(k,t) & = & -ick b(k,t) + g a .
\eqa
Solving the equation for the mode operators in terms of the system operator $a$ we have
\bq
  b(k,t) = b(k,0)e^{-ickt} + g\int_0^t e^{-ick(t-t')}a(t') dt'.
\eq
Substituting this back into the equation of motion for $c$ we obtain
\bq
  \dot{c} = \ioh [c,H_{\ms{sys}}] + [c,a]\left(\smallfrac{\gamma}{2}a^\dagger + \sqrt{\gamma}b_{\ms{in}}^{\dg}(t)\right) -
            \left(\smallfrac{\gamma}{2}a + \sqrt{\gamma}b_{\ms{in}}(t)\right)[c,a^\dagger] , 
\eq
where we have defined $\gamma = \pi g^2/\hbar^2$, and the operator
\bq
  b_{\ms{in}}(t) = \sqrt{\frac{2\pi}{c}} \int_{-\infty}^{\infty} \!\!\!\!\!\! b(k,0)e^{-ickt}.
\eq
We see that the equations of motion for the system operators are driven by $b_{\ms{in}}(t)$, which may therefore be referred to as the field `input' to the system. For a detailed explanation of the relationship between the initial field state and the input and output fields (being those parts of the field travelling to and from the system respectively), the reader is referred to appendix A. We have scaled $b_{\ms{in}}(t)$ specifically so that it satisfies the commutation relations
\bq
  [b_{\ms{in}}(t),b_{\ms{in}}^{\dg}(t)] = \delta(t-t') ,
\label{incom}
\eq
and which is a direct result of the commutation relations of the field mode operators. Note that the expectation value of this `input' operator at all future times is specified completely by the state of the field at $t=0$. If the initial state of the field is the vacuum, then $\langle b_{\ms{in}}^{\dg}(t), b_{\ms{in}}(t) \rangle = 0$. Using the commutation relation Eq.(\ref{incom}), this gives the anti-normally ordered correlation function as
\bq
  \langle b_{\ms{in}}(t), b_{\ms{in}}^{\dg}(t) \rangle = \delta(t-t') .
\eq
We see that the terms that drive the equations of motion for the system are delta correlated! They are therefore a quantum analogue of the classical Langevin noise source that we introduced in section~\ref{stca}. It is useful for us to put this equation in the form of an explicit quantum analogue of the Ito stochastic increment. From this we will be able to derive the master equation of section~\ref{d1me} and in so doing show that this quantum Langevin formalism is equivalent to the master equation.

Defining the stochastic increment operator as $dB(t) = b_{\ms{in}}(t)dt$, we have
\bq
  \langle \int_{0}^{t} dB(t') \int_{0}^{t} dB^\dagger(t'')\rangle =  \int_{0}^{t} \int_{0}^{t} \langle b_{\ms{in}}(t')b_{\ms{in}}^\dagger(t'')\rangle dt' dt'' = t ,
\eq
and in the infinitesimal limit the (quantum) Ito relation $dBdB^\dagger=dt$. Further discussion of these noise operators may be found in reference~\cite{Gardiner2}. 

To derive the Ito equation we move into the interaction picture with respect to $H_{\ms{field}}$, and write the resulting interaction Hamiltonian in terms of $b_{\ms{in}}(t)$. We have
\bq
  H = H_{\ms{sys}} + i\hbar\sqrt{\gamma}(a b_{\ms{in}}^\dagger(t) - b_{\ms{in}}(t) a^\dagger ) .
\eq
The evolution of the system operator $c$ is therefore
\bq
  c(t+dt) = e^{-i(H_{\ms{sys}}/\hbar)dt + \sqrt{\gamma}(a dB^\dagger(t) - dB(t) a^\dagger)} c(t) e^{-i(H_{\ms{sys}}/\hbar)dt - \sqrt{\gamma}(a dB^\dagger(t) - dB(t) a^\dagger)} .
\eq
Expanding this to first order in $t$, using the Ito relation and the fact that the stochastic increment operators commute with the system operators at time $t$, we obtain the Ito equation
\bq
  c(t+dt) = \ioh [c,H_{\ms{sys}}] dt + \gamma (a^\dagger c a - \half a^\dagger a c - \half c a^\dagger a) dt + \sqrt{\gamma} [c,a\; dB^\dagger(t) - dB(t) a^\dagger] .
\eq
From this we may derive the master equation. We do this by calculating the expectation value of the derivative of $c$ using the Ito equation, and by equating this to that calculated using the density operator: $\langle \dot{c}\rangle = \mbox{Tr}[c\dot{\rho}]$. Using $\langle dB\rangle = 0$, the expectation value of the derivative of $c$ from the Ito equation is
\bqa
  \langle \dot{c}\rangle & = & \ioh \langle [c,H_{\ms{sys}}] \rangle + \gamma (\langle a^\dagger c a\rangle - \half \langle a^\dagger a c\rangle  - \half \langle c a^\dagger a \rangle) \nn \\
 & = & \mbox{Tr}\left[\left(\ioh [c,H_{\ms{sys}}] + \gamma (a^\dagger c a - \half a^\dagger a c - \half c a^\dagger a) \right) \rho\right] \nn \\
 & = & \mbox{Tr}\left[c \left( \ioh [H_{\ms{sys}},\rho] + \gamma (a \rho a^\dagger - \half a^\dagger a \rho - \half \rho a^\dagger a)\right)\right] \nn \\
 & = & \mbox{Tr}[c \dot{\rho}] .
\eqa
This is true for any operator $c$, and so we obtain the master equation
\bq
  \dot{\rho} = \ioh [H_{\ms{sys}},\rho] + \gamma (a \rho a^\dagger - \half a^\dagger a \rho - \half \rho a^\dagger a) .
\eq
Every master equation has an equivalent quantum Langevin equation and vice versa.

We may extend the quantum Langevin equation to include other kinds of input. In our analysis above we have set the initial state of the field operators $b(k)$ to the vacuum. An almost trivial extension is to take the initial state of each mode to be coherent, so that $b(k,0)\ket{\beta(k)}_k = \beta(k) \ket{\beta(k)}_k$. Using a displacement operator to transform each of the initial states to the vacuum~\cite{Gardiner2,Mollow,HMWPhD}, the transformed input operator becomes
\bq
  b_{\ms{in}}'(t) = b_{\ms{in}}(t) + \beta(t) ,
\eq
where $\beta(t)$ is a c-number, being an appropriately scaled Fourier transform of $\beta(k)$. As we are free to choose the $\beta(k)$, we are free to choose any $\beta(t)$. For a coherent state input $\beta(k)$ is a delta function about $k=\omega_{\ms{c}}/c$, and consequently $\beta(t)$ is constant. In Chapter~\ref{qnc} we will want to consider the input from a laser. While this is basically coherent, it will naturally contain some noise so that the amplitude and phase will fluctuate. We can include this classical fluctuation by allowing the modes on either side of $k=\omega_{\ms{c}}/c$ to have a non-zero coherent amplitude. Assuming small fluctuations about the mean, we may write the input as
\bq
  \beta + \delta x(t) + i\delta y(t) + b_{\ms{in}}(t) ,
\eq
where $\beta$ is the coherent amplitude of the laser, $\delta x(t)$ are fluctuations in the amplitude quadrature and $\delta y(t)$ are the fluctuations in the phase quadrature. The autocorrelation functions for the fluctuations may be left arbitrary, and can be tailored to match the noise on the laser used in any particular realisation.

\chapter{Evolution Operators for Linear Quantum Trajectories}
\label{evopch}

In this chapter we present a method for obtaining evolution operators for 
linear quantum trajectories, and apply this to a number of physical examples of varying mathematical complexity. In particular we use as examples quantum trajectories describing the continuous projection measurement of physical observables. Using this method we calculate the average conditional uncertainty for the measured observables, being a central quantity of interest in these measurement processes. 

\section{Introduction}

We have seen in the previous two chapters that real continuous measurements may be described by stochastic equations for the quantum state vector. That these may be written in a linear form has been known in the mathematical physics 
literature for some time~\cite{mathLSE}, but has only fairly recently seen 
exposure in the physics literature~\cite{GandG,Strunz,HMWetc}, where it has 
been common to use non-linear stochastic equations~\cite{DCM,PK,solNLSE,BS}. The advantage of writing master equations as LSE's, rather than the more familiar non-linear version, is that in certain cases it has been found that explicit evolution operators corresponding to these equations may be obtained in a straightforward manner. Naturally the non-linear form does not admit a solution in terms of evolution operators, as these operators are by definition linear. 
However, as far as we are aware, the only method that has been used previously to obtain evolution operators for quantum trajectories driven by the Wiener process is to choose an initial state which allows the stochastic equation for the state to be written as a stochastic equation for an eigenvalue, or which simplifies the action of the evolution operator~\cite{GandG,HMWetc}. In this chapter we present a more general method for obtaining explicit evolution operators for 
these equations which makes no reference to the initial state. We also show that this method may be used to derive evolution operators for quantum trajectories driven by the Poisson process, although we include this essentially just to show that a unified treatment may be used for both cases; the resulting operators do not extend significantly those which have been derived elsewhere~\cite{GK}.
We note also that various authors have found solutions to certain non-linear equations for a single particle~\cite{BS,SG,HMWsol,HCsol}. These require however that the initial state takes a Gaussian form, and do not easily generalise to other initial states due to the non-linearity of the equations. For treatments of non-linear stochastic state-vector equations see also~\cite{Belvkn,Belsol}.

Naturally the resulting evolution operators contain classical 
random variables. The complexity of the stochastic equations which govern these 
classical random variables depends upon the complexity of the 
commutation relations between the operators appearing in the LSE. If the 
complexity of the commutation relations is sufficiently high then the 
stochastic 
equations governing the classical random variables become too complex to solve 
analytically. Nevertheless, even if this is the case, the form of the evolution 
operator provides information regarding the type of states produced by the LSE, 
and the problem is reduced to integrating the classical stochastic equations 
numerically. We also note that the solution to an LSE provides additional 
information to that contained in the solution to the equivalent master 
equation, because it gives the state of the system for each trajectory. For 
example, 
the variance of a system operator may be calculated for each final state (ie. 
for each trajectory), and this is referred to as the {\em conditional} variance 
as it is conditional upon the results of the measurement. The overall average 
of these variances may be then be calculated. The solution to the master 
equation allows us to calculate only the variance which is obtained by 
first averaging 
the final states over all trajectories, which is, in general, quite a 
different quantity.

In the following treatment of stochastic equations involving the Wiener 
process we use as examples LSE's corresponding to the continuous 
measurement of physical observables. A term of the form 
\begin{equation}
  \dot{\rho} = \cdots -k[O,[O,\rho]] \cdots
\label{mterm}
\end{equation}
in a quantum master equation for the evolution of a density matrix, $\rho$, 
for a quantum system $S$, describes a continuous projection measurement of 
an observable $O$ of $S$. We refer to this measurement process as a continuous {\em projection} measurement because in the absence of any system evolution, the sole effect of this term is to reduce the off-diagonal elements of the density matrix to zero in the eigenbasis of that observable. That is, it describes, in the long time limit, a projection on to one of the eigenstates of the observable under observation. The rate at which information is gained regarding the observable is determined by $k$ which is a positive constant. Recall that in Chapter~\ref{chapter1} we gave a specific example of a continuous position measurement which took this form. Various examples of measurements of this kind may be found in references~\cite{SHM,MilCav,Barchielli,Imoto}. If, in addition, the observable commutes with the Hamiltonian describing the free evolution of the system under observation, then the free evolution does not interfere with this process of projection, and the measurement is referred to as a continuous Quantum Non-Demolition (QND) measurement~\cite{WandM,QNDdefn}.

\section{The Wiener Process: General Method}
We will explicitly treat here LSE's which contain only one 
stochastic increment. However it will be clear that this treatment 
may be easily extended for multiple stochastic increments. Let us write
a general LSE with a single stochastic increment as
\begin{equation}
d|\tilde{\psi}(t)\rangle = (\tilde{A} \; dt + B \; dW(t))|\tilde{\psi}(t)\rangle . 
\label{lqse}
\end{equation}
In this equation $\tilde{A}$ and $B$ are arbitrary operators. 
We will see that the complexity of the evolution operator will depend 
upon the complexity of the commutation relations between $\tilde{A}$ and $B$.

As a first step in obtaining an evolution operator for the LSE 
in Eq.(\ref{lqse}) we rewrite it in the form
\bq
|\tilde{\psi}(t+dt)\rangle = e^{(\tilde{A}-(B^2/2))dt}e^{BdW(t)}
|\tilde{\psi}(t)\rangle = e^{Adt}e^{BdW(t)}|\tilde{\psi}(t)\rangle 
\label{lqse2} ,
\eq
were we have defined $A=\tilde{A}-B^2/2$.
It is easily verified that this is correct to first order by 
expanding the exponentials to second order and using the Ito calculus 
relation $dW(t)^2=dt$.
To first order the state at time $t+\Delta t$ is therefore
\begin{equation}
|\tilde{\psi}(t+\Delta t)\rangle = e^{A\Delta t}e^{B\Delta W(t)}|\tilde{\psi}(t)\rangle 
\label{lqse3} ,
\end{equation}
so that the state at time $t$ may be written 
\begin{equation}
|\tilde{\psi}(t)\rangle_{\mbox{\scriptsize w}} = \lim_{\Delta t \rightarrow 0} 
\prod_{n=1}^{N} (e^{A\Delta t}e^{B\Delta W_n}) \; |\psi(0)\rangle , 
\label{lqse4}
\end{equation}
where
\begin{equation}
\Delta W_n=\int_{(n-1)\Delta t}^{n\Delta t} \!\!\!\!\!\!\!\!\!
\!\!\! dW(t) ,
\end{equation}
and $N\rightarrow \infty$ as $\Delta t \rightarrow 0$ so that 
$N\Delta t=t$ is always true. We have included the subscript `w' to remind us that the state is dependent on the Wiener process. 
To complete the derivation of the 
evolution operator we must take the limit in Eq.(\ref{lqse4}). 
To do this we must combine the arguments of the exponentials which 
appear in the product, so that we may sum the infinitesimals. We 
will choose to do this by first repeatedly swapping the order of 
the exponentials containing the operator $A$ with those containing 
the operator $B$. The simplest case occurs when $A$ and $B$ commute 
so that the problem essentially reduces to the single variable 
case, and 
we treat this in Sec.{\ref{QNDPH}}. The simplest non-trivial case 
occurs when the commutator $[A,B]$, while non-zero, commutes with 
both $A$ and $B$, and we treat this in subsection~\ref{pmeas}. In the 
final part of this section we examine a more complicated example 
in which the commutator $[A,B]$ does not commute with either $A$ or 
$B$.

\subsection{A QND Measurement of Photon-Number}\label{QNDPH}
The mathematically trivial case occurs when $A$ and $B$ commute. A 
non-trivial
physical example to which this corresponds is a QND measurement of 
the photon number of a single cavity mode. Denoting the annihilation 
operator describing the mode by $a$, the cavity field Hamiltonian 
is given by~\cite{WandM}
\begin{equation}
  H = \hbar\omega(a^\dagger a + \frac{1}{2}),
\end{equation}
in which $\omega$ is the frequency of the cavity mode, and the 
observable to be measured is $O = a^\dagger a$. With this we have
\begin{eqnarray}
  A & = & -i\omega(a^\dagger a + \frac{1}{2}) -  2k(a^\dagger 
a)^2 , \\
  B & = & \sqrt{2k}a^\dagger a ,
\end{eqnarray}
in which $k$ is the measurement constant introduced in 
Eq.(\ref{mterm}).
As $A$ and $B$ commute the exponentials in Eq.(\ref{lqse4}) 
combine trivially and we obtain
\bq
|\tilde{\psi}(t)\rangle_{\mbox{\scriptsize w}} = \lim_{\Delta t 
\rightarrow 0} e^{A N\Delta t}\exp\left[{B \sum_n \Delta W_n}\right] 
\; |\psi(0)\rangle = e^{A t}e^{B W(t)} \; |\psi(0)\rangle . \label{sol1}
\eq 
As the Wiener process, $W(t)$, is a sum of independent Gaussian 
distributed random 
variables, $W_n$, it is naturally Gaussian distributed, the mean 
and variance of $W(t)$ being zero and $t$ respectively. In a 
particular realisation of the stochastic equation Eq.(\ref{lqse}), 
the Wiener process will have a particular value at each time 
$t$, and as we mentioned 
above, the set of all these values corresponds to the trajectory that 
is taken by that particular realisation. The fact that to obtain the 
state at time $t$ we require only the value of the Wiener process 
at that time means that we do not require all the trajectory information, 
but just a single variable associated with that trajectory. For more 
complicated cases, in which the operators do not commute, we will 
find that other variables associated with the trajectory appear in 
the evolution operator.

\begin{figure} 
\begin{center} 
\leavevmode 
\epsfxsize=7cm 
\epsfbox{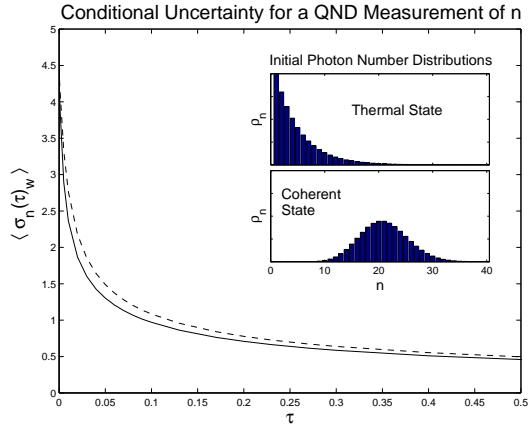}
\caption[Conditional uncertainty for a QND measurement of photon-number.]{The conditional uncertainty in photon number averaged 
over all trajectories, $\langle \sigma_n(t)_{\mbox{\scriptsize w}}\rangle$, 
is plotted here against the dimensionless scaled time, $\tau=kt$. The dotted 
line corresponds to an initial coherent state, and the solid line to an 
initial thermal state. Both initial states were chosen to have $\sigma_n^2=20$, 
giving the thermal state a mean photon number of $\langle n\rangle=4$, and the 
coherent state a mean photon number of $\langle n\rangle=20$. The photon number 
distributions for the two initial states are displayed in the inset.}
\label{nqndfig} 
\end{center} 
\end{figure}

As the situation we consider here is a QND measurement, the phase 
uncertainty introduced by the measurement of photon number does 
not feed back to affect the measurement, so that the result is simply 
to decrease continuously the uncertainty in photon number, and the 
state of the system as $t$ tends to infinity tends to a number state. 
If we denote the evolution operator derived in Eq.(\ref{sol1}) by 
$V(t)$, and start the system in an arbitrary mixed state $\rho(0)$, 
then at time $t$ the normalised state of the system may be written
\begin{equation}
   \rho(t) = \frac{V(t)\rho(0) V^\dagger(t)}{\mbox{Tr}\left\{ 
V(t)\rho(0) V^\dagger(t)\right\} }.
\end{equation}
As $V(t)$ is diagonal in the photon number basis, we only require 
the diagonal elements of the initial density matrix to calculate moments 
of the photon number operator. Denoting the diagonal elements of the 
initial density matrix by $\rho_n$, and the diagonal elements of 
$V(t)V^\dagger(t)$ by $V_n$, the variance of the photon number, for a 
given trajectory, is given by
\begin{equation}
  \sigma_{\scriptsize n}^2(t)_{\mbox{\scriptsize w}} = 
\frac{\sum_n n^2\rho_nV_n}{\sum_n \rho_nV_n} - 
\frac{\left(\sum_n n\rho_nV_n\right)^2}
     {\left(\sum_n \rho_nV_n\right)^2} .
\end{equation}
The uncertainty in our knowledge of the number of photons is the square 
root of this variance. Averaging this uncertainty over all trajectories 
therefore tells us, on average, how accurately we will have determined 
the number of photons at a later time. To calculate the value of the 
uncertainty for each trajectory, averaged over all trajectories we must 
multiply $\sigma_{\scriptsize n}(t)_{\mbox{\scriptsize w}}$ by the 
probability for each final state and average over all the final states. 
The probability measure for the final states, $\rho(t)$, is given by the 
Wiener measure multiplied by the norm of the final state, 
$\mbox{Tr}\left\{ V(t)\rho(0) V^\dagger(t)\right\}$. This probability measure 
is not in general Gaussian in $W(t)$, but a weighted sum of Gaussians, one for 
each $n$. Performing the multiplication, we obtain the average conditional 
uncertainty in photon number as
\begin{equation}
   \langle \sigma_{\scriptsize n}(t)_{\mbox{\scriptsize w}}\rangle = 
   \int \sqrt{\sum_{nm} n(m-n)\rho_n\rho_mV_nV_m} \; 
d P_{\mbox{\scriptsize w}} ,
\end{equation}
in which
\begin{eqnarray}
  V_n & = & e^{-4ktn^2 + 2\sqrt{2k}nW} , \\
  d P_{\mbox{\scriptsize w}} & = & \frac{1}{\sqrt{2\pi t}} 
e^{-W^2/(2t)}d W.
\end{eqnarray}
We note that 
$\langle \sigma_{\scriptsize n}(t)_{\mbox{\scriptsize w}}\rangle$ may be 
written as a function of $\tau =kt$, being the time scaled by the measurement 
constant. Hence, as we expect, increasing the measurement time has the same 
effect on 
$\langle \sigma_{\scriptsize n}(t)_{\mbox{\scriptsize w}}\rangle$ as 
increasing the measurement constant.
We evaluate 
$\langle \sigma_{\scriptsize n}(\tau)_{\mbox{\scriptsize w}}\rangle$ 
numerically for an initial thermal state, and an initial coherent state, and 
display the results in figure~\ref{nqndfig}. We have chosen the initial states 
so that they have the same uncertainty in photon number, with the result that 
the mean number of photons in each of the two states is quite different. The 
results show the decrease in uncertainty with time, which is seen to be only 
weakly dependent upon the initial state. 

\subsection{A Measurement of Momentum in a Linear Potential}\label{pmeas}
The simplest mathematically non-trivial case occurs when the commutator 
between $A$ 
and $B$, while non-zero, is such that it commutes with both $A$ and $B$. 
A physical situation to which this corresponds is a continuous measurement 
of the momentum of a particle in a linear potential. If we denote the 
position and momentum operators for the particle as $Q$ and $P$ respectively, 
then the Hamiltonian is given by
\begin{equation}
H = \frac{1}{2m}P^2 - FQ,
\end{equation}
in which $m$ is the mass of the particle and $F$ is the force on the 
particle from the linear potential. In this case we have
\begin{eqnarray}
  A & = & (\frac{-i}{2\hbar m} - 2k)P^2 + \frac{iF}{\hbar}Q , \\
  B & = & \sqrt{2k}P ,
\end{eqnarray}
in which $k$ is again the measurement constant.

Returning to Eq.(\ref{lqse4}) we see that to obtain a solution we must 
pass all the exponentials containing the operator $B$ to the right through 
the exponentials containing the operator $A$. In order to perform this 
operation we need a relation of the form
\begin{equation}
e^{B}e^{A} = e^{A}e^{C} . \label{oprel}
\end{equation}
For the present case the required relation is simply given by the 
Baker-Campbell-Hausdorff formula~\cite{BCH,BHE}
\begin{equation}
e^{B}e^{A} = e^{A}e^{B}e^{-[A,B]} .
\end{equation}
Using this relation to propagate successively all of the exponentials 
containing $B$ to the right in the product in Eq.(\ref{lqse4}) we 
obtain
\begin{equation}
\prod_{n=1}^{N} (e^{A\Delta t}e^{B\Delta W_n}) = 
\exp\left[{AN\Delta t}\right]\exp\left[{B \sum_{n=1}^N\Delta W_n} 
\right] \exp\left[{-[A,B]\Delta t \sum_{n=1}^N(n-1)\Delta W_n} \right] .
\end{equation}
All that remains is to calculate the joint probability density for the 
random variables. The first is the Wiener process, and the second is
\begin{eqnarray}
Y(t) & = & \lim_{\Delta t \rightarrow 0}\Delta t \sum_{n=1}^N(n-1)
\Delta W_n = \int_0^t t' dW(t') . 
\end{eqnarray}
Clearly these are both Gaussian distributed with zero mean and 
all that we require is to calculate the covariances 
$\langle Y(t)^2\rangle$ and $\langle 
W(t)Y(t)\rangle$ to determine completely the joint density 
at time $t$. Using $\langle \Delta W_n \Delta W_n \rangle = 
\delta_{nm}\Delta t$ these quantities are easily obtained:
\bqa
\langle Y(t)^2\rangle & = & \lim_{\Delta t \rightarrow 0} 
\sum_{n=1}^N((n-1)\Delta t)^2 \Delta t = \int_0^t t'^2 \; dt' = t^3/3 , \\
\langle W(t)Y(t)\rangle & = & \lim_{\Delta t \rightarrow 0} 
\sum_{n=1}^N((n-1)\Delta t) \Delta t = \int_0^t t' \; dt' = t^2/2 .
\eqa
The state at time $t$, under the evolution described by the 
stochastic equation, is 
therefore
\begin{equation}
|\tilde{\psi}(t)\rangle_{\mbox{\scriptsize w}} = e^{At}e^{BW(t)}
e^{-[A,B]Y(t)}|\psi(0)\rangle ,
\end{equation}
where the joint probability density for $W$ and $Y$ at time $t$ 
is given by 
\begin{eqnarray}
  P_{\mbox{\scriptsize w}}(W,Y) = \left( \frac{\sqrt{12}}{2\pi t^2} \right) 
\exp\left[-\frac{2}{t}W^2 - \frac{6}{t^3} Y^2 + \frac{6}{t^2}WY \right] . 
\nonumber 
\end{eqnarray}
Note that to obtain the probability density for the final state, this must be multiplied by the norm of the state at time $t$.                                                                 
Returning to the specific case of a particle in a linear potential, we may 
now obtain results for various quantities of interest. Writing the evolution
operator in terms of the momentum and position operators we have
\bq
|\tilde{\psi}(t)\rangle_{\mbox{\scriptsize w}} = \exp\left[{((\frac{-i}{2\hbar 
m} - 2k)P^2 + \frac{iF}{\hbar}Q)t}\right] \exp\left[\sqrt{2k}
(PW(t)+FY(t))\right] |\psi(0)\rangle .
\eq
Using the Zassenhaus formula~\cite{Witschel} to disentangle the 
argument of the first exponential we may rewrite this in 
the more convenient form
\bq
|\tilde{\psi}(t)\rangle_{\mbox{\scriptsize w}} = \exp\left[{ 
\frac{iF}{\hbar}Qt}\right]  \exp\left[{\eta (-P^2t - PFt^2 - 
F^2 t^3/3)}\right] \exp\left[{ \sqrt{2k}(PW(t)+FY(t))}
\right]  |\psi(0)\rangle ,
\label{psit2}
\eq
in which $\eta = (\frac{i}{2\hbar m} + 2k)$. For those not familiar 
with the Zassenhaus formula, it is complementary to the BCH formula. 
While the BCH formula shows how to write a product of the exponentials of two operators as an exponential of the sum of the operators and their commutator (or in more complicated cases repeated commutators of the operators), the Zassenhaus formula shows how to write an exponential of the sum of two operators as a product of the exponentials of the operators and repeated commutators.

Let us first take an arbitrary initial state, writing it in the momentum 
eigenbasis so that we have
\begin{equation}
|\psi(0)\rangle = \int_{-\infty}^{\infty} \!\!\!\!\! \Psi(p)|p\rangle  
\; d p  \;\; , \;\;\;  \int_{-\infty}^{\infty} \!\!\!\!\! |\Psi(p)|^2 
\; d p = 1 .
\label{psi0}
\end{equation}
Using Eqs.(\ref{psit2}) and (\ref{psi0}) in Eq.(\ref{avo}) to calculate the 
moments of $p$ given by first averaging over all trajectories (that is, 
the moments which would be given by the equivalent master equation) we 
readily obtain
\begin{equation}
 \langle p(t)^n\rangle = \langle (p(0)+Ft)^n\rangle .
\end{equation}
In particular, for any initial state, $|\psi(0)\rangle$, the average value 
of the momentum at time $t$, $\langle p(t)\rangle$, is simply shifted from 
the initial value by the impulse $Ft$. The variance of the momentum at time 
$t$, $\sigma_p^2(t) = \langle p(t)^2\rangle - \langle p(t)\rangle^2$, 
remains equal to its original value. That is, the uncertainty introduced into 
the position of the particle by the momentum measurement does not feed back 
into the momentum, even though the momentum does not commute with the 
Hamiltonian. This is because while the momentum determines the position 
at a later time, the converse is not true. These results for the moments 
are easily checked using the equivalent master equation.

Now let us consider the conditional variance of the momentum at time $t$
averaged over all trajectories. In the previous section we calculated the 
conditional uncertainty, being the square root of the variance, and 
averaged this over all trajectories. Here however, we will find that the 
conditional variance is independent of the trajectory taken, and depends 
only on the measurement time. This will also be true of the example which 
we will treat in the next section. In this case clearly it does not matter 
if we first average the conditional variance over the trajectories, and 
then take the square root, or if instead we average the conditional 
uncertainty, because the averaging procedure is redundant. However, in 
general the two procedures are not equivalent.  We will denote 
the conditional variance by 
$\langle\sigma_p^2(t)_{\mbox{\scriptsize w}}\rangle$. As the uncertainty in 
position does not feed back into the momentum, we expect that this variance 
should steadily decrease to zero. This is because during a trajectory our 
knowledge of the momentum steadily increases so that the distribution over 
momentum becomes increasingly narrow. To perform this calculation 
we take the initial state to be the minimum uncertainty wave-packet given by 
the ground state of a harmonic oscillator of frequency $\omega$. The average 
values of the position and momentum of the particle are both zero in this 
state and the respective variances are 
\begin{eqnarray}
 \langle Q^2\rangle & = & \frac{\hbar}{2m\omega} \equiv \sigma_Q^2(0) \;\; 
, \;\;\; \langle P^2\rangle =  \frac{m\hbar\omega}{2} \equiv \sigma_p^2(0) 
\nonumber ,
\end{eqnarray}
and in  momentum space the state may be written
\begin{equation}
 |\psi(0)\rangle = \left(\frac{1}{\pi m\hbar\omega}\right)^{\frac{1}{4}} 
\int_{-\infty}^{\infty} \!\!\!\!\! e^{-P^2/(2m\hbar\omega)} |p\rangle   
\; d p.
\end{equation}
The moments of momentum for each trajectory are given by
\begin{equation}
 \langle p^n\rangle_{\mbox{\scriptsize w}} = \frac{\langle\tilde{\psi}(t)|p^n|\tilde{\psi}(t)\rangle_{\mbox{\scriptsize w}}}{\langle\tilde{\psi}(t)|\tilde{\psi}(t)\rangle_{\mbox{\scriptsize w}}},
\end{equation}
and we calculate the first and second to give 
$\sigma_p^2(t)_{\mbox{\scriptsize w}} = \langle 
p(t)^2\rangle_{\mbox{\scriptsize w}} - \langle p(t)
\rangle^2_{\mbox{\scriptsize w}}$. We obtain
\begin{equation}
  \langle\sigma_p^2(t)_{\mbox{\scriptsize w}}\rangle = 
\frac{\sigma_p^2(0)}{1 + 8k\sigma_p^2(0)t } .
\end{equation}
This is independent of $W$ and $Y$ and hence independent of the 
trajectory. It is therefore unnecessary to average over the final 
states. 
Indeed $\langle\sigma_p^2(t)_{\mbox{\scriptsize w}}\rangle$ decreases 
steadily from the initial value to zero as $t\rightarrow\infty$ as we 
expect from the discussion above. This means that while the average value 
of momentum is determined by the measurement results, the error in our 
estimate of the momentum at time $t$ is not.

\subsection{A Quadrature Measurement with a General Quadratic 
Hamiltonian}\label{quadm}
We now consider an LSE in which the 
commutator $[A,B]$ does not commute with either $A$ or $B$. As in the 
previous example, let $P$ and $Q$ be, respectively, the canonical 
momentum and position operators for a single particle so that they obey 
the canonical commutation relation $[Q,P]=i\hbar$. With this definition 
we will take $A$ and $B$ to have the following forms:
\begin{eqnarray}
A & = & \alpha P^2 + \gamma Q^2 + \xi QP + \eta P + \zeta Q , 
\label{aexp} \\
B & = & kQ+\kappa P , \label{bexp}
\end{eqnarray}
where $\alpha, \gamma, \eta, \zeta, k$ and $\kappa$ are complex numbers. 
This example applies to an optical mode of the electromagnetic field, 
including classical driving and/or classically driven subharmonic 
generation~\cite{subharm} and for which an arbitrary quadrature is 
continuously measured. It also applies to the situation of a single 
particle, which may feel a linear and/or harmonic potential, and which 
is subjected to continuous observation of an arbitrary linear combination 
of its position~\cite{MJW} and momentum.

To obtain an evolution operator for the LSE with this choice of the 
operators $A$ and $B$, we require, as before, a relation of the form 
given by Eq.(\ref{oprel}). To derive this relation we proceed in the 
following manner.

First we may use the Baker-Campbell-Hausdorff expansion~\cite{BHE}, or 
alternatively solve the equations of motion given by $dB/d\epsilon = 
[A,B]$, to obtain an expression for $e^{\epsilon A}Be^{-\epsilon A}$. 
The result is
\begin{equation}
e^{-\epsilon A}\varepsilon Be^{\epsilon A} = \varepsilon f_1(\epsilon) 
Q +\varepsilon f_2(\epsilon) P + \varepsilon f_3(\epsilon) , 
\label{orel1}
\end{equation}
in which
\begin{eqnarray}
f_1(\epsilon) & = & \frac{1}{\lambda} (-2\kappa\gamma + k\xi) S 
+ k C, \\
f_2(\epsilon) & = & \frac{1}{\lambda} (2k\alpha - \kappa\xi) S 
+ \kappa C, \\
f_3(\epsilon) & = & \frac{1}{\lambda^2} (k\eta\xi + 2k\alpha\zeta
- \kappa\zeta\xi - 2\kappa\gamma\eta) [C-1] + \frac{1}{\lambda} (k\eta 
+ \kappa\zeta) S .
\end{eqnarray}
In these expressions $C=\cosh(i\hbar\lambda\epsilon )$, $S=
\sinh(i\hbar\lambda\epsilon )$ and $\lambda=\sqrt{\xi^2-4\alpha\gamma}$. 
Using the relation
\begin{equation}
e^{-\epsilon A}f(\varepsilon B)e^{\epsilon A} = f(e^{-\epsilon A}
\varepsilon Be^{\epsilon A}) ,
\end{equation}
we obtain from Eq.(\ref{orel1})
\begin{equation}
e^{-\epsilon A}e^{\varepsilon B}e^{\epsilon A} = e^{\varepsilon 
f_1(\epsilon) Q +\varepsilon f_2(\epsilon) P}e^{\varepsilon f_3
(\epsilon)} .
\end{equation}
Multiplying both sides of this equation on the left by 
$e^{\epsilon A}$ we obtain a relation of the form $e^{\varepsilon B}
e^{\epsilon A}=e^{\epsilon A}e^{\varepsilon D(t)}$, as we require.

We see from the above procedure that the relation in Eq.(\ref{oprel}) 
may be obtained so long as a closed form can be found for the solution 
to the operator differential equation $dB/d\epsilon = [A,B]$. Clearly 
this is straightforward if this equation is linear in $B$, which is 
true in the example we have treated here, and is sometimes possible in 
cases in which the equations are non-linear. In addition, for this example we also require the BCH relation in the form
\begin{equation}
e^{A}e^{B}=e^{A+B+\frac{1}{2}[A,B]} . \label{BCH}
\end{equation}
This is so that we can sum up in one exponential the operators that 
result from swapping $e^{\Delta W_n B}$ and $e^{n\Delta t A}$.

Using the expressions derived above, with the replacements 
$\epsilon=n\Delta t$ and $\varepsilon=\Delta W_n$, for each $n$ from 1 
to $N$, by repeatedly swapping the exponentials containing $B$ with 
those containing $A$ as in the previous example, we obtain
\begin{eqnarray}
\lim_{\Delta t \rightarrow 0}\prod_{n=1}^{N} (e^{A\Delta t}
e^{B\Delta W_n}) & = & 
e^{At}e^{X_1(t) Q + X_2(t) P}e^{X_3(t)}e^{i\hbar Z(t)} ,
\end{eqnarray}
in which the classical stochastic variables $X_i$ and $Z$, are 
given by
\begin{eqnarray}
 X_i(t) & = & \int_0^t \!\!\! f_i(t') dW(t'), \\
 Z(t) & = & \int_0^t \!\!\! f_1(t')X_2(t') dW(t') - \int_0^t \!\!
\! f_2(t')X_1(t') dW(t'), \nonumber
\end{eqnarray}
where the expressions for the $f_i$ are given above, and the 
integrals are Ito integrals. The $X_i$ are Gaussian distributed with 
zero mean, and their covariances are easily calculated as in the 
previous example:
\begin{equation}
\langle X_i(t)X_j(t)\rangle = \!\! \int_{0}^{t}\!\!\!\! f_i(t')
f_j(t')\; dt' . 
\end{equation}
In addition, the two-time correlation functions for these variables 
are also easily obtained analytically. In particular we have
\begin{equation}
\langle X_i(t)X_j(\tau)\rangle = \!\! \int_{0}^{\mbox{min}(t,\tau)}
\!\!\!\! f_i(t')f_j(t')\; dt' . 
\end{equation}
However, $Z(t)$ is not Gaussian distributed. We are not aware of an 
analytic expression for this variable, so that its probability 
density may have to 
be obtained numerically. We note in passing, however, that in some 
cases double stochastic integrals of this kind may be written explicitly 
in terms of products of Gaussian variables~\cite{Gardiner1}. We note also that 
$Z$ determines only the normalisation of the final state, and not the state 
itself. The normalised state at time $t$ is therefore independent of $Z$, and 
we discuss the consequences of this in subsection~\ref{openq1}.

We may now write the state at time $t$ as
\begin{equation}
|\tilde{\psi}(t)\rangle_{\mbox{\scriptsize w}} = e^{At}e^{X_1(t)Q + X_2(t)P}
e^{X_3(t) + i\hbar Z(t)} |\psi(0)\rangle .
\label{eop3}
\end{equation}
 Hence even though values for averages 
over all trajectories may in general have to be calculated numerically,  
the evolution operator provides us with information regarding the type 
of states that will occur at time $t$. In particular, if the initial 
state is 
Gaussian in position (and therefore also Gaussian in momentum), then as 
each of the exponential operators in the above equation transform 
Gaussian states to Gaussian states, we see that the state of the system 
remains Gaussian at all times. The mean of the Gaussian in 
both position and momentum change with time in a random way determined 
by the values of the stochastic variables.

We will shortly consider a particular example: that of a harmonic oscillator 
undergoing a continuous observation of position, and use this evolution 
operator to calculate the conditional variance for the position at time 
$t$. We will take the initial state to be a coherent state, which is a 
Gaussian wave packet. This conditional variance does not depend upon the 
trajectory, but simply upon the initial state and the measurement time, 
as indeed we found to be the case for the momentum measurement in 
section~\ref{pmeas}.

Let us first show that for an initial coherent state the conditional 
variance of any linear combination of position and momentum is 
independent of the trajectory for all of the cases covered by the 
evolution operator in Eq.(\ref{eop3}). To do this we must calculate 
the effect of this evolution operator on a coherent state. Clearly 
the effect of the right-most exponential operator is at most to 
change the normalisation, which effects neither the average values 
of position and momentum, nor the respective variances. The effect 
of the next exponential, being linear in $P$ and $Q$, is calculated 
in appendix~\ref{apB}. We find that it changes the mean values of the 
position and momentum, and alters the normalisation, but the state 
remains coherent in that the position variance (and hence the momentum 
variance) is unchanged. Finally, 
the effect of the exponential quadratic in $P$ and $Q$ is calculated 
in appendix~\ref{apB}. We find that this operator modifies the position 
variance. However, as the operator does not contain any stochastic 
variables, and as the manner in which it changes the position variance 
is independent of the mean position and momentum, we obtain the result 
that the effect on the position variance, and hence the variance of 
any linear combination of position and momentum, is trajectory 
independent. 

Let us now consider a harmonic oscillator in which the position is 
continuously 
observed. This situation has been analysed by Belavkin and 
Staszewski using the equivalent non-linear equations~\cite{BS}. 
The operators $A$ and $B$ in this case are given by
\begin{eqnarray}
  A & = & \left( \frac{-i}{2\hbar m} \right)P^2 + \left( \frac{-im\omega^2}
{2\hbar}  - 2k \right)Q^2 , \\
  B & = & \sqrt{2k}Q ,
\end{eqnarray}
in which $m$ is the mass of the particle, $\omega$ is the frequency of 
the harmonic oscillation, and $k$ is the measurement constant for the 
continuous observation of position. Taking the initial state to be 
coherent, and denoting it $|\alpha\rangle$, the initial position 
wave-function is given by 
\begin{equation}
 \langle x|\alpha\rangle = \left( \frac{2s^2}{\pi} \right) ^{1/4} 
e^{-s^2x^2+2sx\alpha -\frac{1}{2}(|\alpha|^2 + \alpha^2)}, 
\end{equation}
where $s^2=m\omega/(2\hbar)$. Using the results in 
appendix~\ref{apB}, we 
find that the coefficient of $x^2$ at a later time $t$ is given by
\begin{equation}
 s'^2 = s^2 \left[ \frac{1-2l}{3-2l} \right] \left[ 1 + 
2\frac{1-2l}{1+2l} \right]
\end{equation}
where
\begin{equation}
  l = \frac{-1/2}{rz\coth(z\omega t) + (1 + ir)} ,
\end{equation}
and we have defined the parameters
\begin{equation}
  z = \sqrt{\frac{2i}{r} - 1} \;\; , \;\;\; r = \frac{m\omega^2}
{2\hbar k} .
\end{equation}
After some algebra this may be written as
\begin{equation}
 s'^2 = s^2 iz \frac{iz\tanh(z\omega t) - 1}{\tanh(z\omega t) - iz} ,
\label{deltax}
\end{equation}
in agreement with that derived by Belavkin and Staszewski.
The conditional variance for $x$ at time $t$ is given by
\begin{equation}
 \sigma_x^2(t)_{\mbox{\scriptsize w}} = 
\frac{1}{4\mbox{\small Re}[s'^2]} .
\end{equation}
As $t$ tends to infinity, Eq.(\ref{deltax}) gives a steady state 
value for the conditional variance, which is
\begin{equation}
 \sigma_x^2 = \frac{1}{4\mbox{\small Im}[z]} =  \left( \sqrt{2}s^2 
\sqrt{\sqrt{4/r^2 + 1} + 1} \right)^{-1}.
\end{equation}
The parameter $r$ is a dimensionless quantity which gives essentially 
the ratio between the frequency of the harmonic oscillator, and the rate 
of the position measurement. We may view the dynamics of the position 
variance as being the result of two competing effects. One is the action 
of the measurement which is continuously narrowing the distribution in 
position, and consequently widening the distribution in momentum. The 
other is the action of the harmonic motion, which rotates the state in 
phase space, so converting the widened momentum distribution into 
position. Depending on the relative strengths of these two processes, 
determined by the dimensionless constant $r$, a steady state is 
reached in which they balance. If the rate of the measurement is very 
fast compared to the frequency of the oscillation (corresponding to 
$r \ll 1$), then the localisation in position is much greater than it 
would be for an unmonitored oscillator, and in that case we succeed 
effectively in tracking the position of the particle. However, if the 
frequency of oscillation is much greater than the rate of localisation 
due to the measurement, then the steady state position variance remains 
essentially that of the unmonitored oscillator.

\subsection{An Open Question}\label{openq1}
The results obtained in the previous section are curious in the light of the solution obtained by various authors for the non-linear equation describing a continuous position measurement for the case of an initial Gaussian state~\cite{BS,SG}. It is clear from this solution that the probability density for the final states is Gaussian. It is surprising therefore that the evolution operator for the equivalent linear equation describing this situation contains random variables which are not Gaussian. These two facts are compatible because the non-Gaussian variable in the evolution operator given in Eq.(\ref{eop3}) affects only the normalisation of the final state, and not the state itself. Because of this it is, at least in theory, possible to eliminate the non-Gaussian variable from the solution. This may be seen as follows.
Let us assume that we have an initial state $|\psi\rangle$, and an 
evolution operator which is a function of the random variables $X$ and 
$Z$ (which may in general be vector valued). We let the random variable 
$Z$ determine only the normalisation of the final state, so that the 
evolution operator may be written
\begin{equation}
   V(X,Z,t) = {\cal O}(X,t)f(Z,t) ,
\end{equation}
where ${\cal O}$ is an operator valued function, and $f$ is simply a 
complex valued function. The unnormalised state at time $t$ is then 
given by
\begin{equation}
  |\tilde{\psi}(t)\rangle_{\mbox{\scriptsize w}} = {\cal O}(X,t)f(Z,t)
|\psi\rangle .
\end{equation}
Clearly once we have normalised that state at time $t$, it is no-longer 
dependent upon $Z$. In particular the normalised state is given by
\begin{equation}
  |\psi(t)\rangle_{\mbox{\scriptsize w}} = 
\frac{ {\cal O}(X,t)|\tilde{\psi}\rangle_{\mbox{\scriptsize w}}}
        {\sqrt{ \langle \tilde{\psi} |{\cal O}^\dagger(X,t)
                              {\cal O}(X,t)
|\tilde{\psi}\rangle_{\mbox{\scriptsize w}} }} .
\end{equation}
The probability density for the final state is 
\begin{equation}
  P(X,Z,t) = \langle \tilde{\psi}(t)|\tilde{\psi}(t)
\rangle_{\mbox{\scriptsize w}} P_{\mbox{\scriptsize w}}(X,Z,t) ,
\end{equation}
in which $P_{\mbox{\scriptsize w}}(X,Z,t)$ is the probability 
density given by the Wiener measure 
for the variables $X$ and 
$Z$. However, seeing as the normalised state depends only upon $X$, 
we require for all calculations only the marginal probability density 
for $X$. Denoting this marginal density also by $P$, we have
\begin{equation}
  P(X,t) = \int P(X,Z,t) \; \mbox{d}Z .
\end{equation}
In certain cases the probability measure for the normalised state may 
therefore be Gaussian, even though the measure for the unnormalised 
state is not. However, as $P(X,Z,t)$ contains a factor of the norm of 
$|\psi(t)\rangle_{\mbox{\scriptsize w}}$, the probability measure for 
the output process will, in general, only be Gaussian if the norm is 
Gaussian in $X$. Clearly the norm is Gaussian in $X$ for initial Gaussian 
states in the case we investigate in section~\ref{quadm}.

In the case considered in subsection~\ref{quadm} it is not clear how to evaluate the marginal probability density in which the non-Gaussian variable has been eliminated because we do not have an expression for the joint distribution of all the variables including $Z$. Whether it is possible to obtain an evolution operator in which the variable $Z$ has been eliminated remains therefore an open question. It seems almost certain, however, that once the non-Gaussian variable has been eliminated the resulting distribution for the final state will be Gaussian in the remaining variables.  

\section{Extension to the Poisson Process}
A linear stochastic equation describing the evolution of a system undergoing quantum jumps takes the form
\begin{equation}
  |\tilde{\psi}(t+dt)\rangle_{\mbox{\tiny N}} = [1 + Ad t + Bd N]|\tilde{\psi}(t)\rangle_{\mbox{\tiny N}}
\end{equation}
in which $d N$ is the increment of a Poisson process, and $A$ and $B$ are arbitrary operators. The Poisson increment, $d N$, is either equal to zero or one, with the result that $(d N)^2=d N$. If during any infinitesimal time interval $d N=0$ then no jump occurs, whereas if $d N=1$ a jump occurs at that instant. The rate of jumps is given by $\lambda$, so that $\langle d N\rangle = \lambda d t$. The average rate of jumps may be chosen so as to render the expressions involved in deriving the solution in their simplest form.
We now note that
\begin{equation}
  1 + Bd N = e^{\ln(B+1)\mbox{\scriptsize d}N}
\end{equation}
where we have used $(d N)^2=d N$. Using this we may write the linear differential equation as
\begin{equation}
  |\tilde{\psi}(t+d t)\rangle_{\mbox{\tiny N}} = e^{A\mbox{\scriptsize d}t}e^{\ln(B+1)\mbox{\scriptsize d}N}|\tilde{\psi}(t)\rangle_{\mbox{\tiny N}}
\end{equation}
where we have used $d td N=0$. The state of the system at time $t$ may now be written as
\begin{equation}
  |\tilde{\psi}(t)\rangle_{\mbox{\tiny N}} = \lim_{M\rightarrow\infty} \prod_{m=1}^{M}e^{A\Delta t}e^{\ln(B+1)\Delta N_{m}}|\psi(0)\rangle
\label{prod}
\end{equation}
where we have cut the time into $M$ slices so that $\Delta t=t/M$. In order to derive a closed form for the evolution operator we must swap the exponentials in $A$ and $B$ repeatedly so as to move all the exponentials containing $\Delta N$ to one side. Once that is achieved the infinitesimals must then be summed. Then, in the limit as $M\rightarrow\infty$ these sums form integrals. The complexity of the result will depend upon the commutation relations between $A$ and $B$, and this procedure will not result in a closed form solution for the evolution operator if the commutation relations are two complex. To illustrate this method we turn now to a concrete example.

\subsection{A Damped Cavity with Coherent Driving}
We consider first an undriven optical cavity mode which is allowed to decay through one of the end mirrors, and then extend this to include coherent driving. A photo-detector is placed in the beam which is output from the cavity. The photons are detected one-by-one as they leak out of the cavity, and a quantum jump in the evolution of the system corresponds to the detection of a photon. A linear stochastic equation describing this situation is
\bq
  |\tilde{\psi}(t+d t)\rangle_{\mbox{\tiny N}} = [1 - ((i\omega + \gamma/2) a^\dagger a - \gamma/2)d t + (a-1)d N ] \;\; |\tilde{\psi}(t)\rangle_{\mbox{\tiny N}} ,
\eq
in which $\omega$ is the cavity resonance frequency, $\gamma$ is the cavity decay rate, $a$ is the cavity annihilation operator, and we have chosen $\lambda=\gamma$ to select the simplest expression.
The operators $A$ and $B$ are therefore
\begin{eqnarray}
  A & = & -(i\omega + \gamma/2) a^\dagger a - \gamma/2, \\
  B & = & a - 1.
\end{eqnarray}
To swap the exponentials in Eq.(\ref{prod}) we now require a relation of the form
\begin{equation}
   ae^{\varepsilon a^\dagger a} = e^{\varepsilon a^\dagger a}f(a) .
\end{equation}
 Multiplying both sides on the left by $e^{-\varepsilon a^\dagger a}$, we see that $f$ is given by
\begin{equation}
   f(a) = e^{-\varepsilon a^\dagger a} a e^{\varepsilon a^\dagger a}.
\end{equation}
This is simply the time evolution of the operator $a$ in the situation in which the Hamiltonian is proportional to $a^\dagger a$. This is easily calculated with standard quantum optical techniques, and we obtain $f(a)=ae^{\varepsilon}$. We see that the first requirement for our procedure to produce an evolution operator in closed form is that we can find the evolution of the annihilation operator when $A$ is taken as the Hamiltonian. The second condition is that  operators that result from performing the swapping (that is, $f(a)$) allow the Poisson increments to be summed together.

Swapping all the terms in Eq.(\ref{prod}), and taking the limit as $M\rightarrow\infty$, we obtain the evolution operator:
\begin{equation}
  |\tilde{\psi}(t)\rangle = e^{[-(i\omega + \gamma/2)a^\dagger a + \gamma/2]t}e^{-(i\omega + \gamma/2)Y(t)}a^{N(t)}|\psi(0)\rangle .
\label{evop}
\end{equation}
Here $N(t)$ is simply the Poisson process, being the sum of the Poisson increments, and therefore the total number of photon detections up until time $t$. The second random variable which appears in the evolution operator, $Y(t)$, is given by
\begin{equation}
  Y(t) = \int_0^t t' d N(t') = \sum_{k=1}^{N(t)}t_k ,
\end{equation}
in which the $t_k$ are the times at which the photo-detections have occurred up until time $t$. The probability distribution for $N(t)$ is simply the Poissonian
\begin{equation}
  \tilde{P}(N,t) = e^{-\lambda t}\frac{(\lambda t)^N}{N!} .
\end{equation}
The probability distribution for $Y(t)$ may be calculated as follows. First note that the probability for each of the $N$ photon-detection times, {\em given} that $N$ detections have occurred, is uniformly distributed in the interval $[0,t]$. The probability density for the sum of the emission times, is the convolution of the probability densities for each emission time. The probability density for $Y(t)$ given that there have been $N$ emissions within time $t$, which is denoted by $\tilde{P}(Y|N,t)$, is therefore the convolution of $N$ probability densities of the form
\begin{equation}
  \tilde{P}(t') = \left\{ \begin{array}{ll} 1/t & 0<t'<t \\  0 & \mbox{otherwise} \end{array} \right. .
\end{equation}
The joint probability density for $N$ and $Y$ is then $\tilde{P}(N,Y,t)=\tilde{P}(N,t)\tilde{P}(Y|N,t)$.

Examining the evolution operator, we see that the first term gives the evolution of the system resulting from the free Hamiltonian, including a smooth decay. The second term, containing $Y$, accounts for the alteration to this evolution from the loss of photons in the quantum jumps. The final term, being a product of annihilation operators, describes the effect of the jumps.

The evolution operator may now be used to obtain the final state of the system for any initial state, given a particular trajectory. In particular, as the operators which appear in this evolution operator act on coherent states in a simple way, the final state of the system may be obtained most readily by writing an initial state in the coherent state basis. Averages over all trajectories may then be calculated by using the distributions for $N$ and $Y$, along with the norm of the final state, which is also a function of $N$ and $Y$. The most complex aspect of the result is the distribution for $Y$. However, due to the fact that $Y$ determines only the norm of the final state, we may eliminate this variable from the solution. We now perform this elimination.

We note that once we have normalised the final state, and ignored any overall phase factor, the final state will not depend upon $Y(t)$. We do not require, therefore, the full probability density for the final states which is a joint probability in $N$ and $Y$, but only the marginal probability obtained by integrating over $Y$. The probability density for the final states is given by
\begin{equation}
  P(N,Y,t) = \langle\tilde{\psi}(t)|\tilde{\psi}(t)\rangle_{\mbox{\tiny N}} \tilde{P}(N,Y,t) .
\end{equation}
We may write $\langle\psi(t)|\psi(t)\rangle_{\mbox{\tiny N}}=h(N,t)e^{-\gamma Y(t)}$. The marginal probability may now be calculated as follows:
\begin{eqnarray}
  P(N,t) & = & h(N,t)\tilde{P}(N,t)\int \tilde{P}(Y|N,t) e^{-\gamma Y(t)} d Y \\
                 & = & h(N,t)\tilde{P}(N,t) \left[ \int_0^t \tilde{P}(t')e^{-\gamma t'} \right]^N \\
                 & = & h(N,t)\tilde{P}(N,t) \left[ \frac{(1-e^{\gamma t})}{\gamma t}\right]^N
\end{eqnarray}
where in the second line we have used the fact that the $t_k$ are all independent. Now we have the probability density for the final states in terms of the norm calculated excluding $Y$. We may now write the state at time $t$ using the simplified evolution operator which excludes $Y$, as
\begin{equation}
  |\hat{\psi}(t)\rangle_{\mbox{\tiny N}} = e^{[-(i\omega + \gamma/2)a^\dagger a + \gamma/2]t}a^{N(t)}|\psi(0)\rangle ,
\label{evop2}
\end{equation}
where the probability for the final state may be written as
\begin{equation}
  P(N,t) = \langle\hat{\psi}(t)|\hat{\psi}(t)\rangle_{\mbox{\tiny N}}
\tilde{P}(N,t) \left[ \frac{(1-e^{\gamma t})}{\gamma t}\right]^N .
\label{probden}
\end{equation}
This evolution operator was first obtained by Garraway and Knight~\cite{GK}, using a somewhat different approach. They applied this to an initial `Schr\"{o}dinger Cat' state, which is a superposition of two coherent states $|\alpha\rangle$ and $|\beta\rangle$. As they did not give the probability density for the final states, we will use this as our example. Setting $\beta=-\alpha$ for simplicity, so that the magnitude of the initial coherent states are the same, the initial state is
\begin{equation}
  |\psi(0)\rangle = \frac{|\alpha\rangle + |-\alpha\rangle}{\sqrt{2(1+e^{-2|\alpha|^2})}} .
\end{equation}
Applying the evolution operator in Eq.(\ref{evop2}), and subsequently normalising, we obtain the state at time $t$ to be
\begin{equation}
  |\psi(t)\rangle_{\mbox{\tiny N}} = \frac{|\alpha e^{-(i\omega + \gamma/2)t}\rangle + (-1)^{N(t)}|-\alpha e^{-(i\omega + \gamma/2)t}\rangle}{\sqrt{2(1+(-1)^{N(t)}e^{-2|\alpha|^2e^{-\gamma t}})}}
\label{catev}
\end{equation}
The probability density for the final states, given in Eq.(\ref{probden}), is found to be
\bq
  P(N,t) = \left(\frac{e^{-|\alpha|^2(1-e^{-\gamma t})}}{1+e^{-2|\alpha|^2}}\right)  \frac{(|\alpha|^2(1-e^{-\gamma t}))^N}{N!} + \left(\frac{e^{-|\alpha|^2(1+e^{-\gamma t})}}{1+e^{-2|\alpha|^2}}\right) \frac{(-|\alpha|^2(1-e^{-\gamma t}))^N}{N!} ,
\eq
which is the sum of two Poissonians.
We see from Eq.(\ref{catev}) that at all times during the evolution the cavity mode remains in a `Cat' state, although the magnitude of the `Cat' decays smoothly away over time. In addition, the state toggles between an even and odd `Cat' at each photon detection.

Now that we have elucidated the procedure, let us extend the model to include coherent driving of the cavity mode. The operators appearing in the linear stochastic equation are now
\begin{eqnarray}
  A & = & -(i\omega + \gamma/2) a^\dagger a + (Ea^\dagger - E^*a) - \gamma/2, \\
  B & = & a - 1 ,
\end{eqnarray}
where $E$ is the amplitude of the driving field in units of $\mbox{s}^{-1}$. We obtain the evolution operator for this equation as
\begin{equation}
  |\tilde{\psi}(t)\rangle_{\mbox{\tiny N}} = e^{At}\prod_{k=1}^{N(t)}((a - \alpha_{\mbox{\scriptsize s}})e^{-\gamma/2 t_k} + \alpha_{\mbox{\scriptsize s}}) |\psi(0)\rangle \; ,
\label{evop3}
\end{equation}
where $\alpha_{\mbox{\scriptsize s}} = (2E/\gamma)$. Note that this time the spontaneous emission times cannot be eliminated as in the previous case. Also, in this case the evolution operator is not strictly in a closed form. Nevertheless, it is sufficiently simple to be useful. If we take a coherent state, $|\alpha\rangle$, as the initial state, then noting that all the parts of the evolution operator which depend on the stochastic variables determine only the normalisation, we see that the evolution is completely deterministic, and we obtain for the normalised state at time $t$,
\begin{equation}
  |\psi(t)\rangle_{\mbox{\tiny N}} =  |(\alpha - \alpha_{\mbox{\scriptsize s}})e^{-\gamma t/2} + \alpha_{\mbox{\scriptsize s}} \rangle \; .
\end{equation}
The state of the cavity therefore changes smoothly from its initial value to the steady state amplitude, $\alpha_{\mbox{\scriptsize s}}$. The jumps in this case have no effect upon the evolution. If the initial state was not coherent, however, the jumps would play a role in the evolution. 

\subsection{A Damped Cavity with a Kerr Non-linearity}
For our final example we consider a damped cavity with a Kerr nonlinearity~\cite{Kerr}. The operators appearing in the stochastic equation are now
\begin{eqnarray}
  A & = & -(i\omega + \gamma/2) a^\dagger a - i\frac{\chi}{2}a^{\dagger 2}a^2 - \gamma/2, \\
  B & = & a - 1 .
\end{eqnarray}
In this case the evolution operator becomes
\begin{equation}
  |\tilde{\psi}(t)\rangle_{\mbox{\tiny N}} = e^{At}e^{-(i\omega + i\chi a^\dagger a + \gamma/2)Y(t)}a^{N(t)}|\psi(0)\rangle .
\end{equation}
In fact, we note in passing that a double integral of the Poisson process appears in the derivation of the evolution operator for this case, but that it may be immediately discarded as it determines only an overall phase factor. Let us now examine the evolution of an initial coherent state, $|\alpha\rangle$, in this system. To proceed we first note that the action of $\exp[i\eta(a^\dagger a)^2]$ on a coherent state results in superpositions of coherent states~\cite{tara}. The number of coherent states in the superposition depends on the value of $\eta$. For $\eta=\pi/n$ the superposition contains $n$ distinct coherent states. If we choose the evolution time so that $\chi t=\pi$ we will obtain a superposition of two coherent states. Applying the evolution operator for this time, we find that the normalised state is
\bq
  |\psi(t)\rangle_{\mbox{\tiny N}} = \left(\frac{1-i}{{\cal N}(t)}\right) |\alpha e^{-(i\omega +\gamma/2)t - i\chi Y(t)} \rangle \nn \\ 
         + \left(\frac{1+i}{{\cal N}(t)}\right) |-\alpha e^{-(i\omega +\gamma/2)t - i\chi Y(t)} \rangle .
\eq
in which
\bq
  {\cal N}(t) = 2\sqrt{1 + \exp(-2|\alpha|^2e^{-\gamma t}) }
\eq
is the normalisation. We see that the final state is a `Schr\"{o}dinger Cat', irrespective of the number of jumps. However, the orientation of the `Cat' in phase space depends on $Y(t)$, the sum of the times at which the jumps occurred. If we record the photo-detections, we can therefore keep track of the location of the `Cat' in phase space. However, if we do not detect the emitted photons, the result of averaging over the emission times will phase average the final state. Consequently, in the limit of many photon emissions the final state will tend towards a mixture of number states.

\section{Possibilities for Future Work}
So far we have examined the use of evolution operators for linear quantum trajectories in calculating final states, their distributions, and the averages of conditional variances over the possible trajectories. More generally, however, quantum trajectories are useful whenever we are interested in  properties of the state of the system during a continuous measurement. One example in particular is the question of the purity of the state of the system, which, to my knowledge, is only just beginning to be explored~\cite{Doherty}. This is interesting because it tells us how much information the continuous measurement is providing us about the state of the system, rather than just certain system variables. If the state of the system is initially impure, then we may view this as characterising a lack of knowledge about the initial state. As the measurement proceeds we would expect the state to purify, reflecting the fact that we are obtaining information. This has applications, among other things, to the preparing of pure states from thermal mixtures. Bose {\em et al.}~\cite{BJK} have presented a scheme for preparing a superposition state of a macroscopic object. This requires, however, that the object be initially prepared in a pure state. As all macroscopic objects exist in thermal states a process of purification would be required, and continuous measurement would provide a possible method for achieving this.

\section{Conclusion}
We have presented in this chapter a method for deriving explicit evolution operators for quantum trajectories when they are cast in a linear formulation. We have shown a number of things regarding these solutions. First, for linear stochastic equations in which the operator multiplying the time infinitesimal and the operator multiplying the Wiener infinitesimal commute with their mutual commutator, it is possible to obtain an explicit evolution operator in which the classical random variables are Gaussian. In general this provides an explicit solution for equations of this form. Second, we have examined the situation in which the corresponding master equation contains measurement operators which are linear in position and momentum, and for which the Hamiltonian is quadratic in these operators. Note that in this case the quantum Langevin equations are linear, and can therefore be solved. The corresponding statement we are able to make about the Wiener process linear quantum trajectories is that in this case explicit evolution operators may be found. However, we have also found the surprising fact that the classical random variable appearing in the evolution operators are not all Gaussian. We have indicated that it may well be possible to eliminate the non-Gaussian variable, and obtain a completely analytic solution to equations of this form, although we have not been able to do this. We have also shown that it is possible to extend the technique to provide a unified approach to deriving evolution operators for both the Wiener process and the Poisson process.

\chapter{Quantum Noise in a Cavity with a Moving Mirror}\label{qnc}

In this chapter we consider an experimental realisation of the position measurement of a moving cavity mirror. We perform this analysis for two reasons. The first is to perform a quantum mechanical treatment of phase modulation detection, which is often used in practice, and which we show has a different shot noise level than homodyne detection. The second is to include the effects of various experimental sources of noise, such as intracavity losses and classical laser noise, so that we may examine how the quantum back-action noise appears among these sources.

\section{Introduction}
Interferometers provide a very sensitive method for detecting small 
changes in the position of a mirror. This has been analysed 
extensively in the context of gravitational wave detection~\cite{gravwd,gravwd2,gravwd3,gravwd4} and atomic force microscopes~\cite{MJW,atmexp}. A key limit to the sensitivity of such 
position detectors comes from the Heisenberg uncertainty principle. 
The reduction in the uncertainty of the position resulting from the 
measurement is accompanied by an increase in the uncertainty in 
momentum. This uncertainty is then fed back into the
position by the dynamics of the object being measured.
This is called the quantum back action of the measurement,
and the limit to sensitivity so imposed is referred to as the 
standard quantum limit. 


In real devices which have been constructed so far, the quantum back
action noise in the measurement record is usually small compared to 
that arising from classical sources of noise. However, 
as the sensitivity of such devices increases it is expected that we 
will eventually obtain displacement sensors that are quantum limited.
The quantum back-action noise has not yet been seen experimentally 
for macroscopic devices, and is therefore a topic of current interest.
Once the standard quantum limit has been achieved, this will not be 
the end of the story, however. Various authors have shown that it is 
possible to use contractive states~\cite{bql1}, or squeezed 
light~\cite{bql2}, to reduce the quantum back action and therefore 
increase the sensitivity of the measurement even further.

The interferometer we consider here for measuring position consists 
essentially of a cavity where one of the mirrors is free to move. This 
system is also of interest from the point of view of cavity QED. 
Usually cavity QED experiments require optical cavities where the
atomic excitations and photons in the optical modes become entangled.
The dynamics follows from the interplay between these quantum variables.
However, a challenging realm for cavity QED experiments
involves instead an {\em empty} cavity (that is, a cavity 
containing no atoms or optical media) where the photons in
the cavity mode interact with the motion of one of the cavity mirrors.
In this scheme, the position of at least one mirror in the optical
resonator is a dynamic variable. The coupling between the photons and 
the mirror position is simply the radiation pressure that stems from 
the momentum transfer of $2 \hbar k$ per one reflected photon
with the wavenumber $k$. It has been been shown that this system may
be used to generate sub-Poissonian light in the output from the 
cavity~\cite{Heidmann94,Fabre94,Pinard95}. The moving mirror alters 
the photon statistics by changing the optical path length in a way 
that is proportional to the instantaneous photon number inside the 
cavity. This system may also be used to create highly non-classical 
states of the cavity field, such as Schr\"{o}dinger 
cats~\cite{BJK,TM}, and might even be used to create cat states of 
the mirror~\cite{BJK}. In addition, it has been shown that such a 
configuration may be used to perform QND measurements of
the light field~\cite{Pinard95,QND,JTCW}, and to detect the decoherence of the
mirror, a topic of fundamental interest in quantum measurement 
theory~\cite{BJK2}. Due to recent technological developments in 
optomechanics, this area is now becoming experimentally accessible. 
Dorsel {\em et al.} have realised optical bistability with this
system \cite{Dorsel}, and other experiments, particularly to probe 
the quantum noise, are now in progress. 

In order to create displacements that are large enough
to be observed, one is tempted to use a mirror having a 
well-defined mechanical resonance with a very high quality 
factor $Q$. Thus, even when excited with weak white
noise driven radiation pressure, the mirror can be 
displaced by a detectable amount
at the mechanical resonance frequency $\nu$.
For such a mirror to behave fully quantum mechanically 
one needs to operate at very low temperatures since the 
thermal energy $kT$ very easily exceeds
$\hbar \nu$. For example, a $\nu/2 \pi=100$ kHz 
resonance is already significantly excited thermally at   
5 $\mu$K. However, it is not necessary to reach the fully quantum
domain to observe the quantum back action.
By simultaneously combining high optical quality factor (ie. by using 
a high-finesse cavity) and a specially designed low mass mirror
with very high mechanical quality factor one can, at typical cryogenic
temperatures, create conditions where the radiation pressure 
fluctuations
(which are the source of the quantum mechanical back-action referred
to earlier) exceed the effects caused by thermal noise. In this chapter 
we discuss considerations for detecting this quantum back-action 
noise. 

There are already a number of publications dealing with 
quantum noise in optical position measurements. Our main purpose here 
is to extend this literature in two ways which are
important when considering the detection of the quantum noise. The
first is the inclusion of the effects of experimental sources of 
noise, such as the classical laser noise and the noise from intracavity
losses. The second is to perform a quantum treatment of phase-modulation 
detection, so that the results may be compared with those for homodyne detection. While this method of phase detection is often used in 
practice, as far 
as we are aware it has not previously been given a quantum mechanical
treatment. In addition to these main objectives, we also show that the 
standard Brownian motion master equation is not adequate to describe 
the thermal damping of the mirror, but that the corrected Brownian 
motion master equation derived by Diosi~\cite{Diosi} rectifies this 
problem. 

In section~\ref{sec2} we describe the configuration of the system. 
In section~\ref{sec3} we perform a quantum mechanical analysis of 
phase modulation detection. In section~\ref{sec4} we
solve the linearised equations of motion for the cavity/mirror system,
using a non-standard Brownian motion master equation which is 
of the Lindblad form~\cite{Diosi}.
In section~\ref{sec5} we use this solution to obtain the noise power 
spectral density (which we refer to simply as the {\em spectrum}) for 
a measurement of the phase quadrature using 
phase modulation detection. In the first part of this section we 
discuss each of the contributions and their respective 
forms. Next we compare the spectrum to that which results if 
the standard (non-Lindblad) Brownian motion master equation is 
used to describe the thermal damping of the mirror, and also to that 
which would have been obtained using homodyne detection rather
than phase modulation detection. Finally we show how the error in a 
measurement of the position of the mirror may be obtained easily
from the spectrum. We evaluate explicitly the contribution to this 
error from various noise sources, and plot these as a function of the
laser power. Section~\ref{sec6} concludes.

\section{The System}\label{sec2}
The system under consideration consists of a coherently driven 
optical cavity with a moving mirror which will be treated as a quantum
mechanical harmonic oscillator. The light driving the cavity reflects 
off the moving mirror and therefore fluctuations in the position
of the mirror register as fluctuations in the light output from the 
cavity. In the limit in which the cavity damping rate is much larger 
than the rate of the dynamics of the mirror (characterised by the 
frequency of oscillation $\nu$ and the thermal damping rate $\gamma$) 
the phase fluctuations of the output light are highly correlated with 
the fluctuations of the position of the mirror and constitute a 
continuous position measurement of the mirror~\cite{MJW}. 

\begin{figure}
\begin{center}
\leavevmode
\epsfxsize=8cm 
\epsfbox{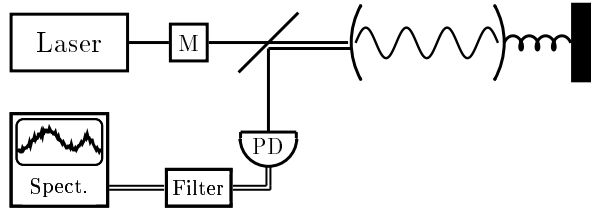}
\caption[A driven optical cavity with a moving mirror]{The light 
output from the laser is modulated at 
frequency $\Delta$, and from there drives the cavity. The front 
mirror 
of the cavity is fixed, while the back mirror is a mechanical 
harmonic 
oscillator. The diagonal line is a shorthand representation for the 
arrangement which isolates the laser from the cavity output, and the 
light that is reflected back off the front mirror. All this light 
falls
upon a photo-detector, and the photo-detection signal is demodulated 
(to pick out the phase quadrature signal) before going to a spectrum 
analyser.}
\label{theoryfig}
\end{center}
\end{figure}

An experimental realisation will therefore involve a continuous 
phase-quadrature measurement of the light output from the cavity to 
determine the output spectrum of the phase-quadrature fluctuations.
The nature of the detection scheme used to measure the phase 
quadrature is of interest to us, as we shall see that it will effect 
the relationship of the shot noise to the other noise sources in the 
measured signal. Quantum theoretical treatments usually assume the 
use of homodyne detection~\cite{MJW,Heidmann94,Fabre94,Pinard95}.
However this is often not used in practice~\cite{HH,HSKO}. Many current 
experiments use instead {\em phase modulation} 
detection~\cite{Tittonen98}, which was developed by 
Bjorklund~\cite{Bjorklund} in 1979. Before we
treat the dynamics of the cavity field/oscillating mirror system, to 
determine the effect of various noise sources, we will spend some time
in the next section performing a quantum mechanical treatment of 
phase modulation
detection. We will focus on this scheme throughout our treatment, and 
compare the results with those for homodyne detection. A diagram of 
the experimental 
arrangement complete enough for the theoretical analysis is given in 
Fig.~\ref{theoryfig}. We note that in practice a feedback scheme is 
used to
lock the laser to the cavity so as to stabilise the laser frequency. 
For an analysis of the method and an expression for the resulting 
classical 
phase noise the reader is referred to references~\cite{HH} 
and~\cite{HSKO}. 
We do not need to treat this feedback explicitly, however. Its 
effect may be
taken into account by setting the value of the classical laser phase 
noise
in our analysis to the level it provides.

\section{Phase Modulation Detection}\label{sec3}
The laser which drives the cavity is isolated from the cavity output,
and the entirety of this output falls upon a photo-detector. In order
that the photo-detection signal contain information regarding the 
phase quadrature, the laser field is modulated at a frequency $\Delta$, 
which is chosen to be much greater than the natural frequency of the 
harmonic mirror. The sidebands that result from this modulation are 
far enough off-resonance 
with the cavity mode that they do not enter the cavity and are simply 
reflected from the front mirror. From there they fall upon the 
photo-detector. The result of this is that the output phase 
quadrature 
signal appears in the photodetection signal as a modulation of the 
amplitude of a `carrier' at frequency $\Delta$. This is then 
demodulated 
(by multiplying by a sine wave at the modulation frequency and time 
averaging) to pick out the phase quadrature signal, and from there 
the 
spectrum may be calculated.

First let us denote the laser output field by 
\begin{equation}
  \beta + \delta a_{\mbox{\scriptsize in}}(t) + \delta x(t) + i\delta 
y(t) .
\end{equation}
In this expression $\beta$ is the average coherent amplitude of the field, 
which we choose to be real. The deviations from this average are 
given by $\delta x(t)$, being the (classical) amplitude noise, 
and $\delta y(t)$, being the (classical) phase noise. The quantum 
noise, which may be interpreted as arising from the vacuum quantum 
field, is captured by the correlation function of the field operator 
$\delta a_{\mbox{\scriptsize in}}(t)$. Here the subscript refers 
to the field's relation to the cavity, 
and not the laser. 
The correlation functions of 
the various noise sources are
\begin{eqnarray}
  & &  \langle \delta a_{\mbox{\scriptsize in}}
(t)\delta a_{\mbox{\scriptsize in}}^\dagger(t+\tau)\rangle 
=\delta(\tau) , \;\;\;\; \langle \delta a_{\mbox{\scriptsize in}}^\dagger(t)
\delta a_{\mbox{\scriptsize in}}(t+\tau)\rangle = 0 ,
\nonumber \\
  & &  \langle \delta x(t)\delta x(t+\tau) 
\rangle = G_{\mbox{\scriptsize x}}(\tau) , \;\;\;\; \langle \delta y(t)\delta y(t+\tau) \rangle = 
G_{\mbox{\scriptsize y}}(\tau) . \nonumber 
\end{eqnarray}
We have left the classical noise sources arbitrary, as this allows 
them to be tailored to describe the output from any real laser source 
at a later time. The units of $\beta$ are such that, if a 
photo-detector 
was placed in the beam, $\beta^2$ would give the average rate of 
photo-detection. The average values of the three noise sources are 
zero, as are all the cross correlation terms.

After modulation the field becomes
\begin{equation}
   (\beta + \delta x(t) + i\delta y(t))(1 + \varepsilon e^{-i\Delta 
t} - 
\varepsilon e^{i\Delta t}) + \delta a_{\mbox{\scriptsize in}}(t) .
\end{equation}
Note that modulation affects the coherent amplitude of the input 
field, 
and has therefore no effect on the quantum contribution. In order to
achieve phase detection it is necessary to modulate the phase 
quadrature, 
which is the reason we choose the modulation to be $\exp(-i\Delta t) 
- \exp(i\Delta t) = 2i\sin(\Delta t)$.
Using now the input-output relations of Collett and 
Gardiner~\cite{inout}, 
the field output from the cavity is
\begin{eqnarray}
   a_{\mbox{\scriptsize out}}(t) & = & - (\beta + \delta x(t) + 
i\delta y(t))
(1 + \varepsilon e^{-i\Delta t} - \varepsilon e^{i\Delta t}) - \delta a_{\mbox{\scriptsize in}}(t) + \sqrt{\gamma} (\delta a(t) + \alpha), 
\end{eqnarray}
in which $a(t)=\delta a(t) + \alpha$ is the operator describing the 
cavity mode and 
$\gamma$ is the decay constant of the cavity. We are interested in 
the steady state 
behaviour, and we choose $\alpha$ to be the average steady state 
field strength 
in the cavity. In addition, in order to solve the equations of motion 
for the cavity 
we will linearise the system about the steady state, which requires 
that 
$\langle \delta a^{\dagger}(t) \delta a(t) 
\rangle \ll |\alpha|^{2}$. The operator describing the 
photo-current
 from the photo-detector is
\begin{eqnarray}
  I(t) & = & a_{\mbox{\scriptsize out}}(t)^\dagger 
a_{\mbox{\scriptsize out}}(t)
\nonumber \\
& = & \tilde{\alpha}^2 + \tilde{\alpha}(\delta X_{\mbox{\scriptsize 
out}} + 2\delta x) +
 2\varepsilon\beta\sin(\Delta t)\left(\delta Y_{\mbox{\scriptsize 
out}} - \frac{2\sqrt{\gamma}\alpha\delta y}{\beta}\right) \nn \\ 
& & + (2\varepsilon\beta)^2\sin^2(\Delta t)\left(1+\frac{2\delta x}{\beta}\right) ,
\end{eqnarray}
in which
\begin{eqnarray}
\delta X_{\mbox{\scriptsize out}}(t) & = & \sqrt{\gamma}\delta X(t) - 
\delta X_{\mbox{\scriptsize in}}(t) \;\; , \;\;\; \delta Y_{\mbox{\scriptsize out}}(t) = \sqrt{\gamma}\delta Y(t) - \delta Y_{\mbox{\scriptsize in}}(t) , \nonumber
\end{eqnarray}
and
\begin{eqnarray}
\delta X(t) & = & \delta a(t) + \delta a^\dagger(t) , \;\;\;\; \delta Y(t) = -i (\delta a(t) - \delta a^\dagger(t)) , \nn \\
\delta X_{\mbox{\scriptsize in}}(t) & = & \delta a_{\mbox{\scriptsize in}}(t) + 
\delta a^\dagger_{\mbox{\scriptsize in}}(t) , \;\;\;\; \delta Y_{\mbox{\scriptsize in}}(t) = -i (\delta a_{\mbox{\scriptsize in}}(t) - \delta a^\dagger_{\mbox{\scriptsize in}}(t)).
\end{eqnarray}
We have also set $\tilde{\alpha}=(-\beta+\sqrt{\gamma}\alpha)$, 
and assumed this to be real. Now, we see that the intensity signal, $I(t)$, contains the phase quadrature multiplied by $\sin(\Delta t)$. That is, the phase quadrature signal appears as a modulation of a carrier at frequency $\Delta$. We therefore choose $\Delta$ to be much larger than the bandwidth of the phase quadrature fluctuations (this bandwidth being determined by the dynamics of the cavity-mirror system), and the classical laser noise, so that the phase quadrature signal appears {\em alone}, centred at the frequency $\Delta$, with the sole exception of the quantum noise, which, being white, also has components at the frequency $\Delta$. To extract the phase quadrature signal, and also, unavoidably, the quantum noise, we perform a demodulation procedure. This involves multiplication by a sine wave at frequency $\Delta$, and subsequent averaging over a time, $T$. The multiplication by the sine wave shifts the signal both up and down by the frequency $\Delta$. As a result of this there is now a component of the phase quadrature signal back at zero frequency. We then extract just the signals around zero frequency by applying a low pass filter, being accomplished by averaging. The averaging time $T$ must be much larger than $1/\Delta$ so as not to pass any of the extraneous signals now appearing centred at the frequencies $\pm\Delta$ and $\pm 2\Delta$, but much smaller that the time-scale of the system dynamics so to pass all of the signal around zero frequency without distortion.
The final signal is therefore given by
\begin{eqnarray}
 R(t) & = &  \frac{1}{T}\int_0^T \sin(\Delta t) I(t+\tau) \;
\mbox{d}\tau .
\end{eqnarray}
We now must evaluate this to obtain $R(t)$ explicitly in terms of the phase quadrature. Writing out the integral, and dropping everything which averages to zero (that is, which is not passed by the low-pass filtering) we obtain
\begin{eqnarray}
 R(t) & = &  \left( \frac{\varepsilon\beta}{T} \right) \left[ \int_0^T \delta Y_{\mbox{\scriptsize out}}(t+\tau) d\tau \right] + \left(\frac{2\varepsilon\sqrt{\gamma}\alpha}{T}\right) \left[ \int_0^T \delta y(t+\tau) \; d\tau \right] + q_1(t) + q_2(t) ,
\end{eqnarray}
where
\begin{eqnarray}
 q_1(t) & = & - \left( \frac{\beta-\sqrt{\gamma}\alpha}{T} \right) \mbox{Re}\left[ \int_0^T ie^{-i\Delta(t+\tau)}\,\delta X_{\mbox{\scriptsize in}}(t+\tau)  \; d\tau \right], \\
 q_2(t) & = &  \left( \frac{\varepsilon\beta}{T} \right) \mbox{Re}\left[ \int_0^T e^{-i2\Delta(t+\tau)}\,\delta Y_{\mbox{\scriptsize in}}(t+\tau)  \; d\tau \right] .\label{defq2}
\end{eqnarray}
Note that we choose $T$ to be much smaller than the time-scale upon which $\delta Y$ and $\delta y$ change, so that the integration is essentially equivalent to multiplication by $T$, an effect which is cancelled by the division by $T$. However, we should also note that $\delta Y_{\mbox{\scriptsize out}}$ contains $\delta Y_{\mbox{\scriptsize in}}$, so that in replacing the first term in $R(t)$ by $\varepsilon\beta \; \delta Y(t)_{\mbox{\scriptsize out}}$ we must remember that this only contains the frequency components of $\delta Y_{\mbox{\scriptsize in}}$ in a bandwidth of $1/T$ around zero frequency. The result of this is that $\delta Y(t)_{\mbox{\scriptsize out}}$ is uncorrelated with $q_1$ and $q_2$, being the quantum noise in the bandwidth $1/T$ around the frequencies $\Delta$ and $2\Delta$ respectively. We need to know the correlation functions of these noise sources, and whether or not they are correlated with any of the other terms in $R(t)$. It is clear that $q_1$ and $q_2$ are not correlated over separation times greater than $2T$. Using Eq.(\ref{defq2}) to evaluate the correlation function of $q_2$, for example, we have
\bq
  \langle q_2(t)q_2(t+\tau)\rangle = \left\{ \begin{array}{cl} 
                                              \frac{(\varepsilon\beta)^2}{2} \left( \frac{T-|\tau|}{T^2} \right) & \mbox{for} \; |\tau| \leq T \\
                                              0 & \mbox{otherwise.}
                                            \end{array}  \right. 
\eq
On the time-scale of the fluctuations of $\delta Y$ we can approximate this as a delta function, so that $q_1$ and $q_2$ (and also $\delta Y_{\mbox{\scriptsize in}}$) are still effectively white noise sources. We may therefore write  
\begin{eqnarray}
 R(t) & = & \varepsilon\beta \; \delta Y(t)_{\mbox{\scriptsize out}} 
+ q_1(t) + q_2(t) - 2\varepsilon\sqrt{\gamma}\alpha\delta y(t) ,
\label{signal}
\end{eqnarray}
and the correlation functions of $q_1$ and $q_2$ are 
\begin{eqnarray}
  & & \langle q_1(t)q_1(t+\tau)\rangle = (1/2)(\beta - 
\sqrt{\gamma}\alpha)^2 \; \delta(\tau) ,\nonumber \\
  & & \langle q_2(t)q_2(t+\tau)\rangle = (1/2)(\varepsilon\beta)^2 \; 
\delta(\tau) .
\end{eqnarray}
The signal therefore contains the phase quadrature of the output field, $\delta 
Y_{\mbox{\scriptsize out}}(t)$, plus three noise terms. While the last term, being the input classical phase noise, is correlated with $\delta Y_{\mbox{\scriptsize out}}(t)$, $q_1$ and $q_2$ are not. Taking the Fourier transform of the signal to give
\begin{equation}
  R(\omega) = \frac{1}{\sqrt{2\pi}}\int_{-\infty}^{\infty} 
\!\!\!\!\!\! R(t) e^{-i\omega t} \; dt ,
\end{equation}
we may write
\begin{equation}
  R(\omega) = \varepsilon\beta \; \delta Y_{\mbox{\scriptsize 
out}}(\omega) + q_1(\omega) + q_2(\omega) - 2\varepsilon\sqrt{\gamma}\alpha\delta 
y(\omega) .
\end{equation}

This is the Fourier transform of the signal in the case of phase 
modulation detection. 
If we were to use ideal homodyne detection this would be 
instead~\cite{SHM}
\begin{equation} \label{above}
  R_{\mbox{\scriptsize h}}(\omega) = 
  \kappa \tilde{\beta}( \delta Y_{\mbox{\scriptsize out}}(\omega) 
- 2 \delta y(\omega)),
\end{equation}
where $\tilde{\beta}$ is the amplitude of the local oscillator and 
$\kappa$ is the reflectivity of the beam splitter used in the homodyne 
scheme. Thus, in the case of phase modulation detection, there are
two white noise sources which do not appear in homodyne detection.
They stem from the fact that the phase quadrature detection method
is demodulating to obtain a signal at a carrier frequency. Because
the quantum noise is broad band (in particular, it is broad
compared to the carrier frequency) the demodulation picks up the
quantum noise at $\Delta$ and $2\Delta$. There is also a term
from the classical phase noise in the sidebands. For real homodyne
detection there will also be an extra contribution from the 
noise on the local oscillator.

Once we have solved the equations of motion for the system operators, 
we will obtain 
$\delta Y(\omega)$ in terms of the input noise sources. We can then 
readily calculate 
$\langle R(\omega)R(\omega')\rangle$, which appears in the form
\begin{equation}
\langle R(\omega)R(\omega')\rangle = S(\omega)\delta(\omega+\omega').
\end{equation}
The delta function in $\omega$ and $\omega'$ is a result of the 
stationarity of $R(t)$, 
and $S(\omega)$ is the {\em power spectral density}, which we will 
refer to from 
now on simply as the {\em spectrum}. This is useful because, when 
divided
by $2\pi$, it gives the average 
of the square of the signal per unit frequency (The square of the 
signal is universally 
referred to as the {\em power}, hence the name power spectral 
density). 
Since the noise has zero mean, the square average is the variance, 
and thus the spectrum provides us with information regarding the 
error 
in the signal due to the noise. The spectrum is 
also a Fourier transform 
of 
the autocorrelation function~\cite{Gardiner1}. The specific relation, 
using
the definitions we have introduced above, is 
\begin{equation}
  S(\omega) = \int_{-\infty}^{\infty} \!\!\!\! \langle 
R(0)R(\tau)\rangle e^{-i\omega\tau} d \tau,
\end{equation}
and as the autocorrelation function has units of $\mbox{s}^{-2}$, the 
spectrum 
has units of $\mbox{s}^{-1}$. To determine the spectrum 
experimentally $R(t)$ is measured for a time long compared to the 
width 
of the auto-correlation function, and the Fourier transform is taken 
of the result. 
Taking the square modulus of this Fourier transform, and dividing by 
the duration 
of the measurement obtains a good approximation to the theoretical 
spectrum. 
We proceed now to calculate this spectrum.

\section{Solving the System Dynamics}\label{sec4}
Excluding coupling to reservoirs, 
the Hamiltonian for the combined system of the 
cavity 
mode and the mirror is~\cite{JTCW}
\begin{eqnarray}
 H & = & \hbar\omega_0 a^{\dagger}a + \frac{p^2}{2m} + 
\frac{1}{2}m\nu^2 q^2 - \hbar ga^\dagger aq  \nonumber \\ 
   &   &  + \hbar
\left\{ i\left[ E + \sqrt{\gamma}\delta x(t) + i\sqrt{\gamma}\delta 
y(t)\right]
a^\dagger + \mbox{H.c.}\right\} .
\label{Ham}
\end{eqnarray}
In this equation $\omega_0$ is the frequency of the cavity mode, $q$ 
and $p$ are 
the position and momentum operators for the mirror respectively, $m$ 
and $\nu$ are 
the mass and frequency of the mirror, $g = \omega_{0}/L$ is the coupling
constant 
between the cavity
mode and the mirror (where $L$ is the cavity length), and $a$ is the
annihilation operator for the 
mode. The classical
 driving of the cavity by the coherent input field is given by $E$ 
which has dimensions 
of $\mbox{s}^{-1}$, and is related to the input laser power $P$ by
 $E=\sqrt{P\gamma/(\hbar\omega_0)}=\sqrt{\gamma}\beta$. 
The classical laser noise appears as noise on this driving term. 

The moving mirror is a macroscopic object at temperature $T$, and as 
such is 
subject to thermal noise. While it is still common to use the 
standard Brownian 
motion master equation (SBMME) \cite{Gardiner2,SBMME} 
to model such noise, as it works well 
in many 
situations, it turns out that it is not adequate for our purposes. 
This is because 
it generates a clearly non-sensical term in the spectrum, which 
will be shown in subsection~\ref{compBM}. As far as we know this is 
the first time that it has been demonstrated to fail in 
the steady state. Discussions regarding the SBMME, and non-Lindblad
master equations may found in references~\cite{MGetc,Haake}.
We will return to this point once we have calculated the spectrum. We 
use instead 
the corrected Brownian motion master equation (CBMME) derived by 
Diosi~\cite{Diosi}, 
to describe the thermal damping of the mirror, as this corrects the 
problems of 
the SBMME. Using this, and the standard master equation for the 
cavity losses 
(both internal and external), the quantum Langevin equations of 
motion 
for the system are given by
\begin{eqnarray} \label{qle1}
\dot{a} & = & -\frac{i}{\hbar}[a,H] - 
\left(\frac{\gamma+\mu}{2}\right)a + 
\sqrt{\gamma}\delta a_{\mbox{\scriptsize in}}(t) + 
\sqrt{\mu}b_{\mbox{\scriptsize in}}(t) ,  \\
\dot{q} & = & -\frac{i}{\hbar}[q,H] + 
\hbar(\Gamma/6mkT)^{\frac{1}{2}} 
\eta(t)\label{qnlang2} , \\
 \dot{p} & = & -\frac{i}{\hbar}[p,H] - \Gamma p + 
(2 m \Gamma kT)^{\frac{1}{2}} \xi(t) , \label{qnlang3}
\label{qle3}
\end{eqnarray}
in which the correlation functions for the Brownian noise sources are
\begin{eqnarray}
\langle \xi(t) \xi(t') \rangle &=& \delta(t-t') , \;\;\;\; \langle \eta(t) \eta(t') \rangle = \delta(t-t'), \\
  \langle \xi(t) \eta(t') \rangle &=& -i(\sqrt{3}/2)\delta(t-t') , \;\;\;\; \langle \eta(t) \xi(t') \rangle = i(\sqrt{3}/2)\delta(t-t') .
\end{eqnarray}
In these equations all internal cavity losses including absorption, 
scattering and transmission
through the movable mirror are included via the input operator 
$b_{\mbox{\scriptsize in}}(t)$ 
and the decay constant $\mu$. The effect of mechanical damping and 
thermal fluctuations of the mirror are given by the 
classical noise sources $\xi(t)$ and 
$\eta(t)$ and the mechanical damping constant $\Gamma$. 

We note here that if we were to use the standard Brownian motion 
master equation~\cite{SBMME,Gardiner2}, 
Eqs.(\ref{qnlang2}) and (\ref{qnlang3}) would instead be given by
\begin{eqnarray}
\dot{q} & = & -\frac{i}{\hbar}[q,H] , \\
\dot{p} & = & -\frac{i}{\hbar}[p,H] - \Gamma p + 
(2 m \Gamma kT)^{\frac{1}{2}} \zeta(t) .
\end{eqnarray}
where $\langle\zeta(t)\zeta(t')\rangle = \delta(t-t')$. These 
Langevin equations do not preserve the commutation relations of the 
quantum mechanical operators, and as a result it is clear that the 
description cannot be entirely correct.

Calculating the commutators in Eqs.(\ref{qle1}) to (\ref{qle3})
\begin{eqnarray}
 \dot{a} & = & E -\left(\frac{\gamma+\mu}{2}\right)a + igaq + 
\sqrt{\gamma}\delta a_{\mbox{\scriptsize in}}(t) + 
\sqrt{\mu}b_{\mbox{\scriptsize in}}(t) + 
\sqrt{\gamma}\delta x(t) + i\sqrt{\gamma}\delta y(t) , \\
 \dot{q} & = & \frac{p}{m} + \hbar(\Gamma/6mkT)^{\frac{1}{2}} 
\eta(t) , \\
 \dot{p} & = & -m\nu^2 q + \hbar g a^\dagger a - \Gamma p + 
(2 m \Gamma kT)^{\frac{1}{2}} \xi(t) ,
\end{eqnarray}
in which $g$ is the strength of the coupling between the cavity
 field and the mirror. 
Introducing a cavity detuning $\Delta$ (that is, setting the cavity
 resonance frequency in the absence of any cavity field to 
$\omega_{\mbox{\scriptsize c}}= \omega_0 + \Delta$), and solving 
these equations for 
the steady state average values we obtain
\begin{eqnarray}
  \langle a\rangle_{\mbox{\scriptsize ss}} & = & 
\frac{2E}{\gamma+\mu} \equiv \alpha ,  \\
  \langle q\rangle_{\mbox{\scriptsize ss}} & = & \frac{\hbar 
g}{m\nu^2}|\alpha|^2 , \\
  \langle p\rangle_{\mbox{\scriptsize ss}} & = & 0 ,
\end{eqnarray}
where we have set the detuning to $\Delta=g\langle 
q \rangle_{\mbox{\scriptsize ss}}$ 
to bring the cavity on resonance with the driving field in the steady 
state. Linearising
 the quantum Langevin equations about the steady state values, and 
writing the result in 
terms of the field quadratures, we obtain the following linear 
equations
\begin{eqnarray}
   \left(  \begin{array}{c} 
                            \delta\dot{X}  \\
                            \delta\dot{Y}  \\ 
                            \delta\dot{Q}  \\ 
                            \delta\dot{P} 
           \end{array} \right) \!\! = \!\!
   \left(  \begin{array}{cccc}   
  -\frac{\gamma+\mu}{2}   & 0                       &  0 & 0  \\
  0      & -\frac{\gamma+\mu}{2}   &  \chi\alpha   & 0   \\
  0       & 0                       &  0            & \nu  \\
  \chi\alpha   & 0             & -\nu       & -\Gamma 
           \end{array} \right)  \!\!
   \left(  \begin{array}{c} 
                         \delta X  \\ 
                         \delta Y  \\ 
                         \delta Q  \\ 
                         \delta P
           \end{array} \right) \!\! + \!\!
   \left(  \begin{array}{c}      
\sqrt{\gamma}\delta X_{\mbox{\scriptsize in}}(t) + 
\sqrt{\mu}\delta X_{\mbox{\scriptsize b,in}}(t) + 
2\sqrt{\gamma}\delta x(t) \\
\sqrt{\gamma}\delta Y_{\mbox{\scriptsize in}}(t) +
\sqrt{\mu}\delta Y_{\mbox{\scriptsize b,in}}(t) + 
2\sqrt{\gamma}\delta y(t)            \\
(\Gamma\hbar\nu/3kT)^{\frac{1}{2}}  \eta(t)  \\
(4\Gamma kT/(\hbar\nu))^{\frac{1}{2}} \xi(t)
           \end{array} \right) . \nn
\end{eqnarray}
In this set of equations we have scaled the position and momentum 
variables using
\begin{eqnarray}
   \delta Q & = & \sqrt{\frac{2m\nu}{\hbar}}(q - \langle 
q\rangle_{\mbox{\scriptsize ss}}) , \\
   \delta P & = & \sqrt{\frac{2}{m\hbar\nu}}(p - \langle 
p\rangle_{\mbox{\scriptsize ss}}) ,
\end{eqnarray}
and we have defined $\chi\equiv g(2\hbar/m\nu)^{1/2}$, which has 
units of s$^{-1}$. 
The quadratures for the input noise due to intracavity losses are 
given by
\begin{eqnarray}
   \delta X_{\mbox{\scriptsize b,in}} & = & b_{\mbox{\scriptsize in}} 
+ 
b_{\mbox{\scriptsize in}}^\dagger , \\
   \delta Y_{\mbox{\scriptsize b,in}} & = & -i(b_{\mbox{\scriptsize 
in}} 
- b_{\mbox{\scriptsize in}}^\dagger) .
\end{eqnarray}
Without loss of generality we have chosen the input field amplitude 
to be
 real ($\mbox{Im}[\beta]=0$), so that the input phase quadrature is 
given
 by $Y_{\mbox{\scriptsize in}}$.
We now solve these in the frequency domain in order to obtain the 
spectrum
 directly from the solution. To switch to the frequency domain we 
Fourier 
transform all operators and noise sources. In particular we have, for 
example
\begin{eqnarray}
  \delta a(\omega) & \equiv & 
\frac{1}{\sqrt{2\pi}}\int_{-\infty}^{\infty}
 \!\!\!\!\!\! \delta a(t) e^{i\omega t} \; dt , \\
  \delta a^\dagger(\omega) & \equiv & 
\frac{1}{\sqrt{2\pi}}\int_{-\infty}^{\infty}
 \!\!\!\!\!\! \delta a^\dagger(t) e^{i\omega t} \; dt = [\delta 
a(-\omega)]^\dagger .
\end{eqnarray}
Rearranging the transformed equations, the solution is given by
\begin{equation}
 (\delta X(\omega), \delta Y(\omega), \delta Q(\omega), \delta 
P(\omega))^T = 
M(\omega) {\bf n}(\omega) ,
\end{equation}
where ${\bf n}(\omega)$ is the vector of transformed noise sources.
If we write the matrix elements of $M(\omega)$ as 
$M_{ij}(\omega)=m_{ij}
(\omega)/D(\omega)$, then $D(\omega) = \left((\gamma+\mu)/2 - i\omega \right)^2 \left( \nu^2 - \omega^2 - i\Gamma\omega \right)$, 
and the non-zero $m_{ij}$ are given by
\begin{eqnarray}
 \begin{array}{rclrcl}
  m_{11} & = & m_{22} = \left((\gamma+\mu)/2 - i\omega \right)  \left( \nu^2 - 
\omega^2 - i\Gamma\omega \right), & m_{21} & = & \chi^2\alpha^2\nu , \\
  m_{23} & = & \chi\alpha (\Gamma - i\omega)\left((\gamma+\mu)/2 - 
i\omega \right), &  m_{24} & = & \chi\alpha\nu \left((\gamma+\mu)/2 - i\omega \right) = m_{31}, \\
  m_{33} & = & (\Gamma - i\omega)\left((\gamma+\mu)/2 - i\omega 
\right)^2, & m_{34} & = & \nu\left((\gamma+\mu)/2 - i\omega \right)^2 = -m_{43}, \\
  m_{41} & = & -i\chi\alpha\omega \left((\gamma+\mu)/2 - i\omega 
\right), & m_{44} & = & -i\omega \left((\gamma+\mu)/2 - i\omega \right)^2  .
\end{array} & & \nn
\end{eqnarray}
We have now solved the equations of motion for the system in 
frequency space. The spectra of the system variables may now be 
calculated in terms of the input noise sources. Using the 
input-output relations, which give the output-field in terms of the 
system variables and the input noise sources,  the spectra of the 
output field, and hence of the measured signal, may be obtained. Note 
that quantum mechanics plays no role in the solution of the motion of 
the system. The linear equations of motion may as well be equations 
for classical variables. The only part that quantum mechanics plays 
in determining the spectra of the system variables is that some of 
the input noise sources are quantum mechanical. That is, their 
correlation functions are determined by quantum mechanics. In fact, 
if all the noise sources had purely classical correlation functions, 
then the SBMME Langevin equations would not lead to any problems, as 
they are perfectly correct as equations of motion for a classical 
system.

\section{The Power Spectral Density}\label{sec5}
To calculate the spectrum of the signal, we require the correlation 
functions of the input noise sources. To reiterate, these are
\begin{eqnarray}
  \langle \delta X_{\mbox{\scriptsize in}}(\omega)
\delta X_{\mbox{\scriptsize in}}(\omega')\rangle & = & \langle \delta 
Y_{\mbox{\scriptsize in}}(\omega)\delta Y_{\mbox{\scriptsize in}}
(\omega')\rangle = \delta(\omega+\omega'), \nonumber \\
   \langle \delta X_{\mbox{\scriptsize in}}(\omega)
\delta Y_{\mbox{\scriptsize in}}(\omega')\rangle & = & - \langle
 \delta Y_{\mbox{\scriptsize in}}(\omega)\delta X_{\mbox{\scriptsize 
in}}
(\omega')\rangle = i\delta(\omega+\omega'), \nonumber
\end{eqnarray}
and similarly for $\delta X_{\mbox{\scriptsize b,in}}(\omega)$ and 
$\delta Y_{\mbox{\scriptsize b,in}}(\omega)$. The correlation 
functions
 for the classical laser noise, and thermal noise sources are 
\begin{eqnarray}
  \langle \delta x(\omega)\delta x(\omega')\rangle & = & 
\tilde{G}_x(\omega)\delta(\omega+\omega'), \nonumber \\
  \langle \delta y(\omega)\delta y(\omega')\rangle & = &  
\tilde{G}_y(\omega)\delta(\omega+\omega'), \nonumber \\
  \langle \xi(t) \xi(t') \rangle &=& \langle \eta(t) \eta(t') \rangle 
= \delta(t-t'), \nonumber \\
  \langle \eta(t) \xi(t') \rangle &=&  -\langle \xi(t) \eta(t') 
\rangle = i(\sqrt{3}/2)\delta(t-t').
\end{eqnarray}
After some calculation we obtain the spectrum of the signal for phase 
modulation
detection as
\begin{eqnarray}
  \frac{1}{(\varepsilon\beta)^2}S(\omega)  
     & = &   \frac{1}{2} \left[ 3 + \left(  \frac{\gamma-\mu}
{\varepsilon(\gamma+\mu)} \right)^2 \right] 
           + \gamma (\gamma + \mu + 4\gamma\tilde{G}_x(\omega)) 
\left[\frac{(\chi^2\alpha^2 \nu)^2}{|D(\omega)|^2}\right] \nn \\
     &   & + \mbox{\normalsize $4\tilde{G}_y(\omega)$} 
             \left[ \frac{4\gamma^2}{(\gamma+\mu)^2}\left( \frac{\omega^2} {\left( \frac{\gamma+\mu}{2} \right)^2 + \omega^2 } \right) 
\right] \nonumber 
\\
     &   & + \gamma (\chi\alpha)^2 \Gamma \left( 4\nu^2 T_{\mbox{s}} 
+ 
\frac{1}{3}(\Gamma^2 + \omega^2) T_{\mbox{s}}^{-1} \right) 
\left[ \frac{\left( \frac{\gamma+\mu}{2} \right)^2 + \omega^2}
{|D(\omega)|^2} \right] ,
\label{eqpsd}
\end{eqnarray}
where
\begin{equation}
  |D(\omega)|^2 = \left[(\gamma+\mu)^2/2 + \omega^2\right]^2
\left[(\nu^2 - \omega^2)^2 +
 \Gamma^2\omega^2\right],
\end{equation}
and $T_{\mbox{s}}$ is a dimensionless scaled temperature given 
by $T_{\mbox{s}}=(k_{\mbox{\scriptsize B}}/(\hbar\nu))T$.
This phase-fluctuation spectrum may be thought of as arising 
in the following way. The mechanical harmonic oscillator, 
which is the moving mirror, is driven by  various noise sources, 
both quantum mechanical and classical in origin, and
 the resulting position fluctuations of the mirror are seen as 
fluctuations 
in the phase of the light output from the cavity.

Let us examine the origin of the various terms in the spectrum in 
turn. The 
first two terms, which appear in the first set of square brackets, 
are 
independent of the frequency, and are the contribution from the 
(quantum 
mechanical) shot noise of the light. 

The next three terms, which multiply the second set of square 
brackets, are 
the back-action 
of the light on the position of the mirror, noise from internal 
cavity losses, and 
the classical amplitude noise on the laser, respectively. Note that 
the only distinction
between the back-action and the internal losses is that the former is 
proportional to the
loss rate due to the front mirror, and the latter is proportional to 
the internal loss
rate. It is easily seen that these noise sources should have the same 
effect upon the
position of the mirror: the back-action is due to the random way in 
which photons
bounce off the mirror, where as the internal losses are due to the 
similarly random 
way in which photons are absorbed by the mirror, (or anything else in 
the cavity). 
The amplitude fluctuations of the laser also affect the mirror in the 
same manner, 
all though since these fluctuations are not white noise (as is the 
case with the 
quantum noise which comes from the photon `collisions'), the response 
function of 
the mirror is multiplied by the spectrum of the amplitude 
fluctuations. 

The term which appears in the third set of square brackets, is 
due to the 
classical phase fluctuations of the laser. Clearly this has quite a 
different form 
from that due 
to the quantum noise and the classical amplitude fluctuations. In 
particular, it is
not dependent upon the coupling constant, $g$, because it is derived 
more or less 
directly from the input phase noise. Conversely, the noise that 
derives from the 
amplitude fluctuations has its origin from the fact that the 
amplitude fluctuations
first drive the mirror, and it is the resulting position fluctuations 
which cause
the phase fluctuations in the output. The classical 
phase noise term has a contribution from the phase noise in the sidebands 
and from the phase noise on the 
light which has passed through the cavity. The final two terms, 
which multiply the
forth set of square brackets, are due to the thermal fluctuations of 
the mirror. Note
that these terms are only valid in the region in which
$k_{\mbox{\scriptsize B}} T \gg \hbar\nu$.

Finally we note that we do not see squeezing in the spectrum of phase 
quadrature fluctuations. This is because squeezing is produced when 
the 
cavity detuning is chosen so that the steady state detuning is 
non-zero~\cite{Fabre94}. We have chosen to set the steady state 
detuning to zero in this treatment as we are not concerned here 
with reducing the quantum noise.

In what follows we examine various aspects of the spectrum which are 
of 
particular interest. Before discussing considerations for detecting 
the 
back-action noise, we compare the spectrum with that which would have 
been 
obtained using the SBMME, and for that which would result from the 
use of 
homodyne detection. We then write the spectrum at resonance as a 
function 
of the laser power, and plot this for current experimental 
parameters. So 
far we have been considering the noise power spectrum, and have made 
no 
particular reference to the limit this implies for a measurement of 
the 
position of the mirror. In the final part of this section we show how 
the 
spectrum tells us the limit to the accuracy of position measurement 
in the 
presence of the noise sources.

\subsection{Comparison with Standard Brownian Motion}\label{compBM}
As we have mentioned above, to obtain the spectrum we have used the 
corrected 
Brownian motion master equation~\cite{Diosi}. This is essential 
because the 
standard Brownian motion master equation produces in this case the 
following 
term in the spectrum:
\begin{equation}
   2\omega\gamma\Gamma\chi^2\alpha^2\nu \left[ \frac{\left( \left( 
\frac{\gamma+\mu}{2}\right)^2 + \omega^2 \right)}  {|D(\omega)|^2} 
\right].
\end{equation}
This is clearly incorrect since it is asymmetric in $\omega$. It 
follows 
readily from the stationarity of the output field, and the fact that 
the 
output field commutes with itself at different times, that the 
spectrum 
must be symmetric in $\omega$. In particular, the stationarity of the 
output
field means that the correlation function of the signal only depends 
on the 
time difference, so that
\begin{equation}
  \langle R(t)R(t+\tau)\rangle = G(\tau) .
\end{equation}
As the output field commutes with itself at different times, $R$ 
commutes with 
itself at different times, and we have
\begin{equation}
  G(-\tau) = \langle R(t)R(t-\tau)\rangle = \langle 
R(t-\tau)R(t)\rangle 
           = G(\tau) .
\end{equation}
The correlation function is therefore symmetric in $\tau$. As the 
spectrum is
the Fourier transform of the correlation function, it follows from the
properties of the Fourier transform that the spectrum is must be 
symmetric in 
$\omega$.

Diosi's corrected Brownian motion master equation removes the term 
asymmetric in 
$\omega$ by adding a noise source to the position (see 
Eq.(\ref{qle3})) which is 
correlated with the noise source for the momentum. In doing so it 
produces an 
additional term in the spectrum proportional to $1/T$. For 
temperatures in which 
$\hbar\nu$ is much smaller than $k_B T$ ($T_s \gg 1$) this term is 
very small and 
can be neglected. However, the question of probing this term 
experimentally is a 
very interesting one, because it would allow the CBMME to be tested.

\subsection{Comparison with Homodyne Detection}
Let us now briefly compare the spectrum derived above for phase 
modulation detection to that which would be obtained with homodyne 
detection. Firstly, if homodyne detection had been used, the 
overall scaling of the spectrum would be different,
as it would be proportional to the strength of the local oscillator. 
Thus the factor of $1/(\varepsilon\beta)^2$ would be 
replaced by $1/(\tilde{\beta}\kappa)^2$, in which $\tilde{\beta}$ 
and $\kappa$ are as defined in Eq.~(\ref{above}). This 
overall factor aside, two terms in the spectrum would change. The 
shot noise component would be reduced to unity, and the classical 
phase noise contribution would become
\begin{equation}
  4\tilde{G}_y(\omega) \left[ \frac{\left( \frac{\gamma -\mu}{2}
\right)^2 + \omega^2}{ \left( \frac{\gamma+\mu}{2} \right)^2 + 
\omega^2} \right] .
\end{equation}
There would also be a contribution from the noise on the local 
oscillator.

\subsection{The Error in a Measurement of Position}
So far we have been considering the noise spectrum of the phase 
quadrature, 
as this is what is actually measured.  However, the reason for 
performing
the phase measurement is that it constitutes essentially a 
measurement of 
the position of the mirror. In this section we show how the error in 
a 
measurement of the position of the mirror may be obtained in a simple 
manner from the spectrum, Eq.(\ref{eqpsd}), and give an example by 
calculating it explicitly for some of the terms.

We can choose to measure the amplitude of position oscillations at 
any frequency, but for the purposes of discussion, a measurement of 
a constant displacement is the simplest. First we must see how the 
position of the mirror appears in the signal, which is the phase 
quadrature measurement (that is, convert from the units of the 
signal into units of the position fluctuations). This is easily 
done by calculating the contribution to the spectrum of the position fluctuations due to one of the noise sources (for the sake of 
definiteness we will take the
thermal noise), and comparing this to the equivalent term in the 
spectrum of the signal.
This gives us the correct scaling. Performing this calculation, we 
find that the
spectrum of position fluctuations of the mirror due to thermal noise 
is given by 
the thermal term in the spectrum (Eq.(\ref{eqpsd})), multiplied by 
the factor 
\begin{equation}
 \frac{\hbar}{2m(\varepsilon\beta)^2\gamma\nu\chi^2\alpha^2} 
\left(\left( 
 \frac{\gamma+\mu}{2} \right)^2 + \omega^2 \right) .
\label{poss}
\end{equation}
From this we see that the scaling factor is frequency dependent. This 
means, 
that the spectrum of the position fluctuations is somewhat different 
from the 
spectrum of the resulting phase quadrature fluctuations. For the 
measurement of 
the phase to correspond to a true measurement of the position the two 
spectra should 
be the same. This is true to a good approximation when $\gamma$ is 
much larger than
the range of $\omega$ over which the spectrum of position 
fluctuations is non-zero, 
and this is why the scheme can be said to constitute a measurement of 
position when 
$\gamma\gg\nu,\Gamma$.
 
In performing a measurement of a constant displacement of the mirror 
(achieved by some constant external force), the signal (after scaling 
appropriately so that it corresponds to position rather than 
photocurrent) 
is integrated 
over a time $\tau_{\mbox{\scriptsize m}}$. The best estimate of the 
displacement 
is this integrated signal divided by the measurement time. The error, 
$\Delta x$, in the case that the measurement time is much greater 
than the 
correlation time of the noise, is given by
\begin{equation}
  \Delta x^2 = \int_{-\infty}^{\infty} \!\!\!\! \langle 
R_x(0)R_x(\tau)\rangle d\tau
               /\tau_{\mbox{\scriptsize m}} = 
S_x(0)/\tau_{\mbox{\scriptsize m}}.
\end{equation}
In this equation $R_x$ and $S_x$ are the appropriately scaled signal and 
spectrum. To
calculate the error in the measurement of a constant displacement, 
all we have to 
do, therefore, is to scale the spectrum using the expression 
Eq.(\ref{poss}),
evaluate this at zero frequency, and divide by the measurement time. 
In general,
the spectrum evaluated at a given frequency gives the error in a 
measurement of the
amplitude of oscillations at that frequency. We calculate
now the contribution to the error in a measurement at zero frequency 
and 
at the mirror resonance frequency, from the shot noise, thermal, and 
quantum 
back-action noise. In the following we write the expressions 
in terms of the parameters usually used by experimentalists. That is, 
the cavity 
finesse, ${\cal F}=\pi c/(2L\gamma)$ (assuming intracavity 
losses are small compared to $\gamma$), and the quality factor for 
the mirror 
oscillator, $Q=\nu/(2\Gamma)$. The contribution due to the shot noise 
is the 
same at all frequencies and is
\begin{equation}
     \Delta x^2_{\mbox{\scriptsize SN}} = \frac{\pi^2}{64} \left[ 
\frac{3}{2} + \frac{1}{2\varepsilon^2} \right] \left( 
\frac{\hbar c^2}{\omega_0} \right) \frac{1}{{\cal F}^2 P \tau_{\mbox{\scriptsize 
m}}} .
\end{equation}
The contribution from the quantum back action for a measurement of a 
constant displacement is
\begin{equation}
      \Delta x^2_{\mbox{\scriptsize BA}}(0) = \frac{4}{\pi^2} \left( 
\frac{\hbar\omega_0}{c^2} \right) \left( \frac{1}{m^2\nu^4} \right) 
\frac{{\cal F}^2 P}{\tau_{\mbox{\scriptsize m}}},
\end{equation}
and for an oscillation at frequency $\nu$ is 
$\Delta x^2_{\mbox{\scriptsize BA}}(\nu) = 4Q^2 
\Delta x^2_{\mbox{\scriptsize BA}}(0)$.
The contribution from the thermal noise is
\begin{equation}
      \Delta x^2_{\mbox{\scriptsize Th}}(0) =  \left( 
\frac{k_{\mbox{\scriptsize B}}T}{m\nu^3 Q \tau_{\mbox{\scriptsize 
m}}} \right) + \left( \frac{\hbar^2}{16m\nu k_{\mbox{\scriptsize 
B}}TQ^3 \tau_{\mbox{\scriptsize m}}} \right),
\end{equation}
for a constant displacement, and is 
$\Delta x^2_{\mbox{\scriptsize Th}}(\nu) = 4Q^2
\Delta x^2_{\mbox{\scriptsize Th}}(0)$ for an oscillation at the 
mirror frequency. The contribution from the other noise sources 
may also be readily evaluated from the terms in the spectrum 
Eq.(\ref{eqpsd}).

Let us examine the total error in a position measurement resulting
from these three contributions for state-of-the-art experimental
parameters. Reasonable values for such parameters are as 
follows~\cite{Tittonen98}. 
The cavity resonance frequency is $\omega_0=2\pi\times 2.82\times 
10^{15}\;\mbox{Hz}$ (assuming a Nd:YAG laser with a wavelength of 
1064 nm), the cavity length is $L=1\;\mbox{cm}$, the mass 
of the oscillating mirror is $m=10^{-5}\;\mbox{kg}$, and the resonant 
frequency of the mirror is $\nu=2\pi\times 2\times 10^4\;\mbox{Hz}$. 
The quality factor of the mirror is two million, which gives $\Gamma 
= 5\times 10^{-2}\;\mbox{s}^{-1}$. With these parameters for the cavity 
we have $\chi=1.49\times 10^{-17}\;\mbox{s}^{-1}$. The cavity damping 
rate through the front mirror is $\gamma = 10^6 \;\mbox{s}^{-1}$, and 
we will take the damping rate through the oscillating mirror to be 
negligible in comparison. This gives $\alpha^2=2.13\times 10^9$, and
we take $\varepsilon=0.2$. The cavity may be cooled to a temperature of 
$T=4.2\;\mbox{K}$, so that
$T_{\mbox{s}}=k_{\mbox{\scriptsize B}}T/(\hbar\nu)=4.37\times 10^6$,
which is certainly in the high temperature regime ($T_{\mbox{s}}\gg 1$).

\begin{figure}
\begin{center}
\leavevmode
\epsfxsize=6.5cm 
\epsfbox{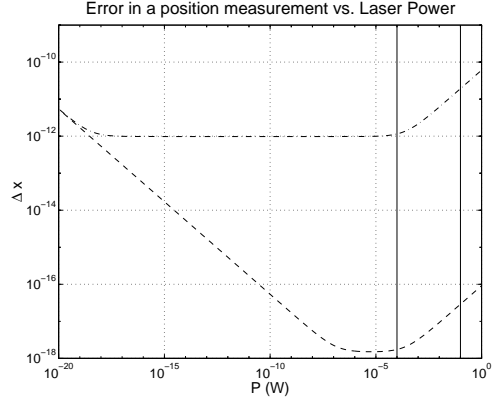}
\caption[Error in a position measurement of a moving cavity mirror]{The 
error in a measurement of the position of 
the mirror, for a measurement time of $\tau_{\scriptsize m}=1$s. 
The dashed curve corresponds to a measurement of a constant
displacement, and the dot-dash curve to a measurement at the
mirror resonance frequency. The contribution of the quantum 
back-action noise to both curves is the sloping section to the 
right. The range of laser powers used in the experiment 
lies between the solid lines. Clearly the experimental parameters 
are in the regime in which the quantum back-action noise may be 
observed.}
\label{theoryfig2}
\end{center}
\end{figure}

In figure~\ref{theoryfig2} we plot the position measurement error as
a function of the laser power, both for the measurement of a constant
displacement, and for a displacement at the mirror resonance
frequency. We have plotted this for a measurement time of one second. 
Conversion to another measurement time merely involves dividing by the
square root of the measurement time given in seconds.

The uncertainty due to the shot noise falls off with laser power,
while that due to thermal noise is independent of laser power, and
that due to the quantum back-action increases with laser power.
These results are already well known. The thermal and shot noise 
contributions are much greater at the 
resonance frequency of the mirror, due to the high mechanical $Q$
factor. Clearly the optimal regime for obtaining a precision
measurement of position is off-resonance, where the back-action 
and thermal noise are suppressed. However, the optimal regime for 
detecting the quantum back-action noise is at resonance, as the 
absolute magnitude of this noise is largest in this case. 
Reasonable experimental values for laser power lie between
the solid lines, where the increase in noise due to the back-action 
is visible. However, our analysis of the spectrum shows us that the
full situation is more complicated. We have shown that the internal 
cavity losses and the classical laser noise have the same dependence
on frequency as the quantum back-action, and the size of these
contributions must therefore be determined accurately for a given 
experimental configuration so that they may be distinguished from the back-action. 

\section{Conclusion}\label{sec6}

We have examined the optomechanical system consisting of an optical cavity containing
a moving mirror to see how the quantum mechanical back-action appears 
among the 
various sources of classical noise. We have shown a number of things 
regarding
this question. First of all, the relationship of the shot noise to 
the noise 
resulting from the oscillating mirror, and hence the limit on a 
position measurement
due to the shot noise, is dependent on the phase measurement scheme. 
In particular, 
the result for phase modulation detection, which is commonly used in 
experiments of this kind, is not the same as that for homodyne 
detection. 
We have found that while the signature of the classical phase noise 
is quite 
different for that of the quantum-back action, the noise due to 
intracavity 
losses and classical amplitude noise has a very similar signature to 
the 
back-action. However, as far as the parameters of the cavity and 
oscillating 
mirror are concerned, realisable experiments are beginning to fall in 
the region 
where the quantum back-action may be observed.

In our treatment of the system we have shown that the standard 
quantum Brownian motion master equation produces a clearly spurious 
term in the steady state noise spectrum for the phase quadrature 
measurement. We have shown that the corrected Brownian motion master 
equation, derived by Diosi, corrects this error. However, it also 
produces a new term in the spectrum which is small in the high 
temperature limit. Testing for the existence of this term now poses 
a very interesting experimental question. 

In the introduction to this chapter we pointed out that many 
authors have shown that there are a large number of
possible uses for optomechanical cavities, including various sensor 
applications, microscopy, gravity wave  detection, the generation of 
non-classical light both internally and in the output beam, basic 
cavity QED experiments, QND  measurements, and for probing the 
decoherence of macroscopic objects. Testing the quantum theory of 
thermal noise, by probing for the existence of the Diosi term in 
the thermal spectrum, provides yet another reason for the importance 
of developing these techniques. The detection of the quantum 
back-action noise in such cavities will represent another
milestone in this development.

\chapter{Determining the State of a Single Cavity Mode}
\label{cmm}
\raggedbottom

In the previous chapters we have been concerned with quantum systems subjected to the continuous measurement of a physical observable. In this chapter we consider an alternative problem in quantum measurement theory: the determination of the {\em entire} quantum state of a system, rather than simply a particular observable. We consider a single cavity mode and show that the quantum state of such a system my be obtained from the photon statistics measured both before and after the interaction with one or more two-level atoms. 


\section{Introduction}

Determining the quantum state of a system is essentially just the process of obtaining the probability distribution of a random variable by taking many samples of that variable. The application of this process to quantum states is, however, complicated by the fact that the state determines a set of probability amplitudes for the values of any given physical observable, rather than just a probability distribution~\cite{qsm}. It is then further complicated if one wants to consider mixed states as well as pure states. To determine the quantum state of a single degree of freedom, given that it is pure, we must obtain the probability distribution for at least two non-commuting physical observables. To see why this is so we need merely note that when expressed in the basis of any physical observable, the wavefunction consists of a complex number for each value of the observable. The probability distribution for that observable gives only the moduli of this set of complex numbers, and consequently we must obtain the distributions for at least two physical observables to obtain the quantum state.

If we wish to determine a {\em mixed} state of a single degree of freedom then we must obtain the distributions for many observables. To gain an insight into why this is so we need to consider how many independent real numbers it takes to specify mixed states. Clearly, to be able to invert the measured data to obtain the quantum state, there must be at least as many independent real numbers in the data as are required to specify the state. To specify a mixed state requires a density matrix or Wigner function. While the wavefunction describing a pure state is merely a function of one variable, these are functions of two variables. As the probability distribution for a single observable is one-dimensional, it only gives us information equivalent to a single slice of the two-dimensional Wigner function. As a consequence we must measure the distributions for many different observables to be able to reconstruct the Wigner function, and hence a mixed state, with any accuracy.

Schemes for determining the quantum state of a system do not necessarily correspond directly to measuring the probability distributions of a set of observables. More generally, we can speak of a set of parameters for which data values are taken. Clearly, to obtain enough independent data values to reconstruct the Winger function, we must obtain data which is parametrised by two continuous co-ordinates. To reconstruct a pure state, on the other hand, only one of the parameters need be continuous (or countably infinite). The other need merely take a small number of discrete values.


There are now a variety of methods which have been suggested for the reconstruction of quantum states. For example the quantum state of a single optical field mode may be determined by tomographic reconstruction from the distributions of all the rotated quadratures obtained using homodyne detection~\cite{Tomo1}. A number of methods for the measurement of a field mode which involve the use of beam splitters have also been suggested~\cite{Konradetc}, and tomographic reconstruction has been suggested and realised for other quantum systems~\cite{otomot,otomoe,atomexp}. In addition, schemes for measuring a single cavity mode by using two-level atoms as probes have been suggested by Wilkens and Meystre (nonlinear atomic homodyne detection)~\cite{atomhom}, and by Bardroff {\em et al.\ }(quantum state endoscopy)~\cite{Betal}. Both these methods involve passing a resonant two-level atom through the cavity containing the state to be measured, and measuring the final atomic excitation probability as a function of the transit time of the atom through the cavity. In the former scheme the atom interacts with a reference field in addition to the cavity mode, and the final atomic excitation probability is also measured as a function of the phase of this reference field. In the latter scheme, while there is no reference field, the measurements are performed for four different initial atomic states. We note also that recently a scheme for measuring the state of a single cavity mode which involves the use of atomic magnetic sublevels has been suggested by Walser {\em et al.\ }\cite{magtom}. In this scheme an atom is sent through the cavity in order to map the cavity state onto the magnetic sublevels, and then the final atomic state is determined from repeated measurements using a tomographic technique.
 
Both Homodyne tomography and nonlinear atomic homodyne detection may be used to measure mixtures as in these schemes measurements are taken for two continuous parameters. On the other hand, in quantum state endoscopy there is one continuous parameter, and one finite discrete parameter (the discrete parameter being the initial atomic state, of which just four are used), so that only pure states may be reconstructed with this scheme. However, quantum state endoscopy requires a somewhat simpler inversion procedure than the other two methods.

We present here various schemes for measuring the state of a cavity mode in
which either practically no inversion, or very simple inversion of the measurement data is required to reconstruct the state; the expansion coefficients of the state in the Fock basis are in an essentially one-to-one correspondence with the measurement data. We consider only pure states, however, and while these schemes are applicable to most pure quantum states currently of interest (number states, coherent states, squeezed states, Schr\"{o}dinger cats) the schemes we present in detail here are not applicable to all pure cavity states. The first two, which involve allowing the cavity mode to interact with one two-level atom, enable the measurement of all pure cavity states which do not contain `zeros' in the photon probability distribution. By a `zero', we mean that one of the photon probabilities, say $P_n$, is zero, while there are both probabilities $P_{m>n}$, and $P_{m<n}$ which are non-zero (ie. the photon distribution is split into two or more parts). An example of a state in this 
category is the squeezed vacuum. The two-atom scheme, which 
involves allowing the cavity to interact with two two-level atoms, allows the 
reconstruction of all pure cavity states which have no contiguous (adjacent) zeros in the photon probability distribution. That is, the photon distribution may have any number of parts, so long as each of the non-zero photon probabilities is separated by no more than one vanishing photon probability. In general if $n$ two-level atoms are sent through the cavity, then pure quantum states with $n-1$ contiguous zeros in their photon probability distribution may be determined. However, as the number of atoms increases, the data inversion that is required increases in complexity. 

The measurement procedure for all these schemes consists of three steps. First the photon statistics of the cavity state are determined by measurements of photon number on a set of identically prepared cavities. This might be achieved by photon-counting the number of photons which leak out of the cavity with a photo-detector (applicable in the optical regime), or by a variety 
of other methods~\cite{PMS}. Next, a second set of cavity modes, all prepared in
the same initial state as before, are allowed to interact for a specified time
with either one or two two-level atoms. The photon statistics of this set is then determined.  In the case of a single atom, we may choose whether or not we measure the final atomic states for each cavity. Finally, the second step is repeated, but the atom(s) are prepared in a different initial state, 
giving a third set of photon statistics. The three sets of photon statistics may
then be used to calculate the phase factors of the coefficients of the initial state in the number basis. Along with the initial photon statistics, which give the amplitudes of the number state expansion coefficients, this determines the
complete quantum state. Clearly this procedure is similar to that required for 
both atomic homodyne detection and quantum state endoscopy. The main difference is that while in the latter two schemes the final atomic excitation probabilities are measured for many different interaction times, in this scheme just one interaction time is chosen, and the final photon probabilities are measured. Thus in this case the cavity state is determined from a measurement of discrete variables only.

In the following section we introduce the general principle on which our
measurement schemes are based, and define some notation. In Section~\ref{M1} we
describe the measurement schemes which involve an interaction with one atom, and in Section~\ref{M2} we describe the schemes involving an interaction with two atoms. Section~\ref{Conc} concludes.

\section{Preliminary Comments} \label{Prelim}
In the Fock (number) basis, the state $|\psi\rangle$ of a single cavity mode of the
electromagnetic field may be written as 
\begin{equation}
  |\psi\rangle = \sum_{n=0}^{\infty} r_n e^{i\theta_n}|n\rangle ,
\end{equation}
where $|n\rangle$ is the quantum state in which the cavity contains $n$ photons, so
that $P_n = r_n^{2}$ is the probability that the single mode will be measured to
contain $n$ photons. To obtain the photon statistics (the photon probabilities
$P_n$), we may measure the number of photons in the cavity for a large number of
cavities prepared in the state $|\psi\rangle$. If we were now to form a new state 
which was the superposition of $|\psi\rangle$ with a known state, for example that 
given by
\begin{equation}
  |\phi\rangle = \sum_{n=0}^{\infty} r_n |n\rangle ,
\end{equation}
then the photon statistics of the resulting superposition state contains 
information about the phase factors $\theta_n$. If we create the superposition 
state
\begin{equation}
  |\chi_1\rangle = \frac{1}{\sqrt{2}}(|\psi\rangle + |\phi\rangle ) ,
\end{equation}
then the photon probabilities for this state,
\begin{equation}
  Q_{1,n} = |\langle\chi |n\rangle |^2 ,
\end{equation}
are given by
\begin{equation}
  Q_{1,n} = P_n(1-\cos{\theta_n}) ,
\end{equation}
so that the cosine of each phase is
\begin{equation}
  \cos{\theta_n} = Q_{1,n}/P_n - 1.
\end{equation}
Alternatively, if we form the superposition
\begin{equation}
  |\chi_2\rangle = \frac{1}{\sqrt{2}}(|\psi\rangle + i|\phi\rangle ) ,
\end{equation}
then the photon probabilities for this state, denoted $Q_{2,n}$, are given by
\begin{equation}
  Q_{2,n} = P_n(1+\sin{\theta_n}),
\end{equation}
so that the sine of each phase factor is given by
\begin{equation}
  \sin{\theta_n} = 1 - Q_{2,n}/P_n.
\end{equation}
As knowledge of the sine and cosine of the angle completely determines the 
angle itself, the three probability distributions, $P_n, Q_{1,n}$ and $Q_{2,n}$, 
which may be measured by photo-counting, allow the complete quantum state to be 
calculated in a simple way.

Unfortunately, it is not possible to create the superposition of an unknown 
quantum state with a known quantum state; both quantum states must be known. 
We can see this from the following argument. It is 
clear that given two states $|\phi\rangle$ and $|\chi\rangle$ we can find a 
unitary evolution operator $U$ with the property
\begin{equation}
  U|\phi\rangle = \frac{1}{\sqrt{2+2\mbox{Re}[\langle\phi |\chi\rangle]}}
(|\phi\rangle + |\chi\rangle) . \label{tran1}
\end{equation}
To construct such a unitary operator we need merely to choose two bases 
${|\phi_i\rangle}$ and ${|\psi_i\rangle}$ such that $|\phi\rangle = 
|\phi_1\rangle$ and $(|\phi\rangle + |\psi\rangle)/ \sqrt{2+2\mbox{Re}[\langle\phi |\chi\rangle]} = 
|\psi_1\rangle$. The desired operator is then $U=\sum_i |\psi_i\rangle 
\langle\phi_i|$. However, the unitary operator which is required to form the 
superposition of $|\phi\rangle$ and $|\chi\rangle$ is dependent in general 
upon the initial state $|\phi\rangle$. This follows immediately from the 
linearity of $U$. If we assume that the same unitary operator will perform 
the operation
\begin{equation}
  U|\psi\rangle = \frac{1}{\sqrt{2+2\mbox{Re}[\langle\psi |\chi\rangle]}}
(|\psi\rangle + |\chi\rangle) , \label{tran2}
\end{equation}
where $|\psi\rangle \not= |\phi\rangle$, then using the linearity of $U$ we 
find that in general
\begin{equation}
  U(\alpha|\phi\rangle + \beta|\psi\rangle) \not= \frac{1}{\cal N}(\alpha|\phi\rangle + \beta|\psi\rangle + |\chi\rangle) , \label{tran3}
\end{equation}
where $N$ is the normalisation constant. It follows that in general for 
different initial states a different evolution operator is required to 
create the superposition of the initial state with a known state $|\chi\rangle$. 
Before we can find the correct evolution operator to create the superposition of
 two states, we must know both of those states. We conclude, therefore, that 
we cannot create the superposition of two unknown quantum states, or even the 
superposition of a known quantum state with an unknown one, just as we cannot 
clone one unknown quantum state~\cite{wootters}.

Nevertheless, with a simple atom-field cavity QED interaction we may perform
measurements along the lines described above. Allowing a cavity mode to interact
with a two-level atom in the Jaynes-Cummings manner~\cite{JCH} forms
superpositions of adjacent number states. In this way we may use a procedure
similar to that described above to measure the phase differences between adjacent
number states in the superposition which makes up the quantum state to be
measured. However, this only allows us to measure the complete quantum state if
there are no zeros in the photon distribution. If a certain number state $|n\rangle$
does not contribute to the state to be measured (ie.\ $r_n=0$) when there are
contributing number states on either side of it, then as we cannot determine the
phase difference between those number states and the number state that does
appear in the superposition, we cannot obtain the phase difference between
the two states on either side. To do this we require to superpose number states
which are separated by more than one photon. This is why sending two atoms
through the cavity allows us to measure quantum states with zeros in the photon
distribution, so long as none of the zeros are adjacent. Coherent states~\cite{WandM} and number states are among those which may be measured with a one-atom interaction, and squeezed states~\cite{WandM} and `Schr\"{o}dinger cats' (being superpositions of two coherent states) are among those which may be measured with a two-atom interaction.

For the calculations in the following sections, it will be convenient to define 
compact notation for certain functions. In the expressions that follow $n$
represents a non-negative integer, and $t$ and $\Omega_n$ represent real numbers. 
We will use the following definitions:
\begin{eqnarray}
 \Gamma^t & = & \cos(\Omega_n t) \nonumber \\ 
 \Gamma^t_{\raisebox{0.3mm}{\tiny $\!\! +$}} & = & 
\cos(\Omega_{n+1} t) , \;\;\; \Gamma^t_{\raisebox{0.3mm}{\tiny $\!\! + \!\! +$}} = \cos(\Omega_{n+2} t) 
\nonumber \\
 \Gamma^t_{\raisebox{0.3mm}{\tiny $\!\! -$}} & = & \cos(\Omega_{n-1} t) \; , \;\; \Gamma^t_{\raisebox{0.3mm}{\tiny $\!\! =$}} = 
\cos(\Omega_{n-2} t) .
\end{eqnarray}

We will use the same notation for sine functions, where $\Gamma$ will be
replaced with $\Upsilon$. When the cosine or sine functions are squared, the
$\Gamma$ and $\Upsilon$ will have tildes added, thus
\begin{equation}
 \tilde{\Gamma}^t = \cos^2(\Omega_n t) \; , \;\; \tilde{\Upsilon}^t =
\sin^2(\Omega_n t).
\end{equation}
When it is convenient we will use a number subscript for $\Gamma$ and $\Upsilon$, 
and in this case the definitions
\begin{equation}
 \Gamma_n^t = \cos(\Omega_n t) , \;\;\; \Upsilon_n^t = 
\sin(\Omega_n t) ,
\end{equation}
will hold. This notation will greatly reduce the space required for the expressions 
derived in the following sections, without introducing any ambiguity.

\section{Measuring quantum states with a one-atom interaction}
\label{M1}
As we stated in section~\ref{Prelim}, the state of the single cavity mode to be 
measured may be written in terms of the number states as
\begin{equation}
  |\psi\rangle = \sum_{n=0}^{\infty} r_n e^{i\theta_n}|n\rangle .
\end{equation}
Once the photon distribution, $P_n$, has been determined, we create a new
state in which the photon distribution depends on the phases $\theta_n$. This
may be achieved by passing a two-level atom through the cavity so that it
interacts with the cavity mode via the familiar Jaynes-Cummings 
Hamiltonian~\cite{JCH,JCsol}
\begin{equation}
  H = \omega\sigma_z + \omega a^\dagger a + \kappa(\sigma^+ a + 
\sigma^- a^\dagger).
\label{Hint}
\end{equation}
Denoting the ground and excited atomic states by $|g\rangle$ and $|e\rangle$
respectively, the operators in this expression are
\begin{equation}
  \sigma_z =  \frac{1}{2}(|e\rangle \langle e| - |g\rangle \langle g|) \; , \mbox{with} \;\;  \sigma^+ = |e\rangle \langle g| \; , \;\; \sigma^- = (\sigma^+)^\dagger ,
\end{equation} 
the latter two being the respective raising and lowering operators for the atom.
The annihilation operator for the cavity mode is denoted by $a$, $\omega$ is the 
frequency of the cavity mode (which is resonant with the atomic transition), 
$\kappa$ is the strength of the (dipole) coupling between the atom and the field, 
and we have set $\hbar=1$ for convenience. As the atom passes through the cavity, 
it interacts with the cavity mode, and depending on its initial state it will 
either `begin' to emit a photon into the mode, or `begin' to absorb a photon from 
the mode. If the atom is sent into the cavity in a superposition of ground and 
excited states, then the state of the mode will be transformed essentially into a 
superposition of the initial mode plus a `partial' extra photon, or the initial 
mode minus a `partial' photon. (The first case if the atom exits in the ground 
state, and the second case if the atom exits in the excited state.) Hence, 
allowing the mode to interact with a resonant two-level atom will create a 
superposition of the mode state with a shifted version of itself. We show now how 
the phases $\theta_n$ may be determined from photon counts performed on the mode 
after the interaction. We will see that forming a superposition state by using two 
copies of the original state is not ideal for the purposes of quantum state 
measurement, in that it only allows a certain set of quantum states to be measured. 
In the following analysis we will work in the interaction picture, where the
interaction Hamiltonian is given by $H_I =  \kappa(\sigma^+ a + \sigma^- a^\dagger)$.

Before we give results for the general case in which the cavity may contain a 
superposition of arbitrarily many number states, we examine the simpler 
situation in which the intra-cavity state is in a superposition of zero and one 
photons. This example contains the basic idea behind our scheme which we simply generalise afterwards. Let the initial cavity state be written as
\begin{equation}
 |\chi(0)\rangle = r_0|0\rangle + r_1e^{i\theta_1}|1\rangle .
\end{equation}
We take the initial atomic state to be $1/\sqrt{2}(|g\rangle + e^{i\phi}|e\rangle )$. 
After the atom has interacted with the cavity
for a time $t$ the joint `cavity--atom' state is (after omitting the $1/\sqrt{2}$ prefactor)
\begin{eqnarray}
 &   & r_0|0,g\rangle +  \left[ r_{0}
e^{i\phi}\Gamma_0^t - ir_{1}
e^{i\theta_{1}}\Upsilon_0^t \right]|0,e\rangle + \left[ r_{1} e^{i\theta_{1}} \Gamma_0^t - ir_{0} e^{i\phi}\Upsilon_0^t \right]|1,g\rangle \nonumber \\ 
 & + & r_{1}
e^{i(\theta_{1}+\phi)}\Gamma_1^t |1,e\rangle -  ir_{1}e^{i(\theta_{1}+\phi)}\Upsilon_1^t |2,g\rangle .
\label{jc}
\end{eqnarray}
If we denote the probability that the cavity will be found to contain zero photons after the interaction by $Q_0/2$, then from Eq.(\ref{jc}) we obtain 
the following equality
\begin{eqnarray}
  Q_0 & = & P_0(1+\tilde{\Gamma}^t_0) + P_{1}\tilde{\Upsilon}^t_0 + 2\sqrt{P_0P_{1}}\Gamma^t_0\Upsilon^t_0 \sin(\theta_1 - \phi) .
\end{eqnarray}
Clearly this contains all the necessary information regarding the phase $\theta_1$, since we can obtain $P_0, P_1$ and $Q_0$ by photon counting techniques. This now completely determines the cavity state. 

We turn now to the general case. Taking the initial state of the cavity mode to be $|\chi(0)\rangle \equiv |\psi\rangle$, and allowing the atom to enter the 
cavity in the state $1/\sqrt{2}(|g\rangle + e^{i\phi}|e\rangle )$, the state of the atom plus cavity system after an interaction time $t$ is~\cite{JCH,JCsol}
\begin{eqnarray}
 \sqrt{2}|\chi(t)\rangle =  r_0e^{i\theta_0} |0,g\rangle  & + & \sum_{n=0}^\infty \left[ r_{n}
e^{i(\theta_{n}+\phi)}\Gamma_n^t - ir_{n+1}
e^{i\theta_{n+1}}\Upsilon_n^t \right] |n,e\rangle \nonumber \\
      &  + &  \sum_{n=0}^\infty \left[ r_{n+1}e^{i\theta_{n+1}} \Gamma_n^t - ir_{n}
e^{i(\theta_{n}+\phi)}\Upsilon_n^t \right] |n+1,g\rangle .
\end{eqnarray}
In this expression the $\Gamma's$ and $\Upsilon's$ are functions of $\Omega_n$
as defined in the previous section, and in addition $\Omega_n=\kappa\sqrt{n+1}$.

Denoting the joint probability that after the interaction both the cavity is 
measured to contain $n$ photons, and the atom is found to be in its excited state 
as $Q_n^e/2$, and the corresponding joint probability for the atom to be found in its 
ground state by $Q_n^g/2$, we have, for $n\geq 1$
\begin{eqnarray}
  Q_{n}^e & = & P_n\tilde{\Gamma}^t + P_{n+1}\tilde{\Upsilon}^t + 2\sqrt{P_nP_{n+1}}\Gamma^t\Upsilon^t \sin(\Delta\theta_{n+1} -\phi) , \\
  Q_{n}^g & = & P_n\tilde{\Gamma}^t_{\raisebox{0.3mm}{\tiny $\!\! 
-$}} + P_{n-1}\tilde{\Upsilon}^t_{\raisebox{0.3mm}{\tiny $\!\! -$}} - 2\sqrt{P_nP_{n-1}}\Upsilon^t_{\raisebox{0.3mm}{\tiny $\!\!
-$}}\Gamma^t_{\raisebox{0.3mm}{\tiny $\!\! -$}} 
\sin(\Delta\theta_{n} - \phi) ,
\end{eqnarray}
where $\Delta\theta_n = \theta_{n+1} - \theta_n$. For $n=0$ we have
\begin{eqnarray}
  Q_{0}^e & = & P_0\tilde{\Gamma}^t_0 + P_{1}\tilde{\Upsilon}^t_0 + 2\sqrt{P_0P_{1}}\Gamma^t_0\Upsilon^t_0 \sin(\Delta\theta_1 - \phi) , \\
  Q_{0}^g & = & P_0 .
\end{eqnarray}
We denote the total probability for the cavity to contain $n$ photons after the 
interaction by $Q_n/2$, which is clearly given by $Q_n/2 = Q_n^g/2+Q_n^e/2$.

As the overall phase factor for the state is unimportant, we may set the phase
$\theta_0$ equal to zero, and the set of phase differences $\Delta\theta_n$
gives us complete information about the phase factors. Solving the above
equations for the sine of the phase differences we see that these are easily
calculated from the photon probabilities $P_n, Q_n^e$ and $Q_n^g$ by
\begin{eqnarray}
 \sin(\Delta\theta_n-\phi) & = & 
\frac{P_n\tilde{\Gamma}^t_{\raisebox{0.3mm}{\tiny
$\!\! -$}}+P_{n-1}\tilde{\Upsilon}^t_{\raisebox{0.3mm}{\tiny $\!\! -
$}}-Q_{n}^g}
                                 {2\sqrt{P_nP_{n-
1}}\Gamma^t_{\raisebox{0.3mm}
{\tiny $\!\! -$}}\Upsilon^t_{\raisebox{0.3mm}{\tiny $\!\! -$}}} , \\
 \sin(\Delta\theta_n-\phi) & = &
\frac{Q_{n-1}^e - P_{n-1}\tilde{\Gamma}^t_{\raisebox{0.3mm}{\tiny
$\!\! -$}}-P_{n}\tilde{\Upsilon}^t_{\raisebox{0.3mm}{\tiny $\!\! -
$}}}
                                 {2\sqrt{P_nP_{n-
1}}\Gamma^t_{\raisebox{0.3mm}
{\tiny $\!\! -$}}\Upsilon^t_{\raisebox{0.3mm}{\tiny $\!\! -$}}} .
\end{eqnarray}
for all $n$. Therefore, choosing the initial atomic state so that $\phi=0$ we 
obtain the sine of every phase difference, and choosing the initial atomic 
state so that $\phi=\pi/2$ we obtain the respective cosines. Naturally once 
we have the sines and cosines the phase differences $\Delta\theta_n$ are well 
determined. In each case we obtain a measurement of all the phase differences 
from both $Q_{n}^e$ and $Q_{n}^g$.

We may also obtain the phase differences without measuring the final atomic 
state. In this case performing photon counting on the final cavity state gives
us the overall photon probabilities $Q_n$. The expressions for these are
\begin{eqnarray}
  Q_0 & = & P_0(1+\tilde{\Gamma}^t_0) + P_{1}\tilde{\Upsilon}^t_0 + 2\sqrt{P_0P_{1}}\Gamma^t_0\Upsilon^t_0 
\sin(\Delta\theta_n -\phi) ,
\label{eqforq0}
\end{eqnarray}
for $n=0$, and
\begin{eqnarray}
  Q_n & = & P_n(\tilde{\Gamma}^t + 
\tilde{\Gamma}^t_{\raisebox{0.3mm}{\tiny $\!\!
-$}}) + P_{n+1}\tilde{\Upsilon}^t +
P_{n-1}\tilde{\Upsilon}^t_{\raisebox{0.3mm}{\tiny $\!\! -$}} + 2\sqrt{P_nP_{n+1}}\Gamma^t\Upsilon^t 
\sin(\Delta\theta_{n+1} -
\phi) \nonumber \\ 
         &   & - 2\sqrt{P_nP_{n-
1}}\Upsilon^t_{\raisebox{0.3mm}{\tiny $\!\!
-$}}\Gamma^t_{\raisebox{0.3mm}{\tiny $\!\! -$}} 
\sin(\Delta\theta_{n} - \phi) ,
\end{eqnarray}
for $n\geq 1$. We may calculate $\Delta\theta_1$ from $Q_0$ using
\begin{eqnarray}
   \sin(\Delta\theta_1-\phi) & = & \frac{ Q_{0} - 
P_{0}(1+\tilde{\Gamma}^t_0)
-P_{1}\tilde{\Upsilon}^t_{0}} {2\sqrt{P_0P_{1}}\Gamma^t_0 
\Upsilon^t_{0}} .
\end{eqnarray}
The other phase differences may then be obtained by recursion, using
\begin{eqnarray}
  \sin(\Delta\theta_{n+1}-\phi) = \frac{Q_n -
P_{n}(\tilde{\Gamma}^t + \tilde{\Gamma}^t_{\raisebox{0.3mm}{\tiny 
$\!\! -$}}) -
P_{n-1}\tilde{\Upsilon}^t_{\raisebox{0.3mm}{\tiny $\!\! -$}} -
P_{n+1}\tilde{\Upsilon}^t_{\raisebox{0.3mm}{\tiny $\!\! +$}}}
{2\sqrt{P_{n}P_{n+1}} \Gamma^t \Upsilon^t} + \sqrt{\frac{P_{n-1}}{P_{n+1}}} \left( 
\frac{\Gamma^t_{\raisebox{0.3mm}{\tiny
$\!\! -$}} \Upsilon^t_{\raisebox{0.3mm}{\tiny $\!\! -
$}}}{\Gamma^t \Upsilon^t}
\right) \sin(\Delta\theta_{n}-\phi) . \nonumber
\end{eqnarray}
It is clear from the above formulae that only phase differences between adjacent
states may be measured in this way. If there is a hole in the photon probability
distribution, so that $P_n=0$ for some $n$, then we cannot measure the phase 
difference between the coefficient for that number state and the two on either 
side. As a consequence the above schemes will allow us to measure only pure 
states which do not contain zeros in the photon distribution.

\section{Measuring quantum states with a two-atom interaction}
\label{M2}
To measure all the phase factors for a state in which there are zeros in the
photon number distribution, we may pass two atoms through the cavity
consecutively. This creates a state which contains superpositions of number
states separated by both 1 and 2 photons. In this case phase differences between
number states separated by 2 photons may be measured by photon counting, so that
zeros in the photon distribution are not an obstacle to the measurement so long as
no two are adjacent.

To analyse this measurement scheme, we may write the initial state of the
system, which consists of three uncorrelated subsystems, the cavity and two
atoms, as 
\begin{eqnarray}
  |\Psi\rangle & = & \sum_{n=0}^\infty \frac{r_n}{2}e^{i\theta_n} 
|n\rangle
\otimes (|g\rangle + e^{i\phi}|e\rangle) \otimes
(|g\rangle + |e\rangle) .
\label{init2}
\end{eqnarray}

We allow the first atom to interact with the cavity for a time $t$, and then, at
some time after that, we allow the second atom to interact with the cavity for a
time $s$. In the interaction picture, the expression for both the state and the
joint probabilities, $Q_n^{gg},Q_n^{ge},Q_n^{eg},Q_n^{ee}$, following  these
interactions are fairly complicated, and are given in Appendix~\ref{app3}.
From these probabilities the phase differences may be calculated. 
If we are measuring the phase factors for a state for which all the 
probabilities for odd numbers of photons are zero (for example a squeezed vacuum), 
then the expressions for the phase differences simplify considerably. Setting 
$s=t$, we have in this case
\begin{eqnarray}
   \cos(\Delta\tilde{\theta}_n-\phi) & = &
\frac{P_n\tilde{\Gamma}^t_{\raisebox{0.3mm}{\tiny
$\!\! -$}}\tilde{\Gamma}^t_{\raisebox{0.3mm}{\tiny $\!\! -$}} + 
P_{n-2}\tilde{\Upsilon}^t\tilde{\Upsilon}^t_{\raisebox{0.3mm}{\tiny 
$\!\! -$}} -
Q^{gg}_n}{2\sqrt{P_nP_{n-
2}}\tilde{\Gamma}^t_{\raisebox{0.3mm}{\tiny$\!\! -$}}
\Upsilon^t\Upsilon^t_{\raisebox{0.3mm}{\tiny $\!\! -$}}} , \\
   \cos(\Delta\tilde{\theta}_n-\phi) & = & \frac{Q_{n-1}^{ge} -
P_{n-2}\tilde{\Gamma}^t_{\raisebox{0.3mm}{\tiny $\!\! -
$}}\tilde{\Upsilon}^t -
P_n\tilde{\Gamma}^t_{\raisebox{0.3mm}{\tiny $\!\!
-$}}\tilde{\Upsilon}^t_{\raisebox{0.3mm}{\tiny $\!\! -
$}}}{2\sqrt{P_nP_{n-2}} 
\tilde{\Gamma}^t_{\raisebox{0.3mm}{\tiny $\!\! -$}} \Upsilon^t
\Upsilon^t_{\raisebox{0.3mm}{\tiny $\!\! -$}} } , \\
   \cos(\Delta\tilde{\theta}_n-\phi) & = & \frac{Q^{eg}_{n-1} -
P_n\tilde{\Gamma}^t_{\raisebox{0.3mm}{\tiny $\!\!
=$}}\tilde{\Upsilon}^t_{\raisebox{0.3mm}{\tiny $\!\! -$}} - 
P_{n-2}\tilde{\Gamma}^t_{\raisebox{0.3mm}{\tiny $\!\!
=$}}\tilde{\Upsilon}^t_{\raisebox{0.3mm}{\tiny $\!\! =$}} 
}{2\sqrt{P_nP_{n-2}}
\tilde{\Gamma}^t_{\raisebox{0.3mm}{\tiny $\!\!
=$}}\Upsilon^t_{\raisebox{0.3mm}{\tiny $\!\!
=$}}\Upsilon^t_{\raisebox{0.3mm}{\tiny $\!\! -$}} } , \\
  \cos(\Delta\tilde{\theta}_n-\phi) & = & \frac{
P_{n-2}\tilde{\Gamma}^t_{\raisebox{0.3mm}{\tiny $\!\! =$}}
\tilde{\Gamma}^t_{\raisebox{0.3mm}{\tiny $\!\! =$}} +
P_n\tilde{\Upsilon}^t_{\raisebox{0.3mm}{\tiny $\!\! =$}} 
\tilde{\Upsilon}^t -
Q^{ee}_{n-2} }{2\sqrt{P_nP_{n-2}} 
\tilde{\Gamma}^t_{\raisebox{0.3mm}{\tiny $\!\!
=$}} \Upsilon^t \Upsilon^t_{\raisebox{0.3mm}{\tiny $\!\! =$}} } ,
\end{eqnarray}
where $\Delta\tilde{\theta}_n = \theta_n-\theta_{n-2}$, and $n$ is even. 
If we do
not measure the final atomic state, then we may use the recursion relationship
\begin{eqnarray}
 \cos(\Delta\tilde{\theta}_2 - \phi) & = & \frac{ P_0(1 + 
2\tilde{\Gamma}^t_0 +
\tilde{\Upsilon}^t_0\tilde{\Upsilon}^t_0 + 
\tilde{\Gamma}^t_0\tilde{\Gamma}^t_0 -
\Gamma^t_0\tilde{\Upsilon}^t_0\cos(\phi)) +
P_2\tilde{\Upsilon}^t_0\tilde{\Upsilon}^t_1 } { 
\sqrt{P_0P_2}\tilde{\Gamma}^t_0
\Upsilon^t_0\Upsilon^t_1 } , \\
\cos(\Delta\tilde{\theta}_{n+2} - \phi) & = & \frac{
P_{n-2}(\tilde{\Upsilon}^t\tilde{\Upsilon}^t_{\raisebox{0.3mm}{\tiny 
$\!\! -$}}) + 
P_n(\tilde{\Gamma}^t\tilde{\Gamma}^t + 
\tilde{\Gamma}^t_{\raisebox{0.3mm}{\tiny $\!\!
-$}}\tilde{\Gamma}^t_{\raisebox{0.3mm}{\tiny $\!\!
-$}}) + 
P_{n+2}(\tilde{\Upsilon}^t\tilde{\Upsilon}^t_{\raisebox{0.3mm}{\tiny 
$\!\!
+$}}) - Q_n } {2\sqrt{P_nP_{n+2}}\tilde{\Gamma}^t \Upsilon^t
\Upsilon^t_{\raisebox{0.3mm}{\tiny $\!\! +$}}} \\ \nonumber
 & & - \frac{\sqrt{P_{n-2}}\tilde{\Gamma}^t_{\raisebox{0.3mm}{\tiny 
$\!\! -$}}
\Upsilon^t_{\raisebox{0.3mm}{\tiny $\!\! -$}} \Upsilon^t } 
{\sqrt{P_{n-2}}
\tilde{\Gamma}^t \Upsilon^t \Upsilon^t_{\raisebox{0.3mm}{\tiny 
$\!\! +$}}
}\cos(\Delta\tilde{\theta}_n - \phi) .
\end{eqnarray}
Similar expressions hold when all the probabilities of even numbers of photons 
are zero, and these are also easily derived by setting the $P_n$ for $n$ even to
zero in the expressions for the $Q_n^{xx}$ given in Appendix~\ref{app3}.

For the general case, in which there are a number of non-adjacent zeros in the
photon probability distribution, we suggest two procedures which may be
used to calculate the phase factors from the joint probabilities. Inspection of 
the expressions for $Q_n^{gg}$, $Q_{n-1}^{ge}$, $Q_{n-1}^{eg}$ and $Q_{n-2}^{ee}$ 
shows that setting $\phi$ to zero, each is a linear combination of
$\sin(\Delta\theta_{n})$,  $\sin(\Delta\theta_{n-1})$ and 
$\cos(\Delta\tilde{\theta}_{n})$. Any three of the expressions for the $Q$'s may
therefore be linearly inverted to obtain these three trigonometric quantities.
Setting $\phi=\pi/2$, the expressions for the four $Q$'s are then a linear
combination of  $\sin(\Delta\theta_{n})$,  $\sin(\Delta\theta_{n-1})$,
$\cos(\Delta\theta_{n})$,  $\cos(\Delta\theta_{n-1})$ and
$\sin(\Delta\tilde{\theta}_{n})$. The first two of these have been determined by 
the previous step, so that there are again three unknowns to determine by linear
inversion of the expressions for three of the $Q$'s. Those two steps determine the
sines and cosines of the phase differences, and consequently the phase differences
themselves. If, for example, $P_m$ is zero, then the calculation of 
$\theta_{m+1} - \theta_{m-1}$ is simpler since the terms containing 
$\Delta\theta_n$ and $\Delta\theta_{n+1}$ vanish in the expressions for the $Q$'s.

Alternatively we may use a recursive procedure. Inspection of the expressions for 
$Q_0^{ge}$ and $Q_0^{eg}$ shows that these may be used to determine the sine and
cosine of $\Delta\theta_1$. Once we know that, there are only two unknowns
rather than three to determine from $Q_2^{gg}$, $Q_{1}^{ge}$, $Q_{1}^{eg}$ and
$Q_{0}^{ee}$, making the inversion simpler. From these four $Q$'s one of the
unknowns we obtain is $\Delta\theta_2$, so that there are again only two
unknowns that we need to determine from the next set of $Q$'s, which is
$Q_3^{gg}$, $Q_{2}^{ge}$, $Q_{2}^{eg}$ and $Q_{1}^{ee}$, and so on up the 
recursion chain.

\section{Conclusion}
\label{Conc}
We have shown that measuring the photon statistics of a single mode cavity field both before and after an interaction with a single two-level atom may be used to reconstruct in a very simple manner pure cavity states which contain no zeros in the photon probability distribution. This uses the fact that when a two-level atom initially prepared in a superposition of ground and excited states interacts with a cavity it creates superpositions of adjacent number states, so that the photon distribution after the interaction contains information about the phase factors of the coefficients in the expansion of the initial state in the Fock basis. We have also shown that this may be extended to measure pure states which contain no adjacent zeros in the photon distribution when two two-level atoms are allowed to interact with the cavity consecutively. In general it is clear from our analysis that sending $n$ consecutive atoms through the cavity will form superpositions of Fock states separated by $n$ photons, and therefor should allow the reconstruction of pure quantum states which contain $n$ contiguous zeros, although the linear inversion that is required becomes increasingly complicated.

\chapter{Conclusion}\label{conc}

In this thesis we have presented work on a number of topics in the field of quantum measurement and quantum noise. In Chapter~\ref{evopch} we were concerned with the linear formulation of quantum trajectories, which describe the evolution of the state of a system subjected to continuous measurement processes. In contrast to the Schr\"{o}dinger equation, which is sufficient to describe the evolution of a closed system, and for which evolution operators may be obtained simply by exponentiating the Hamiltonian, obtaining evolution operators for the linear stochastic equations describing quantum trajectories is not straightforward. We presented a method for doing this, and showed that it would render explicit evolution operators for a variety of stochastic equations describing various measurement processes.

The quantum trajectory formulation, because it describes the evolution of a quantum system for every realisation of the measurement process, may be used to calculate aspects of the dynamics of the measured system not possible with the use of a master equation. One such aspect, which is yet to be investigated in detail, is the purification of an initially mixed state during a continuous measurement. This purification reflects the fact that information about the state of the system is being obtained continually during the measurement, and will undoubtedly be the subject of future work.

Our ability to manipulate and observe individual quantum systems is continually increasing, and continuous quantum measurement theory is therefore likely to become increasingly relevant experimentally. Experiments are now being developed to probe the quantum noise in the position measurement of macroscopic objects, such as the configuration we examined in Chapter~\ref{qnc}, and position measurements of single atoms in a cavity are now beginning to be realised~\cite{Mabuchi}. The continuous tracking and control of the position of a single quantum degree of freedom is unlikely to be far away.

In Chapter~\ref{cmm} we considered the question of the determination of the state of a quantum system from a set of measurements. This subject has been a topic of intense activity in the past few years, and there are now many measurement schemes which have been suggested for this purpose. These are now of  practical importance in determining the states of quantum systems under our control.

It is not by chance that the systems we have been concerned with here are quantum optical. This field has proved a rich ground for both testing the efficacy of, and the development of new ideas in, quantum measurement theory. Probably the most famous example of the use of quantum optics for the former is the testing of Bell's inequalities~\cite{Bell}. Regarding the latter, both the quantum trajectory and quantum noise formulations of open quantum systems were developed in the field of quantum optics~\cite{Carm} (although the former was also developed in a fairly parallel fashion by pure measurement theorists~\cite{Belvkn}). In addition, the extension of continuous measurement theory to include continuous feedback was also developed in this field~\cite{HMWPhD}. Quantum optics has also proved a fertile ground for the development of many kinds of measurement schemes, such as QND measurements, joint measurements on incompatible observables, and the reconstruction of quantum states, among others. Note that in using the term `quantum optics' we mean to include also the study of non-relativistic atom-laser and atom-atom interactions as well, such as the cooling, trapping and manipulation of atoms and ions. These are now providing the most powerful means yet of manipulating quantum states~\cite{otomoe}.

Recently a new focus has arisen in quantum measurement theory, due to the emergence of quantum information theory. The laws of quantum information theory stem from the dynamics of quantum systems, and the rules of quantum measurement theory. As a consequence, the formalism of quantum measurement theory, and in particular the formulation in terms of Positive Operator-Valued Measures, forms the basis of quantum information theory. However, in this case the focus is on the information contained in quantum systems, and ways in which this may be manipulated. 

Clearly a two-state quantum system may contain one classical bit of information. However, whether this is all that such a system may contain is not such a trivial question. A two-state quantum system is actually described by two real parameters. Specifying these parameters with arbitrary accuracy requires an arbitrarily large amount of information, and from this point of view the amount of information which may be stored in a two-state quantum system is arbitrarily large. As a consequence it is the amount of information that may be extracted during a measurement process which gives the maximum storage capacity of a quantum system. Similar arguments apply to the amount of classical information which may be transmitted by using entangled quantum systems, and quantum measurement theory is now being employed to answer questions of this nature.

We may also consider a new kind of information, quantum information, as distinct from the standard `classical' kind defined in Shannon's information theory~\cite{Shannon}. That this is a useful concept is illustrated by the following two examples. The first of these is the discovery that computations may be performed on information contained in quantum two-state systems, or `q-bits', that cannot be perform on the usual classical variety~\cite{Deutsch}. As a consequence, the transmission of quantum states, and therefore quantum information, takes on an importance of its own. A second example is the use of entanglement between two quantum systems for secret communication, a procedure referred to as {\em quantum cryptography}~\cite{QuanCryp}. When two classical systems are correlated, they are referred to as containing `mutual information'. However, it is certainly not possible to use two classically correlated systems for secret communication. The `mutual information' between two entangled quantum systems is therefore distinct from that between two classically correlated systems, and its properties must be studied accordingly. Now that entanglement is seen as a useful resource, quantum measurement theory is being used to investigate its manipulation.

We see that quantum measurement is becoming increasingly relevant, not only to the experimental investigation of individual quantum systems, but also to the exploration of new ways in which the nature of the quantum world can be put to practical use for the manipulation of information. As time goes by no doubt we will begin to see the new theoretical developments becoming a practical reality.


\appendix
\chapter{An Introduction to the Input-Output Relations}\label{appendix1}

Here we present an introduction to the Input-Output formulation of 
open quantum systems. A basic knowledge of Quantum Electrodynamics is 
assumed. The derivation of the one-dimensional Lagrangian, the 
modal decomposition of the reservoir light field, the nature of 
the input and output fields and their relationship to initial and final conditions are discussed.

\section{Introduction}
Input-Output theory, or Input-Output {\em formalism}, is a 
particular formulation of the quantum treatment of open systems. 
Open systems are systems coupled in some way to an 
environment, or  {\em reservoir}, which has many degrees of 
freedom. The result of the coupling is that the system may exchange 
energy with the reservoir, and the reservoir introduces noise into 
the system.

Open systems see extensive use in Quantum Optics; an atom 
coupled to the quantum electro-magnetic field, which as a result    
undergoes spontaneous emission, is an example of an open 
system, and we have found that we may treat the decay of 
optical cavities in essentially the same manner. The Input-Output 
formulation is a one dimensional theory, and is particularly useful 
for treating optical cavities; the modes of the free quantum electric 
field outside the cavity, which form the reservoir, may be decomposed 
into travelling fields which are input to, and output from, the 
cavity. The Input-Output treatment of open optical cavities is 
equivalent to the Master equation treatment~\cite{BHE}, but 
where as the Master equation treats only the evolution of the system, 
the Input-Output formulation allows the calculation of the output, as 
it shows that this may be written in terms of the input and the system operators. In fact, if it is desired, the system evolution may be 
solved using the Master equation (rather than the equivalent quantum 
Langevin equations of the Input-Output formulation), and then the output 
may be calculated using the relations of Input-Output theory. 

Input-Output theory was first formulated in a paper by Collett and 
Gardiner\cite{inout}, and their treatment appears in the texts by 
Gardiner\cite{Gardiner2} and by Walls and Milburn\cite{WandM}. In treating a cavity that experiences damping through one end mirror, we are only interested in the free electromagnetic field in half of one dimensional space; from the lossy cavity mirror out to infinity. The approach taken by Collet and Gardiner is to treat the field in only this half-space. This results in a modal decomposition of the free field in which each mode operator corresponds to a mode travelling in {\em both} directions. Other treatments, which instead use a unidirectional field may be found in the book by Carmichael~\cite{Carm}, and in Wiseman's PhD thesis~\cite{HMWPhD}. Our treatment here is essentially an expansion upon that given by Gardiner. We will, however, follow the notation in 
Cohen-Tannoudji {\em et al.}~\cite{CCT}.

In Section~\ref{LandH} we introduce the Lagrangian and 
Hamiltonian that are used for the treatment of Input-Output 
theory. We show what steps are necessary to arrive at these from 
the standard Lagrangian of non-relativistic quantum 
electrodynamics. In Section~\ref{InOut} we show how the external 
light field may be decomposed into a field travelling towards the 
system, the `input' field, and one travelling away from it, the 
`output' field. By specifying the external light field at some 
initial time, the input field is determined for all later times. 
The `output' field may be written as the sum of the reflected 
input field and a field radiated by the system. In the final 
section we examine the input-output relations for the specific 
example of a damped optical cavity.

\section{The Lagrangian and Hamiltonian} \label{LandH}
\subsection{The Standard Lagrangian}
The starting point for our discussion is the Lagrangian of Quantum 
Electrodynamics in the Coulomb  gauge. We assume the reader is 
familiar with this description of the interaction of light and 
matter. 
For readers unfamiliar with the quantisation of the 
electro-magnetic field the excellent text by Cohen-Tannoudji, 
Dupont-Roc and Grynberg~\cite{CCT} is highly recommended. 
This 
text is also recommended if the reader is unfamiliar with the 
treatment of Maxwell's equations in Fourier space.
The Lagrangian in question is
\begin{eqnarray}
L & = & \frac{1}{2}\sum_{\alpha}m_{\alpha} \dot{\bf 
r}_{\alpha}^2 + V_{\mbox{\scriptsize Coul}} + \int {\bf j}\cdot {\bf A} \mbox{d}^3r + 
\int \mbox{d}^3r 
{\cal L} \label{Leqs1} \\
{\cal L} & = & \frac{\varepsilon_{0}}{2} \left[ \dot{\bf 
A}^2+c^2({\bf \nabla}\times {\bf 
A})^2 \right] .
\end{eqnarray}
In these equations ${\cal L}$ is the Lagrangian density for the fields, and $\int {\bf j}\cdot {\bf A} \mbox{d}^3r$ describes the interaction between the particles and the fields. The index $\alpha$ labels the particles, where  
${\bf r}_{\alpha}=(r^\alpha_x, r^\alpha_y, r^\alpha_z)$ are the 
particle positions and 
$m_{\alpha}$ the particle masses. The vector potential is denoted 
by ${\bf A}$, and ${\bf j}$ is the current. For a discrete number of 
particles this may be written as ${\bf j}({\bf r}) = 
\sum_{\alpha}{\bf q}_{\alpha}{\bf \dot{ r}}_{\alpha}\delta ({\bf 
r}-{\bf r}_{\alpha})$, where ${\bf q}_{\alpha}$ are the charges of 
the particles. The vector potential is transverse, so that in real 
space it satisfies the condition ${\bf \nabla}\cdot {\bf A}=0$. This 
condition is much easier to understand in Fourier space, which 
will be introduced below. The Lagrangian determines the 
equations of motion of the 
positions of the particles and the values of the fields at each point 
in space via Lagrange's equations~\footnote{We use the functional 
derivative here. If the reader is not familiar with the functional 
derivative, then it is only necessary to note that in this case it means 
differentiate what ever appears under the integral sign in the 
expression for the Lagrangian with respect to the variable in 
question.}:
\begin{eqnarray}
\frac{d}{dt}\left( \frac{\raisebox{0.75ex}{-}\!\!\!\!\partial L} 
{\raisebox{0.75ex}{-}\!\!\!\!\partial \dot{A}_i}\right) 
& = & 
\frac{\raisebox{0.75ex}{-}\!\!\!\!\partial L}{\partial A_i} - 
\sum_{j=x,y,z}\partial_j\frac{\raisebox{0.75ex}{-}\!\!\!\!\partial 
{\cal L}}{\raisebox{0.75ex}{-}\!\!\!\!\partial (\partial_j 
A_i)} 
\\
\frac{d}{dt}\left( \frac{\partial L}{\partial \dot{r}_\alpha}\right) 
& = & 
\frac{\partial L}{\partial r_\alpha} .
\end{eqnarray}
The positions of the particles, and the values of the fields at each 
point in space are the variables governed by the Lagrangian.

If we take the spatial Fourier transform of the fields we obtain 
new field variables which are linear combinations of the old 
variables. The spatial Fourier transform of a field ${\bf A}({\bf 
r})$, along with its inverse, is given by
\begin{equation}
   \bar{\bf A}({\bf k}) = \left( 
\frac{1}{\sqrt{2\pi}} \right)^3 \int {\bf A}({\bf r})e^{-i{\bf 
k}\cdot{\bf 
r}}\;\mbox{d}^3r \;\;\; , \;\; {\bf A}({\bf r}) = \left( 
\frac{1}{\sqrt{2\pi}}\right)^3 \int \bar{\bf 
A}({\bf k})e^{i{\bf k}\cdot{\bf 
r}}\;\mbox{d}^3k .
\end{equation}
We will refer to the fields which are functions of position as `real 
space' fields, and the Fourier transformed fields, which are 
functions of the `Fourier space' variable ${\bf k}$, as ${\bf k}$-space 
fields. We will also use an abuse of notation and drop the bar over the 
${\bf k}$-space field, writing it simply as ${\bf A}({\bf k})$. Note 
that the inverse transform relation gives the real field as a sum over 
plane wave `modes' $e^{i{\bf k}\cdot{\bf r}}$. If we were to 
evolve the field in time in the absence of particles, then the plane 
waves would propagate. The direction of propagation of these 
waves is given by the direction of ${\bf k}$. 

The dynamic equations are simplified by the process 
of taking the Fourier transform. The equations which would 
include spatial derivatives in real space contain only time 
derivatives in Fourier space; the equations are now ordinary 
differential equations. The result is that the field at each point in 
Fourier space is a dynamical variable uncoupled to the field at any 
other point (actually in the presence of particles the field 
variables will be coupled to each other indirectly via the motion of 
the particles). 

We also note that transverse fields are easily characterised in 
${\bf k}$-space. A general vector field has three degrees of 
freedom at each point in ${\bf k}$-space. A transverse field on 
the other hand satisfies ${\bf k}\cdot{\bf A}({\bf k})\equiv 0$. 
That is, 
the field at each point ${\bf k}$ in Fourier space is transverse to 
${\bf k}$. It  has therefore only two degrees of freedom at each 
point.

To obtain the correct Lagrangian for the ${\bf k}$-space 
fields, all we have to do is 
substitute in the Lagrangian Eq.(\ref{Leqs1}) for the real space 
fields in 
terms of the ${\bf k}$-space fields. Note that this is not the same
as simply taking the Fourier transform of the Lagrangian. If the
${\bf k}$-space fields 
were real, then Lagrange's equations (with the real-space fields 
replaced by the ${\bf k}$-space fields) would generate the correct 
equations for these new ${\bf k}$-space variables~\footnote{This 
can be 
seen as follows from the fact that Lagrange's 
equations determine the correct fields to produce a stationary 
point of the action integral. If the action is at a stationary point for 
a particular choice of the real fields, then when we substitute for  
the real fields in terms of the corresponding ${\bf k}$-space fields 
it will remain stationary for that particular choice of the ${\bf 
k}$-space fields. The correct equations for  the ${\bf k}$-space 
variables are therefore those equations which produce a 
stationary point of the action integral with the Lagrangian written 
in terms of these fields. These are of course simply Lagrange's 
equations written using the ${\bf k}$-space fields, which is what 
we wanted to show.}. Actually the situation is complicated by the
fact that 
the new fields are complex, but a thorough derivation shows that 
the correct equations are generated by treating the field and its 
complex conjugate as independent variables~\cite{CCT}. The 
correct
Lagrange equations for the fields are now
\begin{eqnarray}
\frac{d}{dt}\left( \frac{\partial {\cal L}}{\partial \dot{A}_i}\right) 
& = & 
\frac{\partial {\cal L}}{\partial A_i} - 
\sum_{j=x,y,z}\partial_j\frac{\partial {\cal L}}{\partial (\partial_j 
A_i)}
\\
\frac{d}{dt}\left( \frac{\partial {\cal L}}{\partial 
\dot{A}_i^*}\right) & = 
& \frac{\partial {\cal L}}{\partial A_i^*} - 
\sum_{j=x,y,z}\partial_j\frac{\partial {\cal L}}{\partial (\partial_j 
A_i^*)} .
\end{eqnarray}

\subsection{The Input-Output Lagrangian}
\subsubsection{Dropping two Dimensions}
For the purposes of Input-Output theory, we are only interested 
in propagation in two directions, one given direction, and its 
reverse. We therefore wish to consider only the fields along a line 
in Fourier space which passes through the origin. (In fact, later we will reduce the description further to just half the real line.) To do this we choose the
$x$-axis as our direction of propagation, and take a finite normalisation area in the $x$ and $y$ directions. Taking this normalisation area to be $L^2$, where $L$ is the normalisation length for both $y$ and $z$, the Fourier decomposition is now partially discrete and partially continuous:
\bq
  {\bf A}({\bf r}) = \frac{1}{\sqrt{2\pi L^2}}\sum_{n,m}\int {\bf A}_{nm}(k_x) \;
        e^{ik_x r_x + i(2\pi n/L) r_y + i(2\pi m/L) r_z} \; dk_{x}
\eq
with
\bq
 {\bf A}_{nm}(k_x) = \frac{1}{\sqrt{2\pi L^2}} 
         \int_{-L/2}^{+L/2}\int_{-L/2}^{+L/2}\int {\bf A}({\bf r}) \;
         e^{-ik_x r_x - i(2\pi n/L) r_y - i(2\pi m/L) r_z} \;d^3r
\eq
We may ignore the degrees of freedom which do not correspond to propagation along the x-direction so long as the equations of motion do not couple these to those which do correspond to propagation in the x-direction. We have noted above that in the 
absence of particles the field is not coupled between different 
points in Fourier space, so that whether or not we may ignore the 
field variables that are not on the line $n=m=0$ will 
depend upon the coupling of the field to the particles.

Let us first consider the part of the Lagrangian which remains in 
the absence of particles: the Lagrangian for the free field. Written in terms of the Fourier variables this is now
\bq
   L =\varepsilon_{\mbox{\tiny 0}}\sum_{nm}\int\!\!\mbox{d}k_x \left[ 
\dot{\bf 
A}^*_{nm}(k_x)\cdot\dot{\bf 
A}_{nm}(k_x) - c^2k_x^2{\bf A}_{nm}^*(k_x)\cdot{\bf A}_{nm}(k_x)\right]
\eq
As the Lagrangian determines the equations of motion for each of the Fourier variables independently, we may ignore the equations of motion for those for which $n$ or $m$ are not equal to zero by simply dropping them from the Lagrangian. The one-dimensional Lagrangian that results is
\begin{eqnarray}
  L =\varepsilon_{\mbox{\tiny 0}}\int\!\!\mbox{d}k_x \left[ 
\dot{\bf 
A}^*(k_x)\cdot\dot{\bf 
A}(k_x) - c^2k_x^2{\bf A}^*(k_x)\cdot{\bf A}(k_x)\right] ,
\label{lang2}
\end{eqnarray}
where ${\bf A}(k_x) \equiv {\bf A}_{00}(k_x)$. In fact the integral 
need  only extend over half the real line because the field for 
negative $k_x$ is determined entirely by the field for positive 
$k_x$. In particular we have ${\bf A}(k_x) = {\bf A}^*(-k_x)$. This 
is due to the fact that the spatial vector potential is purely real. Also, in general there are two polarisations, so ${\bf A}(k_x) = (0, 
A_y(k_x), A_z(k_x))$. 
However, we will consider only one polarisation, the $y$-component, writing this simply as $A(k)$, where $k\equiv k_x$. From Eq.(\ref{lang2}) the 
Lagrangian determining the equation of motion for $A(k)$ is
\begin{eqnarray}
 L = \varepsilon_{\mbox{\tiny 0}}\int\!\!\mbox{d}k \left[ 
|\dot{A}(k)|^2- 
c^2k^2|A(k)|^2 \right] \label{lang3} .
\end{eqnarray}
Writing $x$ for $r_x$, a notation we will keep from now on, the part of the real-space field that interests us is given by $A(x){\bf e_y}$, where ${\bf e_y}$ is a unit vector in the $y$ direction and 
\bq
  A(x) = \frac{1}{\sqrt{2\pi L^2}}\int A(k) \;
        e^{ik x} \; dk .
\eq

\subsubsection{Dropping Half the Real Line}\label{hsp}
\begin{figure}
\begin{center}
\leavevmode
\epsfxsize=10cm
\epsfbox{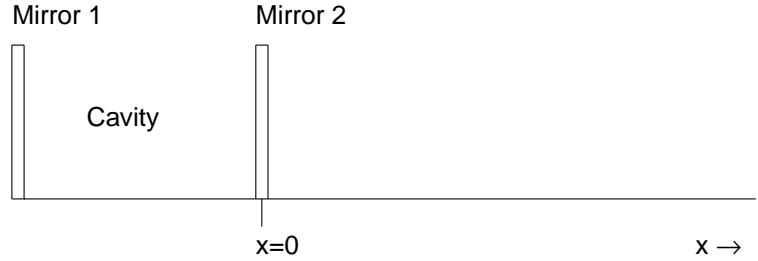}
\caption[A small system interacting with an external field which extends from $x=0$ out to infinity]{A diagram of the small system (in this case a cavity) which interacts with an external field extending from $x=0$ out to infinity. Mirror 1 is totally reflecting so that neither the cavity modes or the external field to the right of the cavity interact with the external field to the left of the cavity. Mirror 2 is partially reflecting, so that there may be an exchange of energy between the cavity and the external field extending to the right.}
\label{iofig1}
\end{center}
\end{figure}

Before adding an interaction Lagrangian we make one further alteration, and that is to treat the real field on only half of the real space only. This makes sense in the case of optical cavities with only one output mirror. Referring to 
Fig.\ref{iofig1}, since mirror 1 is chosen to be completely reflecting, the field to the left of the cavity is coupled neither to the field to the right of the cavity, nor to the field inside the cavity. 

We return briefly to mention the standard treatment of the 
quantum light field. We have indicated above that taking the 
Fourier transform of the field simplifies greatly the equations of 
motion. We may refer to taking the Fourier transform as 
decomposing the field into travelling wave {\em modes}. The 
Fourier fields are not independent for all ${\bf k}$ however, and 
the standard approach is to take linear combinations of the 
Fourier fields $E$ and $B$ (or equivalently $A$ and $\Pi$) to 
obtain a new operator which is independent for all ${\bf k}$. This 
new operator has the advantage that it decouples (diagonalises) the 
equations of motion between $A$ and $\Pi$, and so obeys a very simple 
equation of motion in the case of a free field. The operator in 
question is $a({\bf k})$, which satisfies
\begin{equation}
   [a({\bf k}),a^{\dagger}({\bf k}')] = \delta ({\bf k}-{\bf k}') , 
\label{modecom}
\end{equation}
and is referred to as the mode operator for the plane wave mode 
${\bf k}$. The reader is assumed to be familiar with this standard 
treatment of non-relativistic Quantum Electrodynamics.

Now that we wish to consider only half the real line the Fourier modes we have been using above are not the most convenient. Since a cosine expansion is sufficient to represent the even extension of $A(x,t)$ on the real line, we may 
use instead the transform
\begin{equation}
   A(x,t) = \mbox{\normalsize $\sqrt{\frac{2}{\pi L^2}}$}\int_0^\infty 
\!\!\!\!\!\! \mbox{d}k \; {\cal A}(k,t) \cos (kx) ,
\end{equation}
which we will refer to as the {\em cosine transform}.
The transformed field is real because $A(x,t)$ is real, and is only 
defined for non-negative $k$. Using the standard Fourier 
transform relationship, the inverse transform may be calculated 
as follows:
defining $\bar{\cal A}(k)$ and $\tilde{\cal A}(k)$ by
\begin{equation}
  \tilde{\cal A}(k,t) = \left\{ \begin{array}{rl} {\cal A}(k,t) & k\geq 
0 \\ {\cal A}(-k,t) & k < 0 \end{array} \right. \;\;\; , \;\; \bar{\cal 
A}(k,t) = \left\{ \begin{array}{rl} {\cal A}(k,t) & k\geq 0 \\ 0 & k 
< 0 \end{array} \right. ,
\end{equation}
we may write
\begin{eqnarray}
A(x,t) & = & \mbox{\normalsize $\frac{1}{\sqrt{2\pi L^2}}$}\int_0^\infty 
\!\!\!\!\!\! \mbox{d}k \; {\cal A}(k,t) (e^{ikx}+e^{-ikx}) \\ 
 & = & \mbox{\normalsize $\frac{1}{\sqrt{2\pi L^2}}$}\int_{-\infty}^\infty 
\!\!\!\!\!\! \mbox{d}k \; \bar{\cal A}(k,t) e^{ikx} + \mbox{\normalsize 
$\frac{1}{\sqrt{2\pi L^2}}$}\int_{-\infty}^\infty \!\!\!\!\!\! \mbox{d}k 
\; \bar{\cal A}(-k,t) e^{ikx} \\
 & = & \mbox{\normalsize $\frac{1}{\sqrt{2\pi L^2}}$}\int_{-\infty}^\infty 
\!\!\!\!\!\! \mbox{d}k \; \tilde{\cal A}(k,t) e^{ikx}.
\end{eqnarray}
This is just the standard Fourier transform. Defining 
$\tilde{A}(x,t)$ to be the even extension of $A(x,t)$, we may use 
the inverse Fourier transform relation to obtain
\begin{eqnarray}
\tilde{\cal A}(k,t)  & = & \mbox{\normalsize 
$\frac{1}{\sqrt{2\pi L^2}}$}\int_{-\infty}^\infty \!\!\!\!\!\! \mbox{d}x 
\; \tilde{ A}(x,t) e^{-ikx} = \mbox{\normalsize 
$\frac{1}{\sqrt{2\pi L^2}}$}\int_0^\infty \!\!\!\!\!\!\mbox{d}x 
A(x,t)\cos (kx), 
\end{eqnarray}
giving the inverse cosine transform as
\begin{equation}
{\cal A}(k,t) = \mbox{\normalsize $\sqrt{\frac{2 L^2}{\pi}}$}\int_0^\infty 
\!\!\!\!\!\!\mbox{d}x A(x,t)\cos (kx).
\end{equation}
We shall denote the cosine transform of the canonical momentum 
field $\Pi (x)$ by ${\mbox{\large $\pi$}}(k)$.
The Lagrangian in terms of this cosine variable is the same as given in Eq.(\ref{lang3}), but the integration over $k$ is now only over non-negative $k$.

In choosing to represent the field by a cosine expansion, rather than, for example, a sine expansion, means that we have chosen the boundary condition at $x=0$ such that an incoming wave is reflected without a phase change. This can be seen by noting that in this case the field for positive $x$ is effectively half of a symmetric field extending over all $x$. Therefore, a wave travelling towards $x=0$ from positive $x$ has a twin wave travelling in from negative $x$. As the first wave passes the origin the twin wave also passes the origin, and appears to an observer in the positive $x$ region as a reflection of the original wave with zero phase change. If we had used a sine expansion instead this would have resulted in a $\pi$ phase change upon reflection. However, we may use any phase change we wish, as the correct phase change in any experimental realisation can be correctly accounted for simply by multiplying the output field by the correct phase factor. Because of this, different conventions are used by different authors. Here we follow Gardiner~\cite{Gardiner2} in setting the phase change to zero for the purposes of the derivation. When we consider the input-output relation for an optical cavity in the final section we will return to this point to indicate the various conventions.

\subsubsection{The System-Field Interaction}
We now consider the interaction Lagrangian. We have simplified 
the initial free Lagrangian, in a manner that makes sense for the 
system we are treating. This is 
somewhat phenomenological, since we are assuming that devices 
such as perfect reflecting mirrors exist, without treating their 
interaction with the light field in a detailed manner. In choosing 
an interaction Lagrangian we make a further phenomenological 
departure from the standard electro-magnetic Lagrangian given 
in Eq.(\ref{Leqs1}).  We drop the matter-field coupling given in that 
equation and 
introduce an interaction which will again make sense for optical 
cavities. We assume that there is a small system (perhaps a field 
mode inside an optical cavity) which is coupled solely to the 
degrees 
of freedom of the field that are propagating along our chosen 
direction by the interaction Lagrangian
\begin{eqnarray}
  {\cal O}_{\mbox{\scriptsize s}}\!\!\int_{0}^{\infty}\!\!\!\!\!\!\mbox{d}k\;\mbox{$\kappa$} (k)\dot{{\cal A}}(k) ,
\end{eqnarray}
where ${\cal O}_{\mbox{\scriptsize s}}$ is some operator of the small system.
This interaction was presumably motivated by the coupling of an 
atom with the electro-magnetic field in the electric dipole 
approximation, which leads to spontaneous emission. The 
important feature of this interaction is that it couples a system 
operator to 
many modes of the electric field so that this field becomes a 
reservoir for the system. When we choose a particular operator for ${\cal O}_{\mbox{\scriptsize s}}$ to treat the case of an optical cavity in Section~\ref{Langevin}, it will become clear our choice models the exchange 
of photons between the cavity and the external field. 

The system operator 
${\cal O}_{\mbox{\scriptsize s}}$ is coupled to the electric field 
over a frequency range given the coupling constant 
$\mbox{$\kappa$}(k)$. The frequency 
range (band width) of the modes 
to which the system is coupled is inversely proportional to the 
length over which the system is coupled to the light field in real 
space. We will find it expedient eventually to take the ideal case 
in which $\mbox{$\kappa$}(k)$ is flat, as this 
will simplify the treatment.

The Lagrangian for the system plus reservoir may now be written
\begin{equation}
L = L_{\mbox{\scriptsize sys}}({\bf Z}) +  
\frac{\varepsilon_{\mbox{\tiny 
0}}}{2}\int_{0}^{\infty}\!\!\!\!\!\!\mbox{d}k  \left[ 
\dot{{\cal A}}(k)^2 + c^2k^2{\cal A}(k)^2 \right] + {\cal O}_{\mbox{\scriptsize s}}\!\!\int_{0}^{\infty} 
\!\!\!\!\!\!
\mbox{d}k \;\mbox{$\kappa$} (k)\dot{{\cal A}}(k) ,
\end{equation}
where $L_{\mbox{\scriptsize sys}}({\bf Z})$ is the Lagrangian for 
the system alone.
 
\subsection{The Hamiltonian and Quantisation}
To quantise the field we calculate the Hamiltonian which 
corresponds to the Lagrangian and define the field canonical 
momentum by
\begin{eqnarray}
  \pi (k) = \frac{\raisebox{0.75ex}{-}\!\!\!\!\partial 
L}{\raisebox{0.75ex}{-}\!\!\!\!\partial\dot{{\cal A}}(k)} = 
\varepsilon_{\mbox{\tiny 0}}\dot{{\cal A}}(k) + {\cal O}_{\mbox{\scriptsize s}}\mbox{$\kappa$}(k) .
\end{eqnarray}
The resulting Hamiltonian is given by
\begin{equation}
H = H_{\mbox{\scriptsize sys}}({\bf Z}) + 
\frac{\varepsilon_{\mbox{\tiny 0}}}{2}\int_0^\infty \!\! \left[ 
\frac{1}{\varepsilon_{\mbox{\tiny 0}}^2}({\mbox{\large $\pi$}}(k) 
- {\cal O}_{\mbox{\scriptsize s}}\kappa(k))^2 + c^2k^2{\cal A}^2(k) \right]\mbox{d}k ,
\label{Ham3}
\end{equation}
Note that if we were using the standard Lagrangian (Eq.(\ref{Leqs1})) the canonical 
momentum would be simply the time derivative of the vector 
potential, where as due 
to the interaction we have chosen, the canonical momentum 
contains the extra term ${\cal O}_{\mbox{\scriptsize s}}\mbox{$\kappa$}(k)$. Note also 
that 
for our 
treatment to be valid, we require the interaction we have chosen 
to be a good approximation to the true Lagrangian, ie. for the 
system we are treating, the true, and exceedingly complex, 
Lagrangian should approximately reduce to the one we are using. 
If this is the case, then the conjugate momentum should also be a 
good approximation to the truth, ie. it should be approximately 
the correct thing to impose the commutation relations upon.

The equations of motion for ${\cal A}(k)$ may now be derived either by 
using Lagrange's equations, or by using Hamilton's equations, 
given by
\begin{equation}
\dot{{\cal A}} = \frac{\raisebox{0.75ex}{-}\!\!\!\!\partial 
H}{\raisebox{0.75ex}{-}\!\!\!\!\partial\pi} \;\;\; , \;\; 
\dot{\pi} = -\frac{\raisebox{0.75ex}{-}\!\!\!\!\partial 
H}{\raisebox{0.75ex}{-}\!\!\!\!\partial {\cal A}} .
\end{equation}
Quantisation of the field is accomplished by imposing 
commutation 
relations between the field and its canonically conjugate 
momentum 
such that the equations of motion may be written
\begin{equation}
\dot{{\cal A}} = \frac{-i}{\hbar}[{\cal A},H] \;\;\; , \;\; \dot{\pi} = 
\frac{-i}{\hbar}[\pi,H] .
\end{equation}
That is, the commutators must be chosen to satisfy the 
relations
\begin{equation}
[{\cal A},F({\cal A},\pi)] =  i\hbar \frac{\raisebox{0.75ex}{-}\!\!\!\!\partial 
}{\raisebox{0.75ex}{-}\!\!\!\!\partial\pi}F({\cal A},\pi) \;\;\; , 
\;\; [\pi,F({\cal A},\pi)] = -i\hbar \frac{\raisebox{0.75ex}{-}
\!\!\!\!\partial }{\raisebox{0.75ex}{-}\!\!\!\!\partial {\cal A}}F({\cal A},\pi) ,
\end{equation}
where $F({\cal A},\pi)$ is an arbitrary function of ${\cal A}$ and $\pi$.
The commutator which satisfies the above criteria is
\begin{equation}
[{\cal A}(k,t),\pi(k',t)] = i\hbar\delta(k-k') .
\end{equation}

\subsubsection{Equations of Motion}
There are now three ways to calculate the equations of motion: using 
Lagrange's equations, using Hamilton's equations, or calculating commutators.
It is the third (calculating commutators) which is most often 
used in quantum mechanics. Performing this calculation, the equations of motion for the fields are
\begin{eqnarray}
  \dot{{\cal A}}(k,t) & = & \frac{1}{\varepsilon_{\mbox{\tiny 
0}}}\mbox{\large $\pi$}(k,t) - \frac{1}{\varepsilon_{\mbox{\tiny 
0}}}{\cal O}_{\mbox{\scriptsize s}}(t)\kappa (k)\\ 
  \dot{{\mbox{\large $\pi$}}}(k,t) & = & \varepsilon_{\mbox{\tiny 
0}}\omega^2 {\cal A}(k,t)
\end{eqnarray}
In order to obtain mode operators for the mode functions $\cos 
(kx)$ we need to take a linear combination of the fields ${\cal 
A}(k,t)$ and ${\mbox{\large $\pi$}}(k,t)$ that both decouples this set 
of differential equations, and satisfies delta commutation 
relations.
The correct linear combination is
\begin{equation}
   b(k,t) = \sqrt{\frac{\varepsilon_{\mbox{\tiny 
0}}\omega}{2\hbar}}{\cal A}(k,t) + i 
\frac{1}{\sqrt{2\varepsilon_{\mbox{\tiny 
0}}\hbar\omega}}{\mbox{\large $\pi$}}(k,t) \;\;\; , \;\; \omega = 
k/c ,
\end{equation}
where the dependence on $\varepsilon_{\mbox{\tiny 0}}\omega$ 
is required to decouple the differential equations, and the scaling 
by $\sqrt{2\hbar}$ is required to obtain the correct commutation 
relations. The differential equation governing the mode 
operators is
\begin{equation}
  \dot{b}(k,t) = -i\omega b(k,t) -  
\sqrt{\frac{\omega}{2\varepsilon_{\mbox{\tiny 
0}}\hbar}}{\cal O}_{\mbox{\scriptsize s}}(t)\kappa(k) .
\end{equation}
Since this is a first order differential equation, given a choice for 
$b(k)$ at a particular time $t_0$, the solution is determined for all 
$t$ both prior to, and later than $t_0$. In particular, the solution 
for the above equation is
\begin{equation}
  b(k,t) = b(k,t_0)e^{-i\omega(t-t_0)} -  
\sqrt{\frac{\omega}{2\varepsilon_{\mbox{\tiny 
0}}\hbar}}\,\kappa(k)\!\!\int^t_{t_0}\!\!\!\! e^{-i\omega(t-
\tau)}{\cal O}_{\mbox{\scriptsize s}}(\tau) \; \mbox{d}\tau . \label{fullev}
\end{equation}
In the absence of the system (ie. ${\cal O}_{\mbox{\scriptsize s}}=0$) the field mode operators 
evolve in the simple manner
\begin{equation}
   b(k,t) = b(k,t_0)e^{-i\omega(t-t_0)} ,
\end{equation}
which gives the evolution of the free field. Writing the vector 
potential in terms of the mode operators we have
\begin{eqnarray}
  A(x,t) & = & \int_0^\infty \!\!\!\!\!\! 
\sqrt{\frac{\hbar}{\pi\varepsilon_0\omega L^2}} \left[ b(k,t)\cos 
(kx) + b^\dagger(k,t)\cos (kx) \right]\; \mbox{d}k \\
  & = & \int_0^\infty \!\!\!\!\!\!  
\sqrt{\frac{\hbar/4}{\pi\varepsilon_{\mbox{\tiny 
0}}\omega L^2}}\left[ b(k,t)(e^{ikx}+e^{-ikx}) + \mbox{h.c.} \right]\; 
\mbox{d}k .
\end{eqnarray}
In the absence of the system the solution for the vector potential 
may be written in terms of the mode operators at some arbitrary 
time $t_0$ as
\begin{equation}
 A(x,t) = \int_0^\infty \!\!\!\!\!\!
\sqrt{\frac{\hbar/4}{\pi\varepsilon_{\mbox{\tiny 0}}\omega L^2}} 
\left[ b(k,t_0)(e^{i\left( kx -\omega (t-t_0)\right)}+e^{-i\left( 
kx+\omega (t-t_0)\right)}) + \mbox{h.c.}\right]\; \mbox{d}k 
\label{freefield}.
\end{equation}
Since $e^{i\left( kx \pm \omega (t-t_0)\right)}$ is a wave 
travelling 
in the positive (negative) $x$ direction, we see that each mode 
operator $b(k)$ is associated with {\em two} waves of frequency 
$\omega$ travelling in opposite directions. 

\section{Input and Output Fields}\label{InOut}
Let us first consider the free field which is given by 
Eq.(\ref{freefield}). At $x=0$ this field is
\begin{equation}
 A_{\mbox{\scriptsize free}}(0,t) = \int_0^\infty \!\!\!\!\!\!
\sqrt{\frac{\hbar}{\pi\varepsilon_{\mbox{\tiny 0}}\omega L^2}}
\left[ b(k,t_0)e^{-i\omega\left( t-t_0\right)} + \mbox{h.c.}\right]\; 
\mbox{d}k .
\end{equation}
By a simple rearrangement of $A(x,t)$ it is seen that
\begin{equation}
 A_{\mbox{\scriptsize free}}(x,t) = A(0,t-\mbox{\scriptsize 
$\frac{x}{c}$}) + 
A(0,t+\mbox{\scriptsize $\frac{x}{c}$}) .
\end{equation}
The total field can therefore be written as the sum of an inward 
travelling field and an outward travelling field. These inward and 
outward propagating fields may be written, as we have done, in 
terms of the field at $x=0$. That is, if we know the field at $x=0$ 
for all time, then we may construct the entire field for all time. 
This may be understood as follows. If we know the field at $x=0$, 
then this is the result of a field which has propagated along the 
$x$-line from positive infinity to $x=0$. Also, now that the 
field has this value at $x=0$, this will propagate in the positive 
$x$ direction back out to infinity. The field at a particular position 
$x$ 
and time $t$ is therefore the sum of a field which, at a time $x/c$ 
later will reach $x=0$, and is therefore $A(0,t+\mbox{\scriptsize 
$\frac{x}{c}$})$, and another field which will have come from 
$x=0$ and taken a time $x/c$ to reach position $x$, and so is 
therefore $A(0,t-\mbox{\scriptsize $\frac{x}{c}$})$.

Let us denote the free field at $x=0$ by $A_{\mbox{\scriptsize 0}}(t)$.
With no system interaction, this operator would be equal to the actual field,
$A(0,t)$, for all time. However, once we include the system interaction the 
evolution of the field is no longer given by multiplying the mode operators by an exponential, but is instead given by Eq.(\ref{fullev}). This means that $A_{\mbox{\scriptsize 0}}(t)$ is only equal to the field at $x=0$ when $t=t_0$, and will in general be different from it for all other times.
 
Writing $A(x,t)$ in terms of the mode operators at $t=t_0$, but 
using the full evolution of the field in the presence of the system, it is possible to write, after some careful manipulation~\cite{Gardiner2}
\begin{equation}
  A(x,t) = A_{\mbox{\scriptsize 0}}(t+\mbox{\scriptsize 
$\frac{x}{c}$}) + A_{\mbox{\scriptsize 0}}(t-\mbox{\scriptsize 
$\frac{x}{c}$}) - 1/(2c) \!\!\!\!\!\!\!\!\! \int\limits_{x-c(t-
t_0)}^{x+c(t-t_0)} \!\!\!\!\!\!\!\!\!
\mbox{\scriptsize $K$}(x'){\cal O}_{\mbox{\scriptsize s}}\!\!\left( t-\left|\mbox{\scriptsize 
$\frac{x'-x}{c}$}\right| \right)\;\mbox{d}x', \label{atot}
\end{equation}
where $\mbox{\small $K$}(x)$ is the inverse cosine transform of $\kappa(k)$.
We will now examine closely the last term in this expression, 
which is the contribution of the system to the field. Recall first of 
all that we take the interaction of the system with the field to be 
non-zero only over some small range close to $x=0$. In particular 
let us choose $\mbox{\scriptsize $K$}(x)$ to be zero outside the 
interval $[0,\Delta x]$. Now let us examine the third term for 
values of $x$ outside the interaction region. In this case $x'$ is 
always less than $x$ in the integral, so that we may write the 
system contribution as
\begin{equation}
A_{\mbox{\scriptsize sys}}(x,t) = - 1/(2c) \!\!\!\!\!\!\!\!\!
\int\limits_{x-c(t-t_0)}^{x+c(t-t_0)} \!\!\!\!\!\!\!\!\!
\mbox{\scriptsize $K$}(x'){\cal O}_{\mbox{\scriptsize s}}\!\!\left( t-\mbox{\scriptsize $\frac{x-
x'}{c}$} \right)\;\mbox{d}x'. \label{asys1}
\end{equation}
Taking $t_0$ to be far enough in the past so that the interval $[x-
c(t-t_0),x+c(t-t_0)]$ covers completely the interval $[0,\Delta x]$, 
we may simplify the above expression further to obtain
\begin{equation}
A_{\mbox{\scriptsize sys}}(x,t) = - \frac{1}{2c} 
\int_{0}^{\Delta x}\!\!\!\!\!\! \mbox{\scriptsize $K$}(x'){\cal O}_{\mbox{\scriptsize s}}\!\!\left( 
t-\mbox{\scriptsize $\frac{x-x'}{c}$} \right)\;\mbox{d}x'.
\end{equation}
The system contribution to the field at position $x$ and time $t$ is 
clearly an integral over the range over which the system operator 
${\cal O}_{\mbox{\scriptsize s}}$ 
interacts with the system. The contribution from each point of the 
interaction being the value that the system operator had at a time 
which is earlier than the current time $t$ by the amount required 
for the field to propagate from that point to $x$. Outside the 
interaction region the field generated by the system is 
propagating in the positive $x$ direction, which is clear since we 
may write $A_{\mbox{\scriptsize sys}}$ as a function of $t-
\mbox{\scriptsize $\frac{x}{c}$}$. Returning to Eq.(\ref{asys1}), it 
is clear that the lower bound on the integration, namely $x-c(t-
t_0)$, is there because the contribution from points which are 
further away from $x$  than $c(t-t_0)$ should not be included, as 
the field will not have had time to propagate to $x$ in the time 
interval $t-t_0$.

What Eq.(\ref{atot}) tells us is that if we choose the state of the 
field at $t=t_0$, then for times {\em later} than $t_0$, the field is 
what it would have been from free evolution (given by calculating 
$A_{\mbox{\scriptsize 0}}=A(0,t)$ using free evolution and 
writing 
$A(x,t) = A_{\mbox{\scriptsize 0}}(t+\mbox{\scriptsize 
$\frac{x}{c}$}) + A_{\mbox{\scriptsize 0}}(t-\mbox{\scriptsize 
$\frac{x}{c}$})$), plus a term generated by the system, which is 
propagating outwards. This is a 
fairly intuitive result, given that Maxwell's equations generate 
propagating solutions.

What we have said so far is, in a sense, all we need say about the 
input and output fields. The in-field may be specified 
independently of the system, by giving an initial condition for the 
full external field. The out-field is then given by the in-field 
plus a system contribution.

Note that as the in- and out-fields are propagating, we
may sit at any distance away from the system to observe them.
Changing the position of observation merely effectively shifts
the origin of the time axis for the observations. As this is the
case, we may evaluate the input and output fields at any given
position and call this the time varying input to and output from
the system. For convenience we will choose the position to be 
$x=0$. Naturally any real observation would be at some finite 
distance away from the interaction region.

To obtain the central result of this section we need merely modify 
Eq.(\ref{atot}) by taking the limit as $t_0\rightarrow -\infty$ and 
setting $\mbox{\scriptsize $K$}(x)= \mbox{\scriptsize 
$K$}\delta(x)$. After identifying $A_{\mbox{\scriptsize 0}}(t)$ as 
the input to the system, (the inward travelling field evaluated at 
$x=0$) and 
the other two terms (also evaluated at $x=0$) as the output from 
the system, this gives us 
\begin{equation}
A_{\mbox{\scriptsize out}}(t) = A_{\mbox{\scriptsize in}}(t) - 
\frac{\mbox{\scriptsize $K$}}{2c}{\cal O}_{\mbox{\scriptsize s}}(t) .
\end{equation}
This allows us to calculate the
properties of the output field in terms of the input field and the 
system evolution. 


At this point we have essentially completed our task in that we have 
derived the input-output relation. However, in 
order to gain an understanding of the way in which the standard 
treatment of Input-Output theory links together the output field 
with a {\em final condition} for the external field (as opposed to an 
{\em initial condition}) we will finish with a discussion 
of this point.

To do this we will now examine the field {\em prior} to $t_0$. Of 
course the solution we have written down above is valid for $t$ earlier 
that $t_0$ as well as $t$ later than $t_0$, but we will shortly 
write it in a more intuitive form for the former case.  Note first
that the field at 
$t=t_0$ contains nothing of the field emitted by the system at 
times later than $t_0$, which is reflected in the fact that the 
solution for later times contains the outward propagating system 
contribution {\em in addition} to the freely evolved terms. 
However, if we wish to look at the solution for the field at an 
earlier time $t$, we should note that the field at $t_0$ {\em does} 
contain the contribution radiated by the system for the time 
interval $t_0-t$. 

Recall that when we use only free evolution, the out-going field is 
simply a reflection of the in-going field.  That is, the in-going and 
out-going fields are the same.  Since, as we have mentioned above, 
$A_0(t)$ contains the field emitted by the system (up until $t_0$),
the in-field that we calculate using free evolution also 
contains this field up until $t_0$, but travelling inwards. In the 
full solution, 
however, the in-field will obviously not have this extra inward 
travelling system contribution. The full solution will therefore be 
the sum of the terms due to free evolution and an inward 
travelling term which exactly cancels the extra inward travelling 
term contained in the free solution. Indeed, with a little 
re-arrangement the full solution, given by Eq.(\ref{atot}), 
may be written~\cite{Gardiner2}
\begin{equation}
  A(x,t) = A_{\mbox{\scriptsize 0}}(t+\mbox{\scriptsize 
$\frac{x}{c}$}) + A_{\mbox{\scriptsize 0}}(t-\mbox{\scriptsize 
$\frac{x}{c}$}) + 1/(2c) \!\!\!\!\!\!\!\!\! \int\limits_{x+c(t-
t_0)}^{x-c(t-t_0)} \!\!\!\!\!\!\!\!\!
\mbox{\scriptsize $K$}(x'){\cal O}_{\mbox{\scriptsize s}}\!\!\left( t+\left|\mbox{\scriptsize 
$\frac{x'-x}{c}$}\right| \right)\;\mbox{d}x'. \label{atot2}
\end{equation} 
Note that the last term of this equation is inward propagating. 
Eq.(\ref{atot}) is the intuitive expression to use for the field at 
times later than $t_0$, and Eq.(\ref{atot2}) is the 
intuitive expression to use for the field at times earlier than $t_0$.

To sum up, if we specify a state for the external field at some 
time ($t_1$, say) after the time at which we wish to consider the 
system, then we should note that the out-field given by free 
evolution contains the system contribution up until time $t_1$ 
(and is therefore the correct output field up until that time). 
Alternatively, if we specify the state of the field at a time ($t_0$, 
say) earlier than the time at which we wish to consider the 
system, we should note that the in-field is the correct input field 
for all times later than $t_0$. This prompts the following 
definitions:
\begin{eqnarray}
 A_{\mbox{\scriptsize in}}(t) & \equiv & \int_0^\infty \!\!\!\!\!\! 
\sqrt{\frac{\hbar}{\pi\varepsilon_{\mbox{\tiny 0}}\omega L^2}}
\left[ b(k,t_0)e^{-i\omega\left( t-t_0\right)} + \mbox{h.c.}\right]\; 
\mbox{d}k = A_{\mbox{\scriptsize 0}}(t) \label{infield} , \\
A_{\mbox{\scriptsize out}}(t) & \equiv & \int_0^\infty \!\!\!\!\!\! 
\sqrt{\frac{\hbar}{\pi\varepsilon_{\mbox{\tiny 0}}\omega L^2}}
\left[ b(k,t_1)e^{-i\omega\left( t-t_1\right)} + \mbox{h.c.}\right]\; 
\mbox{d}k , \label{outfield}
\end{eqnarray}
where $t_0$ is a time in the past, and $t_1$ is a time in the 
future. Note that $A_{\mbox{\scriptsize out}}$ is simply
$A_{\mbox{\scriptsize in}}$ with $t_0$ replaced with $t_1$. From
the above discussion we can conclude that 
$A_{\mbox{\scriptsize out}}(t-x/c)$ is the correct output field (up 
until time $t_1$), and $A_{\mbox{\scriptsize in}}(t+x/c)$ is the 
correct input field (for times after $t_0$).

By equating Eqs~(\ref{atot}) and (\ref{atot2}) (with 
$A_{\mbox{\scriptsize 0}}(t)$ replaced by $A_{\mbox{\scriptsize 
in}}(t)$ in Eq.(\ref{atot}), and replaced by $A_{\mbox{\scriptsize 
out}}(t)$ in Eq.(\ref{atot2})) we find, as we expect, that for $x$ 
outside the interaction region
\begin{equation}
A_{\mbox{\scriptsize out}}(t+x/c) = A_{\mbox{\scriptsize in}} 
(t+x/c) -\frac{1}{2c}\int_{-\infty}^\infty \!\!\!\! \mbox{d}
\mbox{\scriptsize $K$}(x'){\cal O}_{\mbox{\scriptsize s}}\!\!\left( t+\left|\mbox{\scriptsize 
$\frac{x-x'}{c}$}\right| \right)\;\mbox{d}x' ,
\end{equation}
where we have taken the limit $t_0\rightarrow - \infty$ and 
$t_1\rightarrow \infty$. Since this is true for any $t$, even 
though $x$ must be outside the interaction region, we can see that
\begin{equation}
  A_{\mbox{\scriptsize out}}(t) = A_{\mbox{\scriptsize in}} (t) -
\frac{1}{2c}\int_{-\infty}^\infty \!\!\!\! \mbox{d}
\mbox{\scriptsize $K$}(x'){\cal O}_{\mbox{\scriptsize s}}\!\!\left( t-\left|\mbox{\scriptsize 
$\frac{x'}{c}$}\right| \right)\;\mbox{d}x' .
\end{equation}
Now $A_{\mbox{\scriptsize in}}(t)$ is the input field at $x=0$, and 
$A_{\mbox{\scriptsize out}}(t)$ is the output field at $x=0$. 
Setting $\mbox{\scriptsize 
$K$}(x)= \mbox{\scriptsize $K$}\delta(x)$ we have
\begin{equation}
   A_{\mbox{\scriptsize out}}(t) = A_{\mbox{\scriptsize in}}(t) - 
\frac{\mbox{\scriptsize $K$}}{2c}{\cal O}_{\mbox{\scriptsize s}}(t)
\end{equation}
In addition, using the previous result and Eq.(\ref{atot}) (with 
$A_{\mbox{\scriptsize 0}}(t)$ replaced by $A_{\mbox{\scriptsize 
in}}(t)$), we find, again as we expect, that
\begin{equation}
 A(x,t) = A_{\mbox{\scriptsize in}}(t+x/c) + A_{\mbox{\scriptsize 
out}}(t-x/c) .
\end{equation}
That is, the total field is the sum of the input field propagating in, 
and the output field (which includes the system contribution) 
propagating out.

\section{Input-Output Relations for a Damped Cavity}
\label{Langevin}
In the previous section we examined the input-output relations for a general open system. To obtain the input-output relation for a lossy optical cavity in their final and most useful form, it is necessary to make the rotating wave approximation. This requires that the frequency of the cavity mode is much larger than the rate at which the system changes due to any other dynamics, such as the interaction with the external field. 

To obtain the correct equations for a damped cavity, we choose the coupling operator to be a quadrature of the cavity mode, which gives dynamics linear in the mode and the field operators. Any quadrature operator will do, as rotating this quadrature simply changes the phase relationship between the input field and the cavity mode. We choose ${\cal O}_s = p = -i(a-a^\dagger)$ as this gives the cavity mode the same phase as the input field. This would be true, for example, if the input mirror was a simple dielectric boundary. At this point we make the rotating wave approximation, which involves moving into the interaction picture and dropping terms from the Hamiltonian (and therefore from the equations of motion) which are oscillating (rotating) very rapidly compared to the remaining dynamics (due to the interaction). These may be dropped because their contribution averages to zero in the time-scale at which the dynamical variables are changing. Thus the terms $ab(k)$ and $a^\dagger b^\dagger (k)$, which oscillate in the interaction picture at frequencies greater than $\omega_0$, are removed from the interaction Hamiltonian. As the remaining terms, $ab^\dagger (k)$ and $a^\dagger b(k)$ will only contribute significantly to the dynamics when the difference between $\omega=ck$ and $\omega_0$ is much smaller than $\omega_0$, again due to this averaging, we perform the further approximation of setting $\omega=\omega_0$ in the integral over the field mode operators, so that it may be removed from the integral. This may be though of as a narrow bandwidth approximation, and assumes that we only need consider frequencies in a bandwidth about $\omega_0$ which is much smaller than $\omega_0$. Before obtaining the final form of the Hamiltonian we make one further approximation which is to extend the integral over the mode wave-number $k$ to negative infinity. While this is physically absurd, the extra terms are once again rapidly oscillating, and contribute negligibly to the dynamics. The resulting Hamiltonian is
\begin{equation}
   H = \hbar\omega_0 a^\dagger a - 
i\sqrt{\frac{\hbar\omega_0}{2\varepsilon_0 L^2}} {\mbox{\small $K$}}\!\! \int_{-
\infty}^{\infty}\!\!\!\!\!\! (b^\dagger(k)a - a^\dagger 
b(k))\;\mbox{d}k + \frac{\varepsilon_{\mbox{\tiny 
0}}}{2}\int_0^\infty \!\! \left[ \frac{1}{\varepsilon_{\mbox{\tiny 
0}}^2}{\mbox{\large $\pi$}}^2(k) + c^2k^2{\cal A}^2(k) 
\right]\mbox{d}k .
\end{equation}
In fact there is another term which comes from the interaction Hamiltonian, but it consists only of system operators, and results merely in a shift of the frequency of the cavity mode. This may always be taken into account by redefining $\omega_0$. The equations of motion for the system and reservoir operators are now derived easily by taking commutators with this Hamiltonian. We have
\begin{eqnarray}
   \dot{a} & = & -i\omega_0 a + 
\sqrt{\frac{\omega_0}{2\varepsilon_0 
\hbar L^2}} \mbox{\small $K$}\!\! \int_{-\infty}^{\infty}\!\!\!\!\!\! 
b(k)\;\mbox{d}k \\
   b(k,t) & = & b(k,t_0)e^{-i\omega(t-t_0) } -  
\sqrt{\frac{\omega_0}{2\varepsilon_0 
\hbar L^2}} \mbox{\small $K$}\!\! \int_{t_0}^t \!\!\!\! e^{-
i\omega(t-t')}a(t')\;\mbox{d}t' ,
\end{eqnarray}
where we have written the solution for the reservoir operators in 
terms of some initial conditions and the system operator. 
Substituting this solution in the equation of motion for the system 
operator, we have
\begin{equation}
   \dot{a}  =  -i\omega_0a - 
\frac{\gamma}{2}a + \sqrt{\gamma}b_{\mbox{\scriptsize in}}(t) 
\end{equation}
where
\begin{equation}
b_{\mbox{\scriptsize in}}(t) = 
\frac{1}{\sqrt{2\pi}}\!\! \int_{-\infty}^{\infty}\!\!\!\!\!\! 
b(k,t_0)e^{-i\omega(t-t_0)}\;\mbox{d}k  \;\; , \;\;\; \gamma = 
\frac{\pi\omega_0}{\varepsilon_0\hbar L^2}\mbox{\small $K$}^2  ,
\end{equation}
which is the negative frequency part of the input vector potential. In particular, continuing to make the narrow bandwidth approximation, we have
\begin{equation}
A_{\mbox{\scriptsize in}}(t) = \sqrt{\frac{\hbar 
}{2\varepsilon_0 \omega_0L^2}}
(b_{\mbox{\scriptsize in}}(t) + b_{\mbox{\scriptsize in}}^\dagger(t) 
) .
\end{equation}
It is clear now that the equation of motion for the cavity mode 
annihilation operator has two terms due to the interaction with the external 
field. The first of these is a damping term, and the second is a 
driving term, being the negative frequency part of the input field.

Defining $b_{out}(t)$ in a similar manner to $b_{in}$, that is, as the negative frequency part of the output vector potential, and using the same scaling, we find that the input-output relations become
\bq
  b_{\ms{out}}(t) = b_{\ms{in}}(t) - \sqrt{\gamma}a(t) .
\eq
Note that to obtain this relation we have used the `convention' that there is no phase change upon reflection at $x=0$ (see subsection~\ref{hsp}). To change this convention, all we need do is to multiply the output field by a phase factor. In particular, if we want a $\pi$ phase change on reflection, we multiply the output field by minus one, which is why one will often see the relations written
\bq
  b_{\ms{out}}(t) = \sqrt{\gamma}a(t) - b_{\ms{in}}(t) .
\eq
It is this latter convention that we use for our analysis in Chapter~\ref{qnc}.

We now wish finally to make a connection with photo-detection theory to see the physical significance of $b_{in}(t)$ and $b_{out}(t)$. Glauber's theory of photo-detection gives the rate of photon detections for a detector placed in the field $E(x,t)$ as proportional to $\langle E^+(x,t)E^-(x,t)\rangle$, where $E^+(x,t)$ is the positive frequency part of the Electric field operator~\cite{Saleh}. We are interested here in detecting the light output from the cavity, so the photon detection rate is given by $\langle E^+_{\ms{out}}(x,t)E^-_{\ms{out}}(x,t) \rangle$. Using $E_{\ms{out}}(x,t) = -\partial_{t} A_{\ms{out}}(x,t)$ we have 
\bq
  E_{\ms{out}}(0,t) =  \sqrt{\frac{\hbar \omega_0
}{2\varepsilon_0 L^2}} (ib_{\ms{out}}(t) - ib_{\ms{out}}^\dagger(t)) .
\eq
The rate of photo-detection is therefore proportional to $\langle 
b_{\ms{out}}(t)b^{\dg}_{\ms{out}}(t) \rangle$, and we just need to determine 
the constant of proportionality. Since $\gamma$ is the rate at which photons 
leave the cavity, for an detector of unit efficiency, and for a vacuum input 
field, the average rate of photo-detection must be $\gamma\langle 
b^{\dg}(t)b(t)\rangle$, being the average number of photons in the cavity field 
at time $t$. Calculating $\langle b^{\dg}_{\ms{out}}(t)b_{\ms{out}}(t) \rangle$ 
gives exactly that result, and hence the constant of proportionality is unity. 
We see now that $b_{\ms{out}}(t)$ is simply the positive frequency part of the 
electric field scaled such that the average photo-counting rate is $\langle 
b^{\dg}_{\ms{out}}(t)b_{\ms{out}}(t) \rangle$. The operators 
$b^{\dg}_{\ms{out}}(t)b_{\ms{out}}(t)$ and 
$b^{\dg}_{\ms{in}}(t)b_{\ms{in}}(t)$ may therefore be interpreted as photon 
flux operators, so that $\langle b^{\dg}_{\ms{in}}(t)b_{\ms{in}}(t)\rangle$ is the rate at which photons are incident on the cavity from the input field. This concludes our discussion of the input-output relations.

\chapter[Exponentials of $P$ and $Q$]{The Action of Exponentials Linear and Quadratic in $P$ and $Q$}
\label{apB}

We first calculate the effect of an operator of the form
\begin{equation}
 e^{\nu P + \mu Q}
\end{equation}
on a coherent state $|\alpha\rangle$. The coherent state is defined as 
the eigenstate of the annihilation operator $a$, such that
\begin{equation}
   a|\alpha\rangle = \alpha |\alpha\rangle ,
\end{equation}
and
\begin{equation}
   a = \sqrt{\frac{m\omega}{2\hbar}} \; x + i\sqrt{\frac{1}{2\hbar 
m\omega}} 
\; p.
\end{equation}
Here $m$ and $\omega$ are the mass and frequency of a harmonic 
oscillator 
which serves for the purposes of defining the coherent state.
 In particular we are interested in the position wave-function of 
the result. 
We therefore wish to calculate
\begin{equation}
 \langle x|\psi\rangle = \langle x|e^{\nu P + \mu Q}|\alpha\rangle ,
\end{equation}
where $|x\rangle$ is an eigenstate of the position operator $Q$ such 
that $Q|x\rangle = x |x\rangle$. Note that in general $|\psi\rangle$ will 
not be normalised. To perform this 
calculation we will need the BCH formula given in Eq.(\ref{BCH}), 
and the 
position wavefunction for a coherent state,
\bq
 \langle x|\alpha\rangle = \left( \frac{2s^2}{\pi} \right) ^{1/4} 
e^{-s^2x^2+2sx\alpha -\frac{1}{2}(|\alpha|^2 + \alpha^2)}
                         = \left( \frac{2s^2}{\pi} \right) ^{1/4} 
e^{-s^2x^2+2sx\alpha - \alpha_r^2 - i\alpha_r\alpha_i}
\eq
where
\begin{eqnarray}
 s & = & \sqrt{\frac{m\omega}{2\hbar}} \\
 \alpha & = & \alpha_r + i\alpha_i .
\end{eqnarray}
Note that this expression contains the phase factor $- i\alpha_r\alpha_i$. 
This is left out in many texts, but is essential for consistency with the 
completeness relations for the position states. We also require the inner 
product of two coherent states,
\begin{equation}
 \langle \alpha |\beta\rangle = e^{-\frac{1}{2}(|\alpha|^2 + |\beta|^2) + 
\alpha^* \beta},
\end{equation}
and the well known integral formula
\begin{equation}
  \int e^{-\alpha x^2 -\beta x} \;  d x = \sqrt{\frac{\pi}{\alpha}} 
e^{\beta^2/(4\alpha)} \;\; , \;\; \mbox{Re}[\alpha] > 0.
\end{equation}
We proceed first by rewriting the exponential in terms of annihilation and 
creation operators, so that we have
\begin{equation}
  e^{\nu P + \mu Q} = e^{\theta a + \phi a^\dagger} = e^{\phi a^\dagger} 
e^{\theta a}e^{\theta\phi/2}
\end{equation}
in which
\bq
 \theta = \left( \nu\sqrt{\frac{\hbar}{2m\omega}} - 
i\mu\sqrt{\frac{m\hbar\omega}{2}} \right) = \phi^* 
\eq
We may now use the completeness relation for the coherent states to obtain
\begin{eqnarray}
 \langle x|\psi\rangle = \langle x|e^{\phi a^\dagger} 
e^{\theta a}|\alpha\rangle e^{\theta\phi/2} & = & \frac{1}{\pi} \int \!\!\!\! \int \langle x|\beta\rangle  
\langle \beta|e^{\phi a^\dagger} e^{\theta a}|\alpha\rangle 
e^{\theta\phi/2} \; d ^2\beta \nonumber \\
 & = & \frac{1}{\pi} \int \!\!\!\! \int \langle x|\beta\rangle  
\langle \beta|\alpha \rangle \; e^{\theta \alpha + \phi \beta^* + 
\theta\phi/2} \; d ^2\beta \nonumber \\
 & = &   \langle x|\alpha + \phi\rangle e^{\frac{1}{2}|\phi|^2+
\mbox{\scriptsize Re}[\alpha\phi^*] + \theta\alpha + \theta\phi/2}.
\end{eqnarray}
We see that the state remains coherent, although it is no longer 
normalised, and is shifted in phase space by $\phi$.

We now wish to calculate the effect of an operator of the form
\begin{equation}
 e^{\eta P^2 + \zeta Q^2 + \xi QP}
\end{equation}
on a coherent state. This time we require to calculate
\begin{equation}
 \langle x|\psi\rangle = \langle x|e^{\eta P^2 + \zeta Q^2 + 
\xi QP}|\alpha\rangle .
\end{equation}
For this calculation we will need the disentangling theorem for 
the exponential of a general quadratic form of the annihilation and 
creation operators, which is given by~\cite{mjcollett}
\begin{equation}
 e^{ua^2 + va^{\dagger 2} + wa^\dagger a} = e^{(w+\chi)/2}e^{la^{\dagger 
2}}e^{\chi a^\dagger a}e^{ma^2},
\end{equation}
in which
\begin{eqnarray}
 l & = & \frac{u}{f\coth(f) - w} \\
 \chi & = & \ln \left( \frac{f}{f\coth(f) - w\sinh(f)} \right)   \\
 m & = & \frac{u}{f\coth(f) - w}   \\
 f & = &  \sqrt(w^2 - 4uv) .
\end{eqnarray} 
First of all rewriting the exponential containing $P$ and $Q$ as an 
exponential in the annihilation and creation operators, we have
\begin{equation}
  \langle x|e^{\eta P^2 + \zeta Q^2 + \xi QP}|\alpha\rangle = 
\langle x|e^{ua^2 + va^{\dagger 2} + wa^\dagger a + u}|\alpha\rangle
\end{equation}
in which
\begin{eqnarray}
 u & = & \left( \frac{\zeta\hbar}{2m\omega} - \frac{\eta m\hbar\omega}{2} 
+ i\frac{\xi\hbar}{2} \right) = v^*  \\
 w & = & \left( \frac{\zeta\hbar}{m\omega} + \eta m\hbar\omega \right) .
\end{eqnarray}
We now proceed by using the disentangling theorem, and employing the 
completeness relation for the coherent states.
\begin{eqnarray}
 \langle x|\psi\rangle & = & \langle x|e^{(w+\chi)/2}e^{la^{\dagger 2}}
e^{\chi a^\dagger a}e^{ma^2}|\alpha\rangle \nonumber \\
 & = & \frac{1}{\pi} \int \!\!\!\! \int \langle x|\beta\rangle  
\langle \beta| e^{(w+\chi)/2}e^{la^{\dagger 2}}e^{\chi a^\dagger a}
e^{ma^2}|\alpha\rangle \; d ^2\beta \nonumber \\
 & = & \frac{1}{\pi} e^{\frac{1}{2}|\alpha|^2(|e^{2\chi}|-1) + m\alpha^2} 
e^{y + (w+k)/2} \int \!\!\!\! \int \langle x|\beta\rangle  
\langle \beta|\alpha e^{\chi}\rangle \; e^{l\beta^{*2}} \;d^2\beta  .
\end{eqnarray}
Performing the integral over the real and imaginary parts of $\alpha$, 
we obtain \begin{eqnarray}
  \langle x|\psi\rangle & = & \frac{1}{\sqrt{1+2l}} \left( \frac{2s^2}
{\pi} \right) ^{\frac{1}{4}} \!\!\! e^{-\frac{1}{2}|\alpha|^2 - m\alpha^2}
 e^{y + (w+k)/2} \nonumber \\
 & & \times \exp \left\{ -s^2x \left[ \frac{1-2l}{3-2l} \right] 
\left[ 1 + 2\frac{1-2l}{1+2l} \right] \right\} \nonumber \\
 & & \times \exp \left\{ 2sx\alpha e^\chi \left[ \frac{1}{3-2l} 
\right] \left[ 1 + 2\frac{1-2l}{1+2l} \right] \right\} \nonumber \\
 & & \times \exp \left\{\alpha^2e^{2\chi} \left[ \frac{1}{3-2l} 
\right] \left[ \frac{1}{2} + \frac{2}{1+2l} \right] \right\}
\end{eqnarray}
It is easily verified that this reduces to $\langle x|\alpha\rangle$ 
as required when we set $l=\chi=m=0$.

\chapter[Atomic Detection Probabilities]{Detection Probabilities Following Two Consecutive Atom-Cavity Interactions}
\label{app3}

We consider a composite system, consisting of a cavity and two atoms, the initial state for which is given by Eq. (\ref{init2}).  We allow the first 
atom to interact with the cavity for a time $t$,  and at some time after this 
allow the second atom to interact with the cavity for a time $s$. Using the 
interaction picture throughout the calculation, so that the Hamiltonian  for the
interaction of each atom with the field is given by the term proportional to $\kappa$ in Eq. (\ref{Hint}), the state of the system after both interactions is given by
\begin{eqnarray}
 &  & |\Psi\rangle = \;\; r_0e^{i\theta_0}|0,g,g\rangle \;\; + \;\;
(\Gamma_0^tr_0e^{i\theta_0+\phi} - 
i\Upsilon_0^tr_1e^{i\theta_1})
\; |0,e,g\rangle
\nonumber \\   
               &   & + \; (\Gamma_0^sr_0e^{i\theta_0} -
i\Upsilon_0^s\Gamma_0^tr_1e^{i\theta_1} -
\Upsilon_0^s\Upsilon_0^tr_0e^{i\theta_0+\phi}) \; \, |0,g,e\rangle 
\\
               &   & + \; (\Gamma_0^s\Gamma_0^tr_1e^{i\theta_1} -
i\Gamma_0^s\Upsilon_0^tr_0e^{i\theta_0+\phi} -
i\Upsilon_0^sr_0e^{i\theta_0}) \; |1,g,g\rangle \nonumber \\
               &   & + \sum_{n=0}^\infty \left[ 
\Gamma^s(\Gamma^tr_ne^{i\theta_n+\phi}-
i\Upsilon^tr_{n+1}e^{i\theta_{n+1}})
-i\Upsilon^s(\Gamma^t_{\raisebox{0.3mm}{\tiny $\!\! +$}}
r_{n+1}e^{i(\theta_{n+1}+\phi)} -
i\Upsilon^t_{\raisebox{0.3mm}{\tiny $\!\! +$}}
r_{n+2}e^{i\theta_{n+2}} ) \right] |n,e,e\rangle \nonumber \\ 
               &   & + \sum_{n=0}^\infty \left[
\Gamma^s(\Gamma^t_{\raisebox{0.3mm}{\tiny $\!\!
+$}}r_{n+1}e^{i(\theta_{n+1}+\phi)}-
i\Upsilon^t_{\raisebox{0.3mm}{\tiny $\!\!
+$}}r_{n+2}e^{i\theta_{n+2}})
-i\Upsilon^s(\Gamma^tr_ne^{i\theta_n+\phi}-
i\Upsilon^tr_{n+1}e^{i\theta_{n+1}} )
\right] |n \!\! + \!\! 1,e,g\rangle \nonumber \\ 
               &   & + \sum_{n=1}^\infty \left[ 
\Gamma^s(\Gamma^t_{\raisebox{0.3mm}{\tiny $\!\!
-$}}r_{n}e^{i\theta_n}-i\Upsilon^t_{\raisebox{0.3mm}{\tiny $\!\!
+$}}r_{n-1}e^{i(\theta_{n-1}+\phi)})
-i\Upsilon^s(\Gamma^tr_{n+1}e^{i\theta_{n+1}}-
i\Upsilon^tr_{n}e^{i(\theta_{n}
+\phi)} ) \right] |n,g,e\rangle \nonumber \\
               &   & + \sum_{n=1}^\infty \left[
(\Gamma^s(\Gamma^tr_{n+1}e^{i\theta_{n+1}}- 
i\Upsilon^tr_{n}e^{i(\theta_{n}
+\phi)}) -i\Upsilon^s(\Gamma^t_{\raisebox{0.3mm}{\tiny $\!\!
-$}}r_{n}e^{i\theta_n}-i\Upsilon^t_{\raisebox{0.3mm}{\tiny $\!\!
+$}}r_{n-1}e^{i(\theta_{n-1}+\phi)} ) \right] |n \!\! + \!\! 1,g,g\rangle 
\nonumber . 
\end{eqnarray}

In this expression, the labels on the kets refer respectively to the cavity photon number, the state of the first atom, and the state of the second atom. The joint probabilities for detecting $n$ photons in the cavity, and the atoms in either their ground or excited states, are given by the following expressions:
\begin{eqnarray}
  Q_0^{gg} & = & P_0 \nonumber \\
  Q_1^{gg} & = & P_0(\tilde{\Upsilon}^s_0 + 
\tilde{\Upsilon}^t_0\tilde{\Gamma}^s_0 +
2\Gamma^s_0\Upsilon^s_0\Upsilon^t_0\cos(\phi)) +
P_1\tilde{\Gamma}^s_0\tilde{\Gamma}^t_0 \nonumber \\
           &   & - 2\sqrt{P_0P_1}\Gamma^s_0\Upsilon^s_0
\Gamma^t_0 \sin(\Delta\theta_1)  \nonumber \\
           &   & + 2\sqrt{P_0P_1}\tilde{\Gamma}^s_0 \Gamma^t_0
\Upsilon^t_0 \sin(\Delta\theta_1 - \phi) \nonumber \\
  Q_{n\geq 2}^{gg} & = & 
P_n\tilde{\Gamma}^t_{\raisebox{0.3mm}{\tiny $\!\! -$}}
\tilde{\Gamma}^s_{\raisebox{0.3mm}{\tiny $\!\! -$}} +
P_{n-2}\tilde{\Gamma}^s_{\raisebox{0.3mm}{\tiny $\!\! -$}} 
\tilde{\Gamma}^s 
  \nonumber \\     &   & +  P_{n-
1}(\tilde{\Gamma}^s_{\raisebox{0.3mm}{\tiny $\!\!
-$}} \tilde{\Upsilon}^t_{\raisebox{0.3mm}{\tiny $\!\! -$}} +
\tilde{\Gamma}^t_{\raisebox{0.3mm}{\tiny $\!\! =$}} +
2\Gamma^s_{\raisebox{0.3mm}{\tiny $\!\! -$}} 
\Upsilon^s_{\raisebox{0.3mm}{\tiny
$\!\! -$}}\Gamma^t_{\raisebox{0.3mm}{\tiny $\!\! =$}}
\Upsilon^t_{\raisebox{0.3mm}{\tiny $\!\! -$}}\cos(\phi)) 
\nonumber \\
            &   & - 2\sqrt{P_nP_{n-1}}
\tilde{\Gamma}^s_{\raisebox{0.3mm}{\tiny $\!\! -$}}
\Gamma^t_{\raisebox{0.3mm}{\tiny
$\!\! -$}} \Upsilon^t_{\raisebox{0.3mm}{\tiny $\!\! -
$}}\sin(\Delta\theta_n -
\phi) \nonumber \\
            &   & - 2\sqrt{P_{n-1}P_{n-2}} 
\Gamma^t_{\raisebox{0.3mm}{\tiny
$\!\! =$}}\Upsilon^t \sin(\Delta\theta_{n-1} - \phi) \nonumber \\
                   &   & - 2\sqrt{P_{n}P_{n-1}}
\Gamma^s_{\raisebox{0.3mm}{\tiny $\!\! -$}} 
\Upsilon^s_{\raisebox{0.3mm}{\tiny
$\!\! -$}} \Gamma^t_{\raisebox{0.3mm}{\tiny $\!\! -$}}
\Gamma^t_{\raisebox{0.3mm}{\tiny $\!\! =$}} 
\sin(\Delta\theta_n)
\nonumber
\\
                   &   & - 2\sqrt{P_{n-1}P_{n-2}}
\Gamma^s_{\raisebox{0.3mm}{\tiny $\!\! -$}} 
\Upsilon^s_{\raisebox{0.3mm}{\tiny
$\!\! -$}} \Upsilon^t_{\raisebox{0.3mm}{\tiny $\!\! -$}} \Upsilon^t
\sin(\Delta\theta_n)
\nonumber
\\ 
                   &   & - 2\sqrt{P_{n}P_{n-2}}
\Gamma^s_{\raisebox{0.3mm}{\tiny $\!\! -$}} 
\Upsilon^s_{\raisebox{0.3mm}{\tiny
$\!\! -$}} \Gamma^t_{\raisebox{0.3mm}{\tiny $\!\! -$}} \Upsilon^t
\cos(\Delta\tilde{\theta}_n - \phi)
\nonumber \\ 
& & \nonumber \\
  Q_0^{ge} & = & P_0(\tilde{\Gamma}^s_0 + 
\tilde{\Upsilon}^s_0\tilde{\Upsilon}^t_0 -
2\Gamma^s_0\Upsilon^s_0\Upsilon^t_0\cos(\phi)) +
P_1\tilde{\Gamma}^t_0\tilde{\Upsilon}^s_0 \nonumber \\
           &   & + 2\sqrt{P_0P_1}\Gamma^s_0\Upsilon^s_0
\sin(\Delta\theta_1) \nonumber \\
           &   & - 2\sqrt{P_0P_1}\Gamma^t_0\Upsilon^t_0
\sin(\Delta\theta_1 - \phi) \nonumber \\
  Q_{n\geq 1}^{ge} & = & P_{n-1} \tilde{\Gamma}^s
\tilde{\Upsilon}^t_{\raisebox{0.3mm}{\tiny $\!\! +$}} + P_{n+1} 
\tilde{\Upsilon}^s
\tilde{\Gamma}^t
\nonumber \\
                   &   & + P_n(\tilde{\Gamma}^s
\tilde{\Gamma}^t_{\raisebox{0.3mm}{\tiny $\!\! -$}} + 
\tilde{\Upsilon}^s
\tilde{\Upsilon}^t - 2\Gamma^s\Upsilon^s 
\Gamma^t_{\raisebox{0.3mm}{\tiny $\!\!
-$}} \Upsilon^t \cos(\phi))   
\nonumber \\
                   &   & - 2\sqrt{P_nP_{n-1}} \tilde{\Gamma}^s
\Gamma^t_{\raisebox{0.3mm}{\tiny $\!\! -$}} 
\Upsilon^t_{\raisebox{0.3mm}{\tiny
$\!\! +$}} \sin(\Delta\theta_n-\phi)
\nonumber \\
                   &   & - 2\sqrt{P_nP_{n+1}} \tilde{\Upsilon}^s \Gamma^t
\Upsilon^t \sin(\Delta\theta_{n+1} -\phi)
\nonumber \\
                   &   & + 2\sqrt{P_nP_{n+1}} \Gamma^s\Upsilon^s
\Gamma^t_{\raisebox{0.3mm}{\tiny $\!\! -$}}
\Gamma^t \sin(\Delta\theta_{n+1})
\nonumber \\
                   &   & + 2\sqrt{P_nP_{n-1}} \Gamma^s\Upsilon^s
\Upsilon^t_{\raisebox{0.3mm}{\tiny $\!\! +$}} \Upsilon^t
\sin(\Delta\theta_n) \nonumber
\\
                   &   & + 2\sqrt{P_{n+1}P_{n-1}} \Gamma^s\Upsilon^s
\Upsilon^t_{\raisebox{0.3mm}{\tiny $\!\! +$}} \Gamma^t
\cos(\Delta\tilde{\theta}_{n+1}-\phi) \nonumber \\
& & \nonumber \\
  Q_0^{eg} & = & P_0\tilde{\Gamma}^t_0 + P_1\tilde{\Upsilon}^t_0 +
\sqrt{P_0P_1}\Gamma^t_0\Upsilon^t_0 \nonumber \\
  Q_{n\geq 1}^{eg} & = & P_{n-1} 
\tilde{\Upsilon}^s_{\raisebox{0.3mm}{\tiny $\!\!
-$}}\tilde{\Gamma}^t_{\raisebox{0.3mm}{\tiny $\!\! -$}} +
P_{n+1}\tilde{\Gamma}^s_{\raisebox{0.3mm}{\tiny $\!\! -
$}}\tilde{\Upsilon}^t
\nonumber \\
                   &   & + P_n(\tilde{\Gamma}^s_{\raisebox{0.3mm}{\tiny 
$\!\!
-$}} \tilde{\Gamma}^t + \tilde{\Upsilon}^s_{\raisebox{0.3mm}{\tiny 
$\!\! -$}}
\tilde{\Upsilon}^t_{\raisebox{0.3mm}{\tiny $\!\! -$}}
-2\Gamma^s_{\raisebox{0.3mm}{\tiny $\!\! -$}} 
\Upsilon^s_{\raisebox{0.3mm}{\tiny
$\!\! -$}} \Gamma^t \Upsilon^t_{\raisebox{0.3mm}{\tiny $\!\! -$}} 
\cos(\phi)
)    
\nonumber \\
                   &   & + 2\sqrt{P_nP_{n+1}}
\tilde{\Gamma}^s_{\raisebox{0.3mm}{\tiny $\!\! -$}} \Gamma^t 
\Upsilon^t
\sin(\Delta\theta_{n+1} - \phi)
\nonumber
\\
                   &   & + 2\sqrt{P_nP_{n-1}} 
\tilde{\Upsilon}^s_{\raisebox{0.3mm}{\tiny $\!\! -$}}
\Gamma^t_{\raisebox{0.3mm}{\tiny $\!\! -$}}
\Upsilon^t_{\raisebox{0.3mm}{\tiny $\!\! -$}} \sin(\Delta\theta_n 
- \phi)
\nonumber \\
                   &   & - 2\sqrt{P_nP_{n-1}} 
\Gamma^s_{\raisebox{0.3mm}{\tiny
$\!\! -$}} \Upsilon^s_{\raisebox{0.3mm}{\tiny $\!\! -$}} 
\tilde{\Gamma}^t
\sin(\Delta\theta_n)
\nonumber \\
                   &   & - 2\sqrt{P_nP_{n+1}} 
\Gamma^s_{\raisebox{0.3mm}{\tiny
$\!\! -$}} \Upsilon^s_{\raisebox{0.3mm}{\tiny $\!\! -$}} \Upsilon^t
\Upsilon^t_{\raisebox{0.3mm}{\tiny $\!\! -$}} 
\sin(\Delta\theta_{n+1})
\nonumber \\
                   &   & + 2\sqrt{P_{n-1}P_{n+1}} 
\Gamma^s_{\raisebox{0.3mm}{\tiny
$\!\! -$}} \Upsilon^s_{\raisebox{0.3mm}{\tiny $\!\! -$}}
\Gamma^t_{\raisebox{0.3mm}{\tiny $\!\! -$}} \Upsilon^t
\sin(\Delta\tilde{\theta}_{n+1} - \phi) \nonumber \\
& & \nonumber \\
  Q_{n\geq 0}^{ee} & = & P_n\tilde{\Gamma}^s\tilde{\Gamma}^t +
P_{n+2}\tilde{\Upsilon}^s \tilde{\Upsilon}^t_{\raisebox{0.3mm}{\tiny 
$\!\! +$}}
\nonumber \\
                   &   & + P_{n+1}(\tilde{\Gamma}^s \tilde{\Upsilon}^t +
\tilde{\Upsilon}^s \tilde{\Gamma}^t_{\raisebox{0.3mm}{\tiny $\!\! 
+$}} +
2\Gamma^s\Upsilon^s \Gamma^t_{\raisebox{0.3mm}{\tiny $\!\! 
+$}} \Upsilon^t
\cos(\phi) )\nonumber \\
                   &   & + 2\sqrt{P_nP_{n+1}}\tilde{\Gamma}^s \Gamma^t 
\Upsilon^t
\sin(\Delta\theta_{n+1} - \phi) \nonumber \\
                   &   & + 2\sqrt{P_{n+1}P_{n+2}}\tilde{\Upsilon}^s
\Gamma^t_{\raisebox{0.3mm}{\tiny $\!\! +$}} 
\Upsilon^t_{\raisebox{0.3mm}{\tiny
$\!\! +$}} \sin(\Delta\theta_{n+2} - \phi) \nonumber \\
                   &   & + 2\sqrt{P_nP_{n+1}}\Gamma^s\Upsilon^s 
\Gamma^t 
\Gamma^t_{\raisebox{0.3mm}{\tiny $\!\! +$}} 
\sin(\Delta\theta_{n+1})
\nonumber \\
                   &   & + 2\sqrt{P_{n+1}P_{n+2}}\Gamma^s\Upsilon^s 
\Upsilon^t
\Upsilon^t_{\raisebox{0.3mm}{\tiny $\!\! +$}} 
\sin(\Delta\theta_{n+2}) \nonumber
\\
                   &   & - 2\sqrt{P_nP_{n+2}}\Gamma^s\Upsilon^s 
\Gamma^t
\Upsilon^t_{\raisebox{0.3mm}{\tiny $\!\! +$}} 
\cos(\Delta\tilde{\theta}_{n+2} -
\phi)
\nonumber 
\end{eqnarray}
In the above expressions $Q_n^{ge}$, for example, denotes the joint probability for
detecting $n$ photons in the cavity and the first and second atoms in their 
ground and excited states respectively.


\end{document}